\documentclass[longauth]{aa}  

\usepackage{graphicx}
\usepackage[flushleft]{threeparttable}
\usepackage{caption}
\usepackage{xcolor}
\usepackage[varg]{txfonts}
\usepackage[breaklinks=true,colorlinks=true,urlcolor=blue,linkcolor=blue,citecolor=blue,bookmarks=true,bookmarksopen=true]{hyperref}
\usepackage{siunitx}
\usepackage{upgreek}
\usepackage{amsmath}
\usepackage{nicefrac}
\DeclareMathAlphabet{\mathbbmsl}{U}{bbm}{m}{sl}

\usepackage{amssymb}
\usepackage{subfig}

\newcommand\T{\rule{0pt}{2.6ex}}
\newcommand\B{\rule[-1.2ex]{0pt}{0pt}}

\newcommand{\Lagr}{\mathcal{L}}

\usepackage{natbib,twoopt}
\bibpunct{(}{)}{;}{a}{}{,} 					
\makeatletter
\newcommandtwoopt{\citeads}[3][][]{\href{http://adsabs.harvard.edu/abs/#3}%
{\def\hyper@linkstart##1##2{}%
\let\hyper@linkend\@empty\citetlp[#1][#2]{#3}}}
\newcommandtwoopt{\citepads}[3][][]{\href{http://adsabs.harvard.edu/abs/#3}%
{\def\hyper@linkstart##1##2{}%
\let\hyper@linkend\@empty\citep[#1][#2]{#3}}}
\newcommandtwoopt{\citetads}[3][][]{\href{http://adsabs.harvard.edu/abs/#3}%
{\def\hyper@linkstart##1##2{}%
\let\hyper@linkend\@empty\citet[#1][#2]{#3}}}
\newcommandtwoopt{\citeyearads}[3][][]%
{\href{http://adsabs.harvard.edu/abs/#3}
{\def\hyper@linkstart##1##2{}%
\let\hyper@linkend\@empty\citeyear[#1][#2]{#3}}}
\makeatother

\begin{document} 


\title{A survey of the linear polarization of directly imaged exoplanets and brown dwarf companions with SPHERE-IRDIS\thanks{Based on observations collected at the European Southern Observatory under ESO programs 098.C-0790, 0101.C-0502, 0101.C-0635, 0101.C-0855, 0102.C-0453, 0102.C-0466, 0102.C-0871, 0102.C-0916, and 0104.C-0265.}}
\subtitle{First polarimetric detections revealing disks around DH~Tau~B and GSC~6214-210~B}

\titlerunning{A survey of the linear polarization of young, directly imaged planets and brown dwarf companions}
\authorrunning{R.\,G. van Holstein et al.}

\author{
R.\,G.~van~Holstein\inst{\ref{inst:leiden},\ref{inst:esosantiago}} 
\and T.~Stolker\inst{\ref{inst:eth},\ref{inst:leiden}} 
\and R.~Jensen-Clem\inst{\ref{inst:santacruz}} 
\and C.~Ginski\inst{\ref{inst:pannekoek},\ref{inst:leiden}} 
\and J.~Milli\inst{\ref{inst:ipag}} 
\and J.~de~Boer\inst{\ref{inst:leiden}} 
\and J.\,H.~Girard\inst{\ref{inst:stsi}} 
\and Z.~Wahhaj\inst{\ref{inst:esosantiago}} 
\and A.\,J.~Bohn\inst{\ref{inst:leiden}} 
\and M.\,A.~Millar-Blanchaer\inst{\ref{inst:jpl},\ref{inst:cit}} 
\and M.~Benisty\inst{\ref{inst:ipag},\ref{inst:calan}} 
\and M.~Bonnefoy\inst{\ref{inst:ipag}} 
\and G.~Chauvin\inst{\ref{inst:ipag},\ref{inst:calan}} 
\and C.~Dominik\inst{\ref{inst:pannekoek}} 
\and S.~Hinkley\inst{\ref{inst:exeter}} 
\and C.\,U.~Keller\inst{\ref{inst:leiden}} 
\and M.~Keppler\inst{\ref{inst:maxplanck}} 
\and M.~Langlois\inst{\ref{inst:lyon}} 
\and S.~Marino\inst{\ref{inst:cambridge}} 
\and F.~M\'enard\inst{\ref{inst:ipag}} 
\and C.~Perrot\inst{\ref{inst:lesia},\ref{inst:valparaiso},\ref{inst:valparaiso_nucleo}} 
\and T.\,O.\,B.~Schmidt\inst{\ref{inst:hamburg}} 
\and A.~Vigan\inst{\ref{inst:lam}} 
\and A.~Zurlo\inst{\ref{inst:diegoportales_nucleo},\ref{inst:diegoportales}} 
\and F.~Snik\inst{\ref{inst:leiden}} 
}

\institute{
Leiden Observatory, Leiden University, PO Box 9513, 2300 RA Leiden, The Netherlands \newline
e-mail: \texttt{vanholstein@strw.leidenuniv.nl} \label{inst:leiden}
\and European Southern Observatory, Alonso de C\'{o}rdova 3107, Casilla 19001, Vitacura, Santiago, Chile \label{inst:esosantiago}
\and Institute for Particle Physics and Astrophysics, ETH Zurich, Wolfgang-Pauli-Strasse 27, 8093 Zurich, Switzerland \label{inst:eth}
\and University of California, Santa Cruz, 1156 High Street, Santa Cruz, CA 95064, USA \label{inst:santacruz}
\and Anton Pannekoek Institute for Astronomy, University of Amsterdam, Science Park 904, 1098 XH Amsterdam, The Netherlands \label{inst:pannekoek}
\and Universit\'e Grenoble Alpes, CNRS, IPAG, 38000 Grenoble, France \label{inst:ipag}
\and Space Telescope Science Institute, Baltimore 21218, MD, USA \label{inst:stsi}
\and Jet Propulsion Laboratory, 4800 Oak Grove Drive, Pasadena, CA 91109, USA \label{inst:jpl}
\and Department of Astronomy, California Institute of Technology, 1200 East California Boulevard, Pasadena, CA 91125, USA \label{inst:cit}
\and Unidad Mixta Internacional Franco-Chilena de Astronomía (CNRS, UMI 3386), Departamento de Astronom\'ia, Universidad de Chile, Camino El Observatorio 1515, Las Condes, Santiago, Chile \label{inst:calan}
\and University of Exeter, Physics Building, Stocker Road, Exeter, EX4 4QL, UK \label{inst:exeter}
\and Max Planck Institute for Astronomy, K\"onigstuhl 17, 69117 Heidelberg, Germany \label{inst:maxplanck}
\and Universit\'e de Lyon, Universit\'e Lyon1, ENS de Lyon, CNRS, Centre de Recherche Astrophysique de Lyon UMR 5574, 69230 Saint-Genis-Laval, France \label{inst:lyon}
\and Institute of Astronomy, University of Cambridge, Madingley Road, Cambridge CB3 0HA, UK \label{inst:cambridge}
\and LESIA, Observatoire de Paris, Universit\'e PSL, CNRS, Sorbonne Universit\'e, Universit\'e Paris Diderot, Sorbonne Paris Cit\'e, 5 place Jules Janssen, 92195 Meudon, France \label{inst:lesia}
\and Instituto de F\'isica y Astronom\'ia, Facultad de Ciencias, Universidad de Valpara\'iso, Av. Gran Breta\~na 1111, Valpara\'iso, Chile \label{inst:valparaiso}
\and N\'ucleo Milenio Formaci\'on Planetaria -- NPF, Universidad de Valpara\'iso, Av. Gran Breta\~na 1111, Valpara\'iso, Chile \label{inst:valparaiso_nucleo}
\and Hamburger Sternwarte, Gojenbergsweg 112, D-21029 Hamburg, Germany \label{inst:hamburg}
\and Aix Marseille Univ, CNRS, CNES, LAM, Marseille, France \label{inst:lam}
\and N\'ucleo de Astronom\'ia, Facultad de Ingenier\'ia, Universidad Diego Portales, Av. Ejercito 441, Santiago, Chile \label{inst:diegoportales_nucleo}
\and Escuela de Ingenier\'ia Industrial, Facultad de Ingenier\'ia y Ciencias, Universidad Diego Portales, Av. Ejercito 441, Santiago, Chile \label{inst:diegoportales}
}


\date{Received 30 August 2020 / Accepted 18 December 2020}

\abstract{
Young giant planets and brown dwarf companions emit near-infrared radiation that can be linearly polarized up to several percent. 
This polarization can reveal the presence of an (unresolved) circumsubstellar accretion disk, rotation-induced oblateness of the atmosphere, or an inhomogeneous distribution of atmospheric dust clouds.
}
{
We aim to measure the near-infrared linear polarization of 20 known directly imaged exoplanets and brown dwarf companions.
}
{
We observed the companions with the high-contrast imaging polarimeter SPHERE-IRDIS at the Very Large Telescope.
We reduced the data using the IRDAP pipeline to correct for the instrumental polarization and crosstalk of the optical system with an absolute polarimetric accuracy ${<}0.1\%$ in the degree of polarization.
We employed aperture photometry, angular differential imaging, and point-spread-function fitting to retrieve the polarization of the companions.
}
{
We report the first detection of polarization originating from substellar companions, with a polarization of several tenths of a percent for DH~Tau~B and GSC~6214-210~B in $H$-band.
By comparing the measured polarization with that of nearby stars, we find that the polarization is unlikely to be caused by interstellar dust.
Because the companions have previously measured hydrogen emission lines and red colors, the polarization most likely originates from circumsubstellar disks.
Through radiative transfer modeling, we constrain the position angles of the disks and find that the disks must have high inclinations.
For the 18 other companions, we do not detect significant polarization and place subpercent upper limits on their degree of polarization.
We also present images of the circumstellar disks of DH~Tau, GQ~Lup, PDS~70, $\beta$~Pic, and HD~106906. 
We detect a highly asymmetric disk around GQ~Lup and find evidence for multiple scattering in the disk of PDS~70. Both disks show spiral-like features that are potentially induced by GQ~Lup~B and PDS~70~b, respectively.
}
{
The presence of the disks around DH~Tau~B and GSC~6214-210~B as well as the misalignment of the disk of DH~Tau~B with the disk around its primary star suggest in situ formation of the companions. 
The non-detections of polarization for the other companions may indicate the absence of circumsubstellar disks, a slow rotation rate of young companions, the upper atmospheres containing primarily submicron-sized dust grains, and/or limited cloud inhomogeneity.
}


\maketitle

%
%

\section{Introduction}
\label{sec:introduction}

Understanding the formation and evolution of young, self-luminous exoplanets and brown dwarf companions is one of the main goals of high-contrast imaging at near-infrared wavelengths~\citep[e.g.,][]{nielsen_gpies, vigan_shine}.
Only a few of these directly imaged substellar companions have been detected close to the parent star and within a circumstellar disk~\citep[e.g.,][]{lagrange_betapic, keppler_pds70, haffert_pds70}; most companions are found at much larger separations (${\gtrsim}100$~au; see e.g.,~\citealp{bowler_exoplanetimaging}).
Close-in planets and companions are generally believed to form through core accretion~\citep{pollack_coreaccretion, alibert_coreaccretion} or gravitational instabilities in the circumstellar disk~\citep{cameron_gi, boss_gi1}.
Companions at larger separations may form through direct collapse in the molecular cloud~\citep{bate_corecollapse} or disk gravitational instabilities at an early stage~\citep{kratter_fragmentation}.
Alternatively, companions may form close to the star and subsequently scatter to wide orbits through dynamical encounters with other companions~\citep[e.g.,][]{veras_scattering}.

In all formation scenarios, the companion is generally expected to form its own circumsubstellar accretion disk~\citep[e.g.,][]{stamatellos_fragmentation, szulagyi_cpdformation}.
Indeed, a handful of substellar companions show evidence for the presence of an accretion disk through hydrogen emission lines, red near-infrared colors, and excess emission at mid-infrared wavelengths~\citep[e.g.,][]{seifahrt_gqlup, bowler_gsc6214nir, bailey_companions, kraus_companions, zhou_pmcaccretion, haffert_pds70}.
Interestingly, whereas ALMA and other radio interferometers have been successful at detecting the dust and gas of disks around isolated substellar objects~\citep[e.g.,][]{ricci_browndwarfdisks, vanderplas_browndwarfdisks, bayo_planetdisk}, attempts to detect such disks around substellar companions have almost exclusively yielded non-detections~\citep{bowler_gsc6214, macgregor_gqlup, wu_radiodisk, wu_gqlupdisk, wolff_dhtau, ricci_2m1207, perez_planetdisk, wu_pmcdisk}.
The only detection of a disk around a substellar companion at mm-wavelengths is that of PDS~70~c with ALMA by~\citet{isella_pds70}.
ALMA has also detected a disk around FW~Tau~C~\citep{kraus_fwtau, caceres_fwtau}, but, from models of the Keplerian rotation of the gas, the companion appears to be a ${\sim}0.1M_\odot$ star~\citep{wu_fwtau, more_fwtau}.
To explain their non-detections, \citet{wu_radiodisk} and \citet{wu_pmcdisk} suggest that the disks around substellar companions must be very compact (${\lesssim}1000~R_\mathrm{Jup}$ or ${\lesssim}0.5$~au) and optically thick to be able to sustain several million years of accretion.
Alternatively, there might be a dearth of large dust grains in circumsubstellar disks because the observed mid-infrared excess could also be explained by a gaseous disk with small micron-sized dust grains.

Although compact circumsubstellar disks cannot be spatially resolved with current 8-m class telescopes, they can create a measurable, integrated linear polarization at near-infrared wavelengths~\citep{stolker_exopol}.
The polarization can be introduced through scattering of the companion's thermal photons by dust within the disk, (partial) obscuration of the companion's atmosphere by the disk, or self-scattering in the case of a high-temperature disk.
In all cases, the disk must have a nonzero inclination because the polarization of a face-on viewed, rotationally symmetric disk integrates to zero and a low-inclination disk cannot obscure the companion's atmosphere.
Measuring polarization originating from circumsubstellar disks enables us to study the structure and physical properties of the disks.

Planets and brown dwarf companions without a disk can also be linearly polarized at near-infrared wavelengths.
Late-M- to mid-L-type dwarfs are expected to have dusty atmospheres because their temperatures are sufficiently low for refractory material to condense~\citep{allard_bddust, ackerman2001}.
This atmospheric dust scatters the thermal radiation emanating from within the object, linearly polarizing the light.
Whereas the spatially integrated polarization signal of a spherical, horizontally homogeneous dusty atmosphere is zero, a net polarization remains when this symmetry is broken~\citep{sengupta_bdpol1}.
Examples of these asymmetries are rotation-induced oblateness and an inhomogeneous distribution of atmospheric dust clouds~\citep{sengupta_bdpol2, dekok_exopol, marley_exopol, stolker_exopol}, or even a large transiting moon~\citep{sengupta_exomoonpol}. 
Based on the models, the degree of linear polarization due to circumsubstellar disks and atmospheric asymmetries can be several tenths of a percent up to several percent in favorable cases.

Spatially unresolved polarimetric observations have already been used to study disks around pre-main sequence stars~\citep[e.g.,][]{rostopchina1997, bouvier1999, grinin2000, menard2003}.
In addition, optical and near-infrared polarization has been detected for dozens of field brown dwarfs~\citep{menard_bdpol, zapatero_bdpol1, tata_bdpol, zapatero_bdpol2, milespaez_bdpol1, milespaez_bdpol2}.
In most cases, the polarization of these brown dwarfs is interpreted as being caused by rotation-induced oblateness or circumsubstellar disks, whereas an inhomogeneous cloud distribution has appeared harder to prove.
However, \citet{millar_luhman16} recently measured the near-infrared polarization of the two L/T transition dwarfs of the Luhman~16 system and found evidence for banded clouds on the hotter, late-L-type object.

With the adaptive-optics-fed high-contrast imaging instruments Gemini Planet Imager~\citep[GPI; ][]{macintosh_gpi} and SPHERE-IRDIS~\citep{beuzit_sphere, dohlen_irdis} at the Very Large Telescope (VLT), we now have access to the spatial resolution and sensitivity required to measure the near-infrared polarization of substellar companions at small separations.
After correction for instrumental polarization effects, the polarimetric modes of both instruments can reach absolute polarimetric accuracies of ${\lesssim}0.1$\% in the degree of polarization~\citep{wiktorowicz_gpicalib, millar_gpicalib, vanholstein_irdis2}.
Early attempts to measure the polarization of substellar companions by \citet{millar_betapic} and \citet{jensenclem_exopol1} with GPI and by \cite{vanholstein_exopol} with SPHERE-IRDIS have been unsuccessful.
Nevertheless, \citet{vanholstein_exopol} showed that SPHERE-IRDIS can achieve a polarimetric sensitivity close to the photon noise limit at angular separations ${>}0.5\arcsec$. 
\citet{ginski_cscha} detected a companion to CS~Cha using SPHERE-IRDIS and measured the companion's polarization to be 14\%, suggesting that it is surrounded by a highly inclined and vertically extended disk. 
However, recent optical spectroscopic observations with MUSE show that the companion is not substellar in nature, but is a mid M-type star that is obscured by its disk~\citep{haffert_cscha}.

In this paper, we present the results of a survey of 20 planetary and brown dwarf companions with SPHERE-IRDIS, aiming to detect linear polarization originating from both circumsubstellar disks and atmospheric asymmetries.
Our study is complemented by a similar survey of seven companions using GPI and SPHERE by~\citet{jensenclem_exopol2}. 

The outline of this paper is as follows.
In Sect.~\ref{sec:target_sample_observations} we present the sample of companions and the observations.
Subsequently, we describe the data reduction in Sect.~\ref{sec:data_reduction} and explain the extraction of the polarization signals in Sect.~\ref{sec:method_dh_tau}.
In Sect.~\ref{sec:results} we discuss our detections of polarization and the upper limits on the polarization for the non-detections.
In the same section, we present images of five circumstellar disks that we detected in our survey. 
Because the most plausible explanation for the polarization of the companions is the presence of circumsubstellar disks, we perform radiative transfer modeling of a representative example of such a disk in Sect.~\ref{sec:modeling_disks}. 
Finally, we discuss the implications of our measurements in Sect.~\ref{sec:discussion} and present conclusions in Sect.~\ref{sec:conclusions}.



%
%

\section{Target sample and observations}
\label{sec:target_sample_observations}

\subsection{Target sample}
\label{sec:target_sample}

The sample of this study consists of 20 known directly imaged planetary and brown dwarf companions, out of the approximately 140 such companions that are currently known\footnote{From The Extrasolar Planets Encyclopaedia, \url{http://exoplanet.eu}, \citep{schneider_exoplaneteu}, consulted on January 5, 2021.}.
Because the expected polarization of the companions is around a few tenths of a percent or less, our primary selection criterion was whether SPHERE-IRDIS can reach a high signal-to-noise ratio (S/N) in total intensity without requiring an excessive amount of observing time. Therefore, the selected companions are relatively bright, are at a moderate companion-to-star contrast, are at a large angular separation from the star, and/or have a bright star for good adaptive-optics (AO) performance~\citep[see][]{vanholstein_exopol}. Our sample contains the majority of the approximately two dozen known companions that match these requirements. 
Three of the remaining companions have been observed by~\citet{jensenclem_exopol2} in their survey of seven companions. 

\begin{figure}[!htbp]
\centering
\includegraphics[width=\hsize, trim={8 15 5 0}]{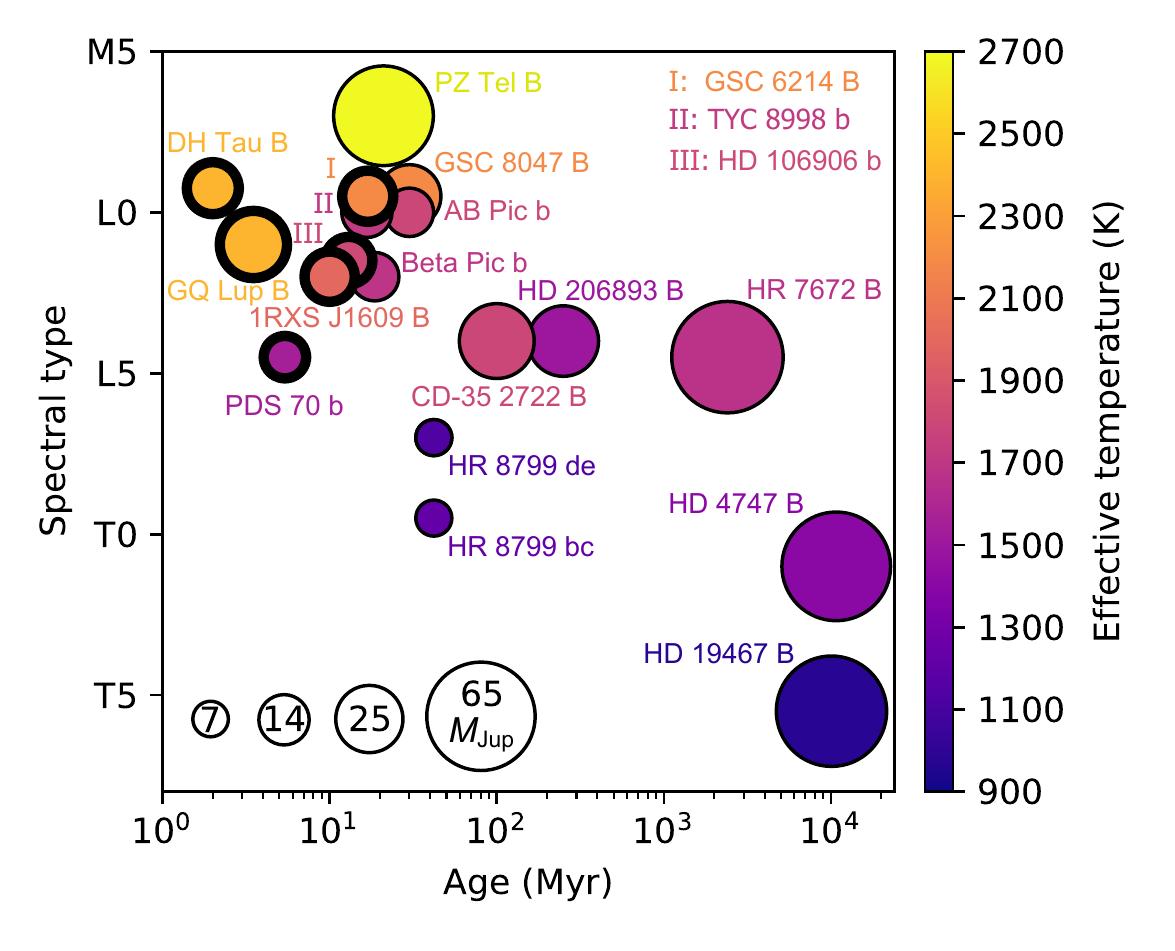} 
\caption{Properties of the companions of our sample showing the age, spectral type, mass (surface area of data points), effective temperature, and possible existence of a circumsubstellar disk (thick border). The data points of HR~8799~b and c, and of HR~8799~d and e, overlap.}
\label{fig:companions_properties} 
\end{figure}

An overview of the properties of the companions of our sample is shown in Fig.~\ref{fig:companions_properties}, with the full details presented in Table~\ref{tab:companions_properties}\footnote{Throughout this paper we use the short names GSC~8047, GSC~6214, 1RXS~J1609, and TYC~8998 for the stars GSC~08047-00232, GSC~06214-00210 (or GSC~6214-210), 1RXS~J160929.1-210524, and TYC~8998-760-1, respectively.}.
The sample is diverse, with the companions spanning spectral types from T5.5 to M7, masses between approximately 6 and 70~$M_\mathrm{Jup}$, and ages between approximately 2~Myr and 11~Gyr. The companions orbit stars of spectral types A5 to M1. Six companions show evidence of hosting a circumsubstellar disk, mostly in the form of red near-infrared colors, excess emission at mid-infrared wavelengths, and hydrogen emission lines that reveal ongoing accretion. As can be seen particularly well from Fig.~\ref{fig:companions_properties}, the overall sample ranges from young, hot, accreting companions with spectral types between late M and early L, to old, cold, and massive companions of later spectral types. For the six companions that show evidence of hosting a circumsubstellar disk, we expect any polarization to be primarily due to this (spatially unresolved) disk, whereas for the other companions polarization would most likely be due to an inhomogeneous cloud distribution or rotation-induced oblateness.
%
%
%
\begin{table*}[!htbp]
\caption{Properties of the companions of our sample.} 
\centering
\addtolength{\tabcolsep}{-2pt}
\resizebox{\textwidth}{!}{%
\begin{threeparttable}
\begin{tabular}{l c l r @{\hspace{3pt}} l c c c c c c c} 
\hline\hline 
\T\B 
Target &
\begin{tabular}{@{}c@{}} $d$ \T \\ (pc) \B \end{tabular} &
\begin{tabular}{@{}c@{}} SpT \T \\ star \B \end{tabular} &
\multicolumn{2}{c}{~~~~~Age} &
\begin{tabular}{@{}c@{}} $\rho$ \T \\ ($\arcsec$) \B \end{tabular} &
\begin{tabular}{@{}c@{}} Mass \T \\ ($M_\mathrm{Jup}$) \B \end{tabular} &
\begin{tabular}{@{}c@{}} SpT \T \\ comp. \B \end{tabular} &
$T_\mathrm{eff}$ (K) &
$\log g$ &
\begin{tabular}{@{}c@{}} Evid. \T \\ CSD \B \end{tabular} &
References \\
\hline 
HR 8799 b & \phantom{0}41.2 & A5 & $42^{+6}_{-4}$ & Myr & 1.7 & \phantom{1}5.8 $\pm$ 0.5\phantom{0} & ${\sim}$L/T\phantom{${\sim}$} & 1175 $\pm$ 125\phantom{0} & ${\sim}3.5$ & - & 1,2,3\T \\
HR 8799 c & \phantom{0}41.2 & A5 & $42^{+6}_{-4}$ & Myr & 0.9 & \phantom{0}$7.2^{+0.6}_{-0.7}$\phantom{0} & ${\sim}$L/T\phantom{${\sim}$} & 1225 $\pm$ 125\phantom{0} & 3.5 \hspace{2pt}-\hspace{2pt} 3.9 & - & 1,2,3 \\
HR 8799 d & \phantom{0}41.2 & A5 & $42^{+6}_{-4}$ & Myr & 0.7 & \phantom{0}$7.2^{+0.6}_{-0.7}$\phantom{0} & L7 $\pm$ 1\phantom{L} & 1200 $\pm$ 100\phantom{0} & 3.0 \hspace{2pt}-\hspace{2pt} 4.5 & - & 1,2,3 \\
HR 8799 e & \phantom{0}41.2 & A5 & $42^{+6}_{-4}$ & Myr & 0.4 & \phantom{0}$7.2^{+0.6}_{-0.7}$\phantom{0} & L7 $\pm$ 1\phantom{L} & 1150 $\pm$ 50\phantom{00} & 4.3 $\pm$ 0.3 & - & 1,2,3,4 \\
PZ Tel B & \phantom{0}47.0 & G9 & $21 \pm 4$ & Myr & 0.5 & \phantom{.0}38 \hspace{2pt}-\hspace{2pt} 72\phantom{.0} & M7 $\pm$ 1\phantom{M} & 2700 $\pm$ 100\phantom{0} & ${<}\hspace{1.5pt}4.5$ & - & 5,6 \\
HR 7672 B & \phantom{0}17.7 & G0 & $2.4^{+0.6}_{-0.7}$ & Gyr & 0.8 & $68.7^{+2.4}_{-3.1}$\phantom{0} & L4.5 $\pm$ 1.5\phantom{L} & 1510 \hspace{2pt}-\hspace{2pt} 1850 & 5.0 \hspace{2pt}-\hspace{2pt} 5.5 & - & 7,8,9,10 \\
GSC 8047 B & \phantom{0}86.0 & K2 & ${\sim}30$ & Myr & 3.2 & \phantom{0.}$22^{+4\phantom{.0}}_{-7\phantom{.0}}$\phantom{0} & M9.5 $\pm$ 0.5\phantom{M} & 2200 $\pm$ 100\phantom{0} & 4.0 $\pm$ 0.5 & - & 5,11,12,13 \\
HD 19467 B & \phantom{0}32.0 & G3 & 10 $\pm$ 1 & Gyr & 1.6 & $67.4^{+0.9}_{-1.5}$\phantom{0} & T5.5 $\pm$ 1\phantom{.5T} & $\phantom{0}978^{+20\phantom{0}}_{-43\phantom{0}}$ & ${\sim}5$\phantom{.0} & - & 14,15,16 \\
GQ Lup B & 151.2 & K7 & 2 - 5 & Myr & 0.7 & \hspace{1.5pt}${\sim}$10 \hspace{2pt}-\hspace{2pt} 40\phantom{.0} & L1 $\pm$ 1\phantom{L} & 2400 $\pm$ 100\phantom{0} & 4.0 $\pm$ 0.5 & H,N & 17,18,19,20 \\
HD 206893 B & \phantom{0}40.8 & F5 & $250^{+450}_{-200}$ & Myr & 0.3 & \phantom{.0}15 \hspace{2pt}-\hspace{2pt} 40\phantom{.0} & L3 - L5 & 1300 \hspace{2pt}-\hspace{2pt} 1700 & 3.5 \hspace{2pt}-\hspace{2pt} 5.0 & - & 7,21,22 \\
HD 4747 B & \phantom{0}18.8 & G9 & 11 $\pm$ 7 & Gyr & 0.6 & $65.3^{+4.4}_{-3.3}$\phantom{0} & T1 $\pm$ 2\phantom{T} & $1407^{+134}_{-140}$ & $5.2^{+0.5}_{-0.6}$ & - & 15,23,24 \\
CD-35 2722 B & \phantom{0}22.4 & M1 & 100 $\pm$ 50 & Myr & 3.1 & \phantom{.0}31 $\pm$ 8\phantom{.00} & L4 $\pm$ 1\phantom{L} & 1700 \hspace{2pt}-\hspace{2pt} 1900 & 4.5 $\pm$ 0.5 & - & 5,25 \\
AB Pic b & \phantom{0}50.0 & K1 & ${\sim}30$ & Myr & 5.5 & \phantom{0.}$13^{+1\phantom{.0}}_{-2\phantom{.0}}$\phantom{0} & L0 $\pm$ 1\phantom{L} & $1800^{+100}_{-200}$ & 4.5 $\pm$ 0.5 & - & 5,13,26,27 \\
HD 106906 b & 103.0 & F5 & 13 $\pm$ 2 & Myr & 7.1 & 12.5 $\pm$ 1.5\phantom{0} & L1.5 $\pm$ 1.0\phantom{L} & 1820 $\pm$ 240\phantom{0} & ${\sim}3.5$ & N,P & 28,29,30 \\
GSC 6214 B & 108.5 & K5 & $17^{+2}_{-3}$ & Myr & 2.2 & 14.5 $\pm$ 2.0\phantom{0} & M9.5 $\pm$ 1\phantom{.9M} & 2200 $\pm$ 100\phantom{0} & \ldots & H,N,M & 19,31 \\
PDS 70 b & 113.0 & K7 & $5.4 \pm 1.0$ & Myr & 0.2 & ${\sim}10$ & ${\sim}$L\phantom{${\sim}$} & 1500 \hspace{2pt}-\hspace{2pt} 1600 & ${\sim}4$\phantom{.0} & H & 32,33,34,35,36 \\
1RXS J1609 B & 139.1 & M0 & ${\sim}10$ & Myr & 2.2 & 14.0 $\pm$ 1.5\phantom{0} & L2 $\pm$ 1\phantom{L} & 2000 $\pm$ 100\phantom{0} & ${\sim}4$\phantom{.0} & N,M,$A_V$ & 19,37,38 \\
DH Tau B & 134.8 & M1 & ${\sim}2$ & Myr & 2.3 & \phantom{0.}$15^{+7\phantom{.0}}_{-4\phantom{.0}}$\phantom{0} & M9.25 $\pm$ 0.25 & 2400 $\pm$ 100\phantom{0} & 3.5 $\pm$ 0.5 & H,N & 13,19,39 \\
$\beta$~Pic b & \phantom{0}19.7 & A6 & $18.5^{+2.0}_{-2.4}$ & Myr & 0.3 & 13 $\pm$ 3 & L2 $\pm$ 1\phantom{L} & 1694 $\pm$ 40\phantom{00} & $4.17^{+0.10}_{-0.13}$ & - & 40,41,42,43,44 \\
TYC 8998 b & \phantom{0}94.6 & K3 & $16.7\pm1.4$ & Myr & 1.7 & 14 $\pm$ 3 & ${\sim}$L0\phantom{${\sim}$} & $1727^{+172}_{-127}$ & $3.91^{+1.59}_{-0.41}$ & - & 32,45 \B \\
\hline 
\end{tabular}
\begin{tablenotes}
\vspace{5pt}
\item \hspace{-3pt}\textbf{Notes.} $d$ is the distance from Earth, SpT stands for spectral type, $\rho$ is the approximate angular separation of the companion from the host star at the time of observation, $T_\mathrm{eff}$ is the effective temperature, and $\log g$ is the surface gravity. The second column from the right indicates the evidence for the existence of a circumsubstellar disk (CSD), which includes hydrogen emission lines (H), red near-infrared colors (N), excess emission at mid-infrared wavelengths (M), a radially extended point spread function in Hubble Space Telescope images (P), and significant extinction by dust ($A_V$). HR~7672~B, HD~19467~B, HD~4747~B, and $\beta$~Pic~b have also been observed by~\citet{jensenclem_exopol2}.
\vspace{5pt}
\item \hspace{-3pt}\textbf{References.} Distances from Gaia DR2~\citep{gaia_dr2, bailerjones_gaia}. Other properties from: (1)~\citet{star_spt1}, (2)~\citet{wang_hr8799}, (3)~\citet{bonnefoy_hr8799}, (4)~\citet{gravity_hr8799}, (5)~\citet{star_spt2}, (6)~\citet{maire_pztel}, (7)~\citet{star_spt3}, (8)~\citet{crepp_hr7672}, (9)~\citet{liu_hr7672}, (10)~\citet{boccaletti_hr7672}, (11)~\citet{chauvin_gsc8047-232b}, (12)~\citet{ginski_gsc8047-232b_orbit}, (13)~\citet{bonnefoy_planetspectra}, (14)~\citet{crepp_hd19467}, (15)~\citet{wood_hd4747_hd19467age}, (16)~\citet{crepp_hd19467_2}, (17)~\citet{kharchenko_vizier}, (18)~\citet{donati_gqlup}, (19)~\citet{wu_radiodisk} and references therein, (20)~\citet{wu_gqlupdisk} and references therein, (21)~\citet{delorme_hd206893}, (22)~\citet{milli_hd206893}, (23)~\citet{montes_spt}, (24)~\citet{crepp_hd4747_2}, (25)~\citet{wahhaj_cd352722}, (26)~\citet{bonnefoy_abpic}, (27)~\citet{chauvin_abpic}, (28)~\citet{star_spt6}, (29)~\citet{kalas_hd106906}, (30)~\citet{daemgen_hd106906}, (31)~\citet{pearce_gsc6214}, (32)~\citet{pecaut_scocen}, (33)~\citet{muller_pds70}, (34)~\citet{keppler_pds70}, (35)~\citet{christiaens_pds70}, (36)~\citet{haffert_pds70}, (37)~\citet{star_spt8}, (38)~\citet{wu_1rxsj1609}, (39)~\citet{star_spt5}, (40)~\citet{star_spt9}, (41)~\citet{miretroig_betapic}, (42)~\citet{stolker_miracles}, (43)~\citet{chilcote_betapic2}, (44)~\citet{dupuy_betapic}, (45)~\citet{bohn_yses_companion}.
\end{tablenotes}
\end{threeparttable}}
\label{tab:companions_properties} 
\end{table*} 
%

%
%

\subsection{Observations}
\label{sec:observations}

All our observations were performed with the dual-beam polarimetric imaging (DPI) mode of SPHERE-IRDIS~\citep{deboer_irdis1, vanholstein_irdis2}. In this mode, linear polarizers are inserted in the left and right optical channels of IRDIS to simultaneously create images of the two orthogonal linear polarization states on the detector. A rotatable half-wave plate (HWP) modulates the incident linear polarization with switch angles $0^\circ$, $45^\circ$, $22.5^\circ$, and $65.5^\circ$ (a HWP cycle) to measure Stokes $Q$ and $U$. The observations were carried out between October 10, 2016, and February 16, 2020, under generally good to excellent atmospheric conditions. An overview of the observations is shown in Table~\ref{tab:observations}.
%
\begin{table*}[!htbp]
\caption{Overview of the observations performed.} 
\centering 
\begin{threeparttable}
\begin{tabular}{l l c c c c c c c c} 
\hline\hline 
\T\B Target & \hspace{13pt}Date & \begin{tabular}{@{}c@{}} Tracking \T \\ mode \B \end{tabular} & Filter & DIT (s) & NDIT & $t_\mathrm{exp}$ (min) & \begin{tabular}{@{}c@{}} Parallactic \T \\ rotation (\si{\degree}) \B \end{tabular} & Seeing (") & \begin{tabular}{@{}c@{}} Coherence \T \\ time (ms) \B \end{tabular} \\
\hline 
HR 8799 & 2016-10-11 & Pupil & BB\_H\phantom{$_\mathrm{s}$} & 16 & 3 & 137.6 & 50.5 & 0.41 - 0.93 & \phantom{1}2.4 - \phantom{1}6.1\hspace{-3pt} \T \\
PZ Tel & 2016-10-10 & Pupil & BB\_H\phantom{$_\mathrm{s}$} & 12 & 4 & \phantom{1}32.0 & 14.3 & 0.57 - 1.21 & \phantom{1}3.6 - \phantom{1}6.4 \\
 & 2016-10-12 & Pupil & BB\_J\phantom{$_\mathrm{ss}$} & 12 & 4 & \phantom{1}32.0 & 12.9 & 0.86 - 1.24 & \phantom{1}2.8 - \phantom{1}6.1 \\
HR 7672 & 2018-06-08 & Pupil & BB\_H\phantom{$_\mathrm{s}$} & \phantom{1}4 & 1 & \phantom{1}12.8 & \phantom{1}9.8 & 0.51 - 0.68 & \phantom{1}3.5 - \phantom{1}5.9 \\
 & 2018-07-13 & Pupil & BB\_H\phantom{$_\mathrm{s}$} & \phantom{1}4 & 1 & \phantom{1}12.8 & 11.0 & 0.36 - 0.46 & \phantom{1}7.2 - 11.0 \\
 & 2018-07-14 & Pupil & BB\_H\phantom{$_\mathrm{s}$} & \phantom{1}4 & 1 & \phantom{1}12.8 & 10.9 & 0.44 - 0.56 & 11.7 - 15.2 \\
GSC 8047 & 2018-08-07 & Pupil & BB\_H\phantom{$_\mathrm{s}$} & 64 & 1 & \phantom{1}42.7 & 13.0 & 0.40 - 0.65 & \phantom{1}3.9 - \phantom{1}6.8 \\
 & 2018-08-09A & Pupil & BB\_H\phantom{$_\mathrm{s}$} & 64 & 1 & \phantom{1}42.7 & 13.7 & 0.43 - 0.82 & \phantom{1}3.1 - \phantom{1}7.3 \\
 & 2018-08-09B & Pupil & BB\_H\phantom{$_\mathrm{s}$} & 32 & 2 & \phantom{1}38.4 & 17.7 & 0.39 - 0.56 & \phantom{1}4.0 - \phantom{1}9.2 \\
HD 19467 & 2018-08-07 & Field & BB\_H\phantom{$_\mathrm{s}$} & 12 & 1 & \phantom{1}25.6 &  & 0.47 - 0.66 & \phantom{1}2.2 - \phantom{1}3.9 \\
 & 2018-08-10A & Field & BB\_H\phantom{$_\mathrm{s}$} & 12 & 1 & \phantom{1}25.6 &  & 0.42 - 0.53 & \phantom{1}6.4 - 11.2 \\
 & 2018-08-10B & Field & BB\_H\phantom{$_\mathrm{s}$} & 12 & 1 & \phantom{1}32.0 &  & 0.52 - 0.78 & \phantom{1}4.3 - \phantom{1}9.5 \\
GQ Lup & 2018-08-15 & Pupil & BB\_H\phantom{$_\mathrm{s}$} & 32 & 1 & \phantom{1}38.4 & \phantom{1}6.0 & 0.48 - 0.72 & \phantom{1}3.9 - \phantom{1}7.9 \\
HD 206893 & 2018-09-06 & Pupil & BB\_K$_\mathrm{s}$ & 32 & 1 & \phantom{1}36.3 & 31.6 & 0.46 - 0.64 & \phantom{1}6.5 - 10.4 \\
 & 2018-09-08 & Pupil & BB\_K$_\mathrm{s}$ & 32 & 1 & \phantom{1}40.5 & 39.3 & 0.48 - 0.84 & 11.2 - 19.5 \\
HD 4747 & 2018-09-10 & Pupil & BB\_K$_\mathrm{s}$ & 12 & 1 & \phantom{1}25.6 & \phantom{1}1.1 & 1.19 - 1.77 & \phantom{1}2.0 - \phantom{1}3.5 \\
 & 2018-09-11 & Pupil & BB\_K$_\mathrm{s}$ & 12 & 1 & \phantom{1}25.6 & \phantom{1}1.2 & 0.53 - 0.75 & \phantom{1}2.2 - \phantom{1}4.4 \\
CD-35 2722 & 2018-11-22 & Pupil & BB\_H\phantom{$_\mathrm{s}$} & 16 & 1 & \phantom{1}16.0 & \phantom{1}3.3 & 0.56 - 0.68 & \phantom{1}2.7 - \phantom{1}5.1 \\
AB Pic & 2019-01-12 & Field & BB\_H\phantom{$_\mathrm{s}$} & 32 & 1 & \phantom{1}46.9 &  & 0.59 - 0.91 & \phantom{1}2.7 - \phantom{1}5.3 \\
HD 106906 & 2019-01-17 & Field & BB\_H\phantom{$_\mathrm{s}$} & 32 & 1 & \phantom{1}29.9 &  & 0.40 - 0.86 & \phantom{1}5.1 - 11.8 \\
 & 2019-01-18 & Field & BB\_H\phantom{$_\mathrm{s}$} & 32 & 1 & \phantom{1}29.9 &  & 0.40 - 0.96 & \phantom{1}8.4 - 14.4 \\
 & 2019-01-20 & Field & BB\_H\phantom{$_\mathrm{s}$} & 32 & 1 & \phantom{1}29.9 &  & 0.44 - 0.78 & 11.5 - 16.7 \\
 & 2019-01-26 & Field & BB\_H\phantom{$_\mathrm{s}$} & 32 & 1 & \phantom{1}29.9 &  & 0.36 - 0.48 & 13.9 - 20.1 \\
GSC 6214 & 2019-02-22 & Pupil & BB\_H\phantom{$_\mathrm{s}$} & 32 & 1 & \phantom{1}29.9 & \phantom{1}1.3 & 0.43 - 0.99 & 11.2 - 21.0 \\
 & 2019-08-06 & Pupil & BB\_H\phantom{$_\mathrm{s}$} & 32 & 1 & \phantom{1}33.1 & \phantom{1}1.6 & 0.34 - 0.53 & \phantom{1}5.1 - 11.4 \\
 & 2019-08-07 & Pupil & BB\_H\phantom{$_\mathrm{s}$} & 32 & 1 & \phantom{1}29.9 & \phantom{1}0.8 & 0.43 - 0.58 & \phantom{1}5.8 - \phantom{1}8.8 \\
PDS 70 & 2019-07-12 & Pupil & BB\_K$_\mathrm{s}$ & 64 & 1 & 135.5 & 85.2 & 0.37 - 0.79 & \phantom{1}2.8 - \phantom{1}5.4 \\
 & 2019-08-09 & Pupil & BB\_H\phantom{$_\mathrm{s}$} & 64 & 1 & \phantom{1}38.4 & 13.5 & 1.28 - 1.67 & \phantom{1}1.8 - \phantom{1}2.5 \\
1RXS J1609 & 2019-08-06 & Pupil & BB\_H\phantom{$_\mathrm{s}$} & 32 & 1 & \phantom{1}29.9 & \phantom{1}1.5 & 0.33 - 0.50 & \phantom{1}8.0 - 13.2 \\
 & 2019-08-29 & Field & BB\_H\phantom{$_\mathrm{s}$} & 64 & 1 & \phantom{1}46.9 &  & 0.55 - 0.81 & \phantom{1}2.6 - \phantom{1}3.6 \\
 & 2019-08-31 & Field & BB\_H\phantom{$_\mathrm{s}$} & 64 & 1 & \phantom{1}11.7 &  & 0.89 - 1.13 & \phantom{1}2.2 - \phantom{1}3.0 \\
 & 2019-09-17A & Field & BB\_H\phantom{$_\mathrm{s}$} & 64 & 1 & \phantom{1}12.8 &  & 0.58 - 0.73 & \phantom{1}3.4 - \phantom{1}4.1 \\
 & 2019-09-17B & Field & BB\_H\phantom{$_\mathrm{s}$} & 64 & 1 & \phantom{1}38.4 &  & 0.52 - 0.80 & \phantom{1}2.7 - \phantom{1}3.9 \\
 & 2019-09-23 & Field & BB\_H\phantom{$_\mathrm{s}$} & 64 & 1 & \phantom{1}38.4 &  & 0.71 - 1.03 & \phantom{1}3.0 - \phantom{1}5.1 \\
DH Tau & 2019-08-17 & Pupil & BB\_H\phantom{$_\mathrm{s}$} & 32 & 1 & \phantom{1}14.9 & \phantom{1}4.3 & 0.48 - 0.56 & \phantom{1}3.9 - \phantom{1}4.8 \\
 & 2019-09-16 & Field & BB\_H\phantom{$_\mathrm{s}$} & 64 & 1 & \phantom{1}38.4 &  & 0.90 - 1.60 & \phantom{1}1.6 - \phantom{1}2.9 \\
 & 2019-10-24 & Field & BB\_H\phantom{$_\mathrm{s}$} & 64 & 1 & \phantom{1}38.4 &  & 0.20 - 0.32 & \phantom{1}5.5 - 12.0 \\
 & 2019-10-25A & Field & BB\_H\phantom{$_\mathrm{s}$} & 64 & 1 & \phantom{1}38.4 &  & 0.50 - 0.99 & \phantom{1}5.9 - 10.4 \\
 & 2019-10-25B & Field & BB\_H\phantom{$_\mathrm{s}$} & 64 & 1 & \phantom{1}38.4 &  & 0.47 - 0.64 & \phantom{1}5.3 - 11.7 \\
$\beta$~Pic & 2019-10-29 & Pupil & BB\_H\phantom{$_\mathrm{s}$} & \phantom{1}4 & 8 & \phantom{1}29.9 & 20.9 & 0.34 - 0.60 & \phantom{1}3.3 - \phantom{1}5.8 \\
 & 2019-11-26 & Pupil & BB\_H\phantom{$_\mathrm{s}$} & \phantom{1}4 & 8 & \phantom{1}29.9 & 19.8 & 0.37 - 0.53 & \phantom{1}2.9 - \phantom{1}7.8 \\
TYC 8998 & 2020-02-16 & Pupil & BB\_H\phantom{$_\mathrm{s}$} & 32 & 4 & \phantom{1}34.1 & 12.8 & 0.46 - 0.75 & \phantom{1}7.1 - 11.2\hspace{-3pt} \B \\
\hline 
\end{tabular} 
\begin{tablenotes}
\vspace{5pt}
\item \hspace{-3pt}\textbf{Notes.} The date is in the format year-month-day, DIT stands for detector integration time, NDIT is the number of detector integrations per HWP switch angle and $t_\mathrm{exp}$ is the total on-source exposure time. The parallactic rotation is only indicated for observations performed in pupil-tracking mode. The seeing and coherence time are retrieved from measurements by the DIMM (Differential Image Motion Monitor) and from the MASS-DIMM (Multi-Aperture Scintillation Sensor), respectively.
\end{tablenotes}
\end{threeparttable}
\label{tab:observations} 
\end{table*} 

The observation strategy was as follows. We generally observed each target multiple times with typically over 30~min of on-source exposure time per visit. However, for some targets a single visit was enough to detect the companion with high S/N in total intensity. We mainly observed in broadband $H$, but sometimes used broadband $J$ or $K_s$ when we wanted to obtain data in an additional filter or in the case the companion was brighter in $K_s$ than $H$. We used the apodized Lyot coronagraph with a mask diameter of 185~mas (for $J$ and $H$) or 240~mas (for $K_s$) to suppress the starlight~\citep{carbillet_spherealc1, guerri_spherealc2}. This allowed us to use longer integration times per frame to minimize the effects of read noise. However, we did not use integration times longer than 64~s to limit the effect of changing atmospheric conditions during a HWP cycle. In addition to the polarimetric science frames, we took star center frames to accurately determine the position of the star behind the coronagraph and star flux frames to measure the total stellar flux. We also took sky frames with the same instrument setup as the science and star flux frames to subtract the sky background from the respective frames.

For the majority of the observations, we used the pupil-tracking mode~\citep{vanholstein_exopol}. In this mode the image derotator (K-mirror) rotates such that the telescope pupil is kept fixed with respect to the detector while the on-sky field of view rotates with the parallactic angle. The pupil-tracking mode has numerous advantages. With sufficient parallactic rotation we can apply angular differential imaging~\citep[ADI;][]{marois_adi} to suppress speckle noise and accurately determine the total intensity of the companions located at small angular separations from the star. Furthermore, because the speckles are quasistatic, they are more effectively removed in the polarimetric data-reduction steps (and can be further suppressed by applying ADI to the polarimetric images). In addition, the diffraction spikes created by the support structure of the telescope’s secondary mirror are suppressed by a mask added to the Lyot stop. Finally, the loss of signal due to the crosstalk produced by the image derotator is limited~\citep[see][]{vanholstein_irdis2}. As a result, the polarimetric efficiency, that is, the fraction of the linearly polarized light incident on the telescope that is actually measured, is always high (typically $\gtrsim90\%$).

For a few targets, we used the field-tracking mode to be able to offset the derotator position angle and control the orientation of the image on the detector. For instance, the companions of AB~Pic and HD~106906 are at such large angular separations (see Table~\ref{tab:companions_properties}) that we needed to place them in one of the corners of the $11\arcsec \times 11\arcsec$ field of view to make them visible. In the case of 1RXS~J1609 and DH~Tau we switched to field-tracking mode after we discovered that both companions crossed a cluster of bad pixels during the pupil-tracking observations. In all cases, we chose the orientation of the image derotator such that the polarimetric efficiency was high~\citep[see][]{deboer_irdis1}.


%
%

\section{Data reduction}
\label{sec:data_reduction}

We reduced the data with the publicly available and highly automated pipeline \mbox{\texttt{IRDAP}}\footnote{\url{https://irdap.readthedocs.io}} (IRDIS Data reduction for Accurate Polarimetry), version 1.2.2~\citep{vanholstein_irdis2}. 
IRDAP preprocesses the raw data by subtracting the sky background, flat fielding, correcting for bad pixels, extracting the images of IRDIS' left and right optical channels, and centering using the star center frames. 
It then subtracts the right images from the left images (the single difference) for each of the measurements taken at HWP switch angles equal to $0\degr$, $45\degr$, $22.5\degr$, and $67.5\degr$ to obtain the $Q^+$-, $Q^-$-, $U^+$-, and $U^-$-images, respectively.
IRDAP also adds these same left and right images (the single sum) to obtain the total-intensity $I_{Q^+}$-, $I_{Q^-}$-, $I_{U^+}$-, and $I_{U^-}$-images.
Subsequently, IRDAP computes cubes of $Q$- and $U$-images from the double difference and the corresponding cubes of total-intensity $I_Q$- and $I_U$-images from the double sum, as:
\begin{align}
    Q &= \frac{1}{2}\left(Q^+ - Q^-\right), \\
    I_Q &= \frac{1}{2}\left(I_{Q^+} + I_{Q^-}\right),
\end{align}
and similar for $U$ and $I_U$.
For the two data sets of HD~4747 and the data set of PZ~Tel in $J$-band, strongly varying atmospheric seeing prevents the double difference from fully removing the signal created by transmission differences between the two orthogonal polarization directions downstream of the image derotator.
To remove this spurious polarization, we used the normalized double difference~\citep[see][]{vanholstein_irdis2} instead of the conventional double difference for these three data sets.

After computing the double difference and double sum, \mbox{IRDAP} uses a fully validated Mueller matrix model to correct for the instrumental polarization (created upstream of the image derotator) and crosstalk of the telescope and instrument with an absolute polarimetric accuracy of ${\lesssim}0.1$\% in the degree of polarization.
IRDAP also derotates the images and corrects them for true north~\citep[see][]{maire_astrometry}. 
This results in a total of four images: $Q$, $U$, $I_Q$, and $I_U$, that constitute our best estimate of the linear polarization state incident on the telescope.
Finally, IRDAP computes images of the linearly polarized intensity $PI = \surd{(Q^2 + U^2)}$, and, following the definitions of~\citet{deboer_irdis1}, images of $Q_\phi$ and $U_\phi$.
Positive (negative) $Q_\phi$ indicates linear polarization in the azimuthal (radial) direction, and $U_\phi$ shows the linear polarization at $\pm\SI{45}{\degree}$ from these directions.
In Sect.~\ref{sec:circumstellar_disks} we use the polarized intensity and $Q_\phi$- and $U_\phi$-images to show the five circumstellar disks that we detected.

The model-corrected $Q$- and $U$-images often contain a halo of polarized light from the star. This polarization can originate from interstellar dust, (unresolved) circumstellar material, and spurious or uncorrected instrumental polarization. 
With IRDAP we can therefore determine the stellar polarization from the \mbox{$I_Q$-,} $I_U$-, and model-corrected $Q$- and $U$-images by measuring the flux in these images in a user-defined region that contains only starlight and no signal from a companion, background star, or circumstellar disk. 
For most data sets we measured the stellar polarization using a star-centered annulus placed over the AO residuals, or in the case that region contains little flux, a large aperture centered on the star. 
IRDAP then determines the corresponding uncertainty by measuring the stellar polarization for each HWP cycle individually and computing the standard error of the mean over the measurements.
Finally, IRDAP creates an additional set of $Q$- and $U$-images with the stellar polarization subtracted.
To this end, it scales the $I_Q$- and $I_U$-images with the measured fractional stellar polarization and subtracts the resulting images from the model-corrected $Q$- and $U$-images. 
Whenever discussing data in this paper, we always mean the reduction without the stellar polarization subtracted, unless explicitly stated.

For the observations taken in pupil-tracking mode, IRDAP additionally performs classical ADI and ADI with principal component analysis~\citep[PCA;][]{soummer_klip, amara_pynpoint} to suppress the stellar speckle halo and detect the companions in total intensity. 
IRDAP also processes the star flux frames by performing sky subtraction, flat fielding, bad-pixel correction, and registering through fitting the frames to a 2D Gaussian function. 
We obtained the final images of the stellar point spread function (PSF) by mean-combining the left and right processed star flux frames and scaling the pixel values to the integration time and system transmission (i.e.,~due to neutral-density filters) of the science frames.
We separately reduced the data sets of targets that we observed multiple times and then used IRDAP to mean-combine the final images produced in each reduction.

The final $Q$- and $U$-images of most data sets still contain a small amount of speckle noise close to the star. 
For the data sets of HR~8799, HD~206893 and $\beta$~Pic, which have companions at small separations from the star, we therefore performed additional reductions in which we apply classical ADI on the polarimetric images to further suppress these speckles~\citep[see][]{vanholstein_exopol}. 
To this end, we added a reduction step to IRDAP in which we median-combine the instrumental-polarization-subtracted $Q$-frames (and $U$-frames) and subtract the resulting median image from each of the frames before derotating them. 
In these reductions we skip the later step of determining and subtracting the stellar polarization because the ADI step has already removed the halo of polarized starlight.


%
%

\section{Extraction of polarization of companions: Detection of polarization of DH~Tau~B}
\label{sec:method_dh_tau}

With the data of all targets reduced, we can determine the polarization of the companions, or, in the case we do not detect significant polarization, place upper limits on the degree of polarization of the companions.
For this we have developed a method similar to that employed by~\citet{jensenclem_exopol2}, which, in turn, is based on the method used by~\citet{jensenclem_exopol1}.
In this method, we use aperture photometry to estimate the probability distributions of the companion signals in the $I_Q$-, $I_U$-, $Q$-, and $U$-images. 
We then use these distributions to calculate the probability distributions of the degree and angle of linear polarization, from which we retrieve the median values, uncertainties, and upper limits.
We applied this method to the data sets of GSC~8047, CD-35 2722, AB~Pic, HD~106906, GSC~6214, 1RXS~J1609, DH~Tau, and TYC~8998.
In this section, we demonstrate the method using the \mbox{2019-10-24} $H$-band data set of DH~Tau and exemplify the detection of the polarization of DH~Tau~B, a companion at a large angular separation from its star. 
For companions at close separations or with large star-to-companion contrasts, we have slightly adapted the method and determine the distributions in $I_Q$ and $I_U$ through ADI with negative PSF injection or fitting of the companion PSF.
In Appendices~\ref{sec:method_beta_pic} and \ref{sec:method_hd_19467} we demonstrate the two respective methods and show how we set upper limits on the polarization of $\beta$~Pic~b and HD~19467~B.

\begin{figure}[!htbp]
\centering
\includegraphics[width=\hsize, trim={15 10 10 5}, clip]{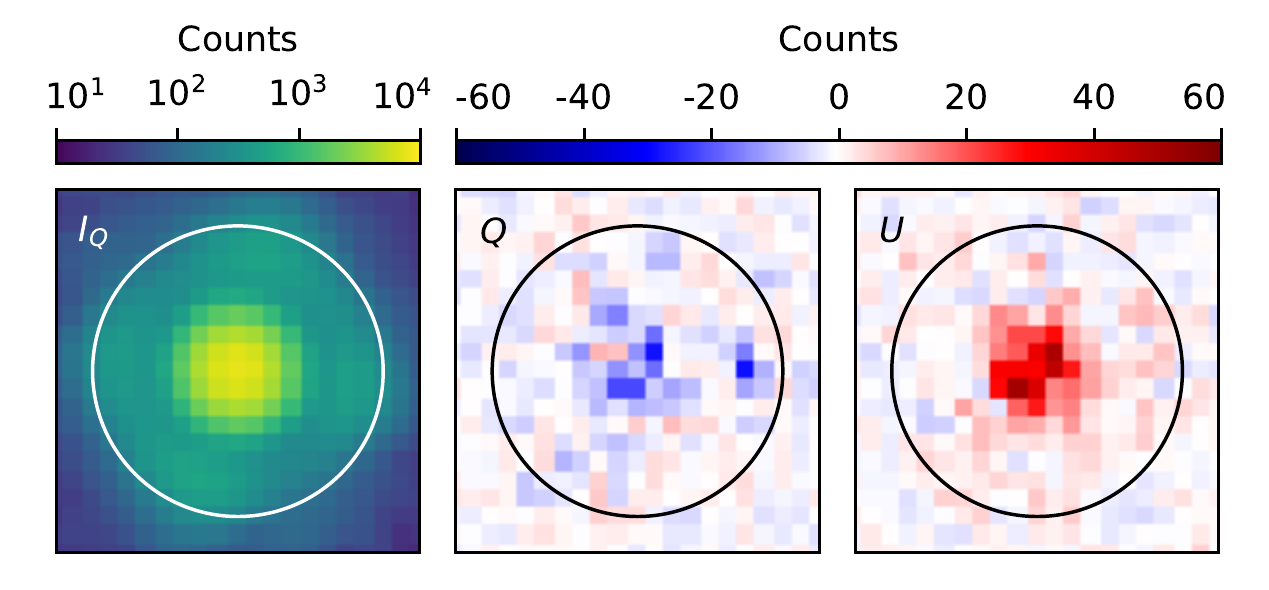}
\caption{Reduced $I_Q$-, $Q$-, and $U$-images (after applying the cosmetic correction described in Appendix~\ref{app:yinyang}) at the position of the companion DH~Tau~B of the \mbox{2019-10-24} data set of DH~Tau, showing an aperture of radius 8~pixels centered on the companion. The $I_U$-image, which is not shown, is very similar to the $I_Q$-image.}
\label{fig:dh_tau_signals} 
\end{figure}

To start the analysis of the \mbox{2019-10-24} data set of DH~Tau, we determine the center coordinates of the companion DH~Tau~B by mean-combining the $I_Q$- and $I_U$-images and fitting a 2D Moffat function to the resulting image at the position of the companion.
We then make a cosmetic correction to the $Q$- and $U$-images (if necessary) to remove spurious structures that result from imperfect relative centering of the images, image motion, and parallactic rotation (see Appendix~\ref{app:yinyang}).
The $I_Q$-, $Q$-, and $U$-images (after the cosmetic correction) at the companion position are shown in Fig.~\ref{fig:dh_tau_signals}.
The signals in $Q$ and (in particular) in $U$ are clear indications that DH~Tau~B is polarized.
%

To determine the probability distributions of the companion signals in $I_Q$, $I_U$, $Q$, and $U$, we define a range of aperture radii from 1 to 10~pixels to be used for the photometry.
For each aperture radius we perform the following five steps, after which we select the final aperture radius to be used for our results.
Because at the end of this section we select a final aperture radius of 8~pixels, we use this radius in the examples of the five steps below.

\begin{figure}[!htbp]
\centering
\includegraphics[width=\hsize, trim={5 5 5 5}, clip]{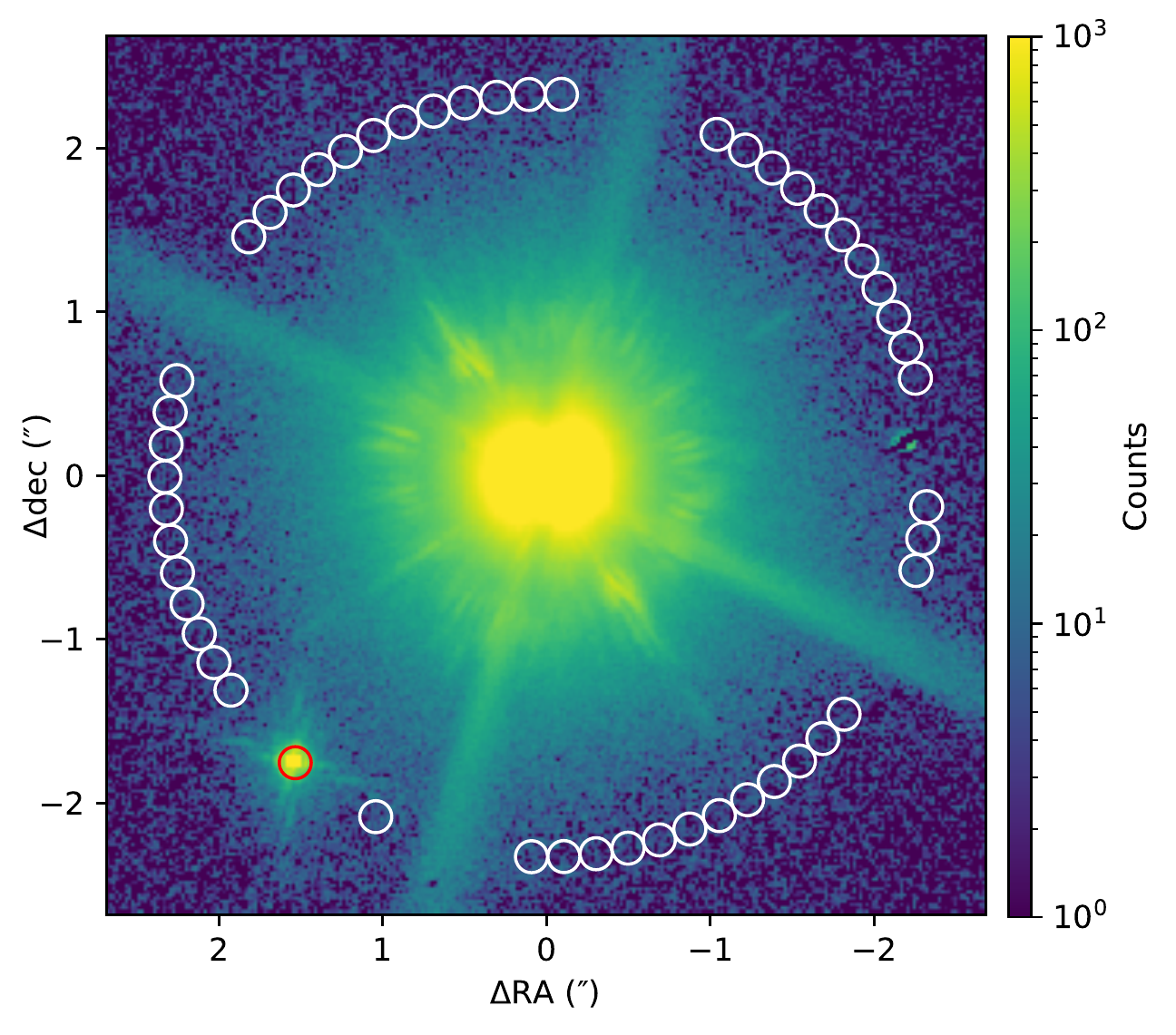}
\caption{Reduced $I_Q$-image of the \mbox{2019-10-24} data set of DH~Tau, showing an aperture of radius 8~pixels at the position of the companion DH~Tau~B (red) and the ring of comparison apertures of the same radius around the star (white).}
\label{fig:dh_tau_apertures} 
\end{figure}

As the first step, we place an aperture of the given radius at the position of the companion in each of the $I_Q$-, $I_U$-, $Q$-, and $U$-images (see Fig.~\ref{fig:dh_tau_signals}) and sum the flux in the aperture.
In the same images we then place a ring of comparison apertures around the star at the same separation as the companion to sample the background.
We exclude those apertures that contain the first Airy ring of the companion, diffraction spikes from the star and the companion, and clusters of bad pixels.
The resulting ring of apertures for an aperture radius of 8~pixels is shown superimposed on the $I_Q$-image in Fig.~\ref{fig:dh_tau_apertures}.
In this figure the first Airy ring and the diffraction spikes created by the Lyot stop mask are clearly visible at the companion position, which is evidence of the extremely good atmospheric conditions during the observations (see Table~\ref{tab:observations}).
Finally, we sum the flux in each of the comparison apertures and compute the mean background as the mean of the aperture sums.
%

\begin{figure}[!htbp]
\centering
\includegraphics[width=\hsize, trim={5 5 5 5}, clip]{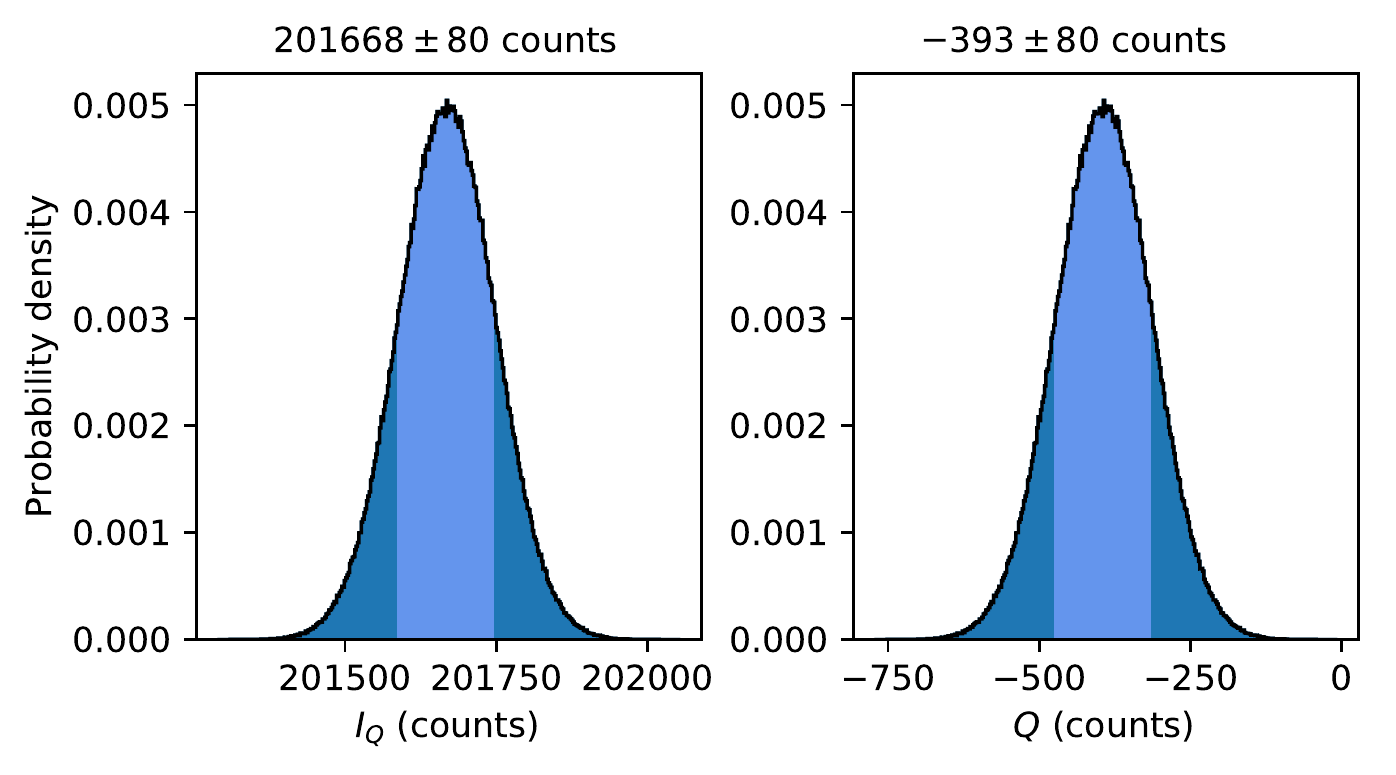}
\caption{PDF of the signal of DH~Tau~B in $I_Q$ (left) and $Q$ (right) from the \mbox{2019-10-24} data set of DH~Tau, using an aperture radius of 8~pixels and taking into account only the photon noise of the companion. The mean and standard deviation of the distributions are shown above the graphs, with the latter also indicated by the light-blue shaded area.}
\label{fig:dh_tau_photon_noise} 
\end{figure}

In step two, we calculate the probability density function (PDF) of the companion signal in $I_Q$, $I_U$, $Q$, and $U$, taking into account only the photon noise of the companion.
To this end, we compute the companion signals in $I_Q$, $I_U$, $Q$, and $U$ by subtracting the mean background from the summed flux of the companion aperture.
We then compute the PDFs of $I_Q$ and $I_U$ from a Gaussian distribution with the mean and variance equal to the respective companion signals, while accounting for the conversion from counts to total number of detected photoelectrons and back to counts (using a detector gain of 1.75~e$^-$/count).
The resulting PDF of $I_Q$ for an aperture radius of 8~pixels is shown in Fig.~\ref{fig:dh_tau_photon_noise} (left).
For large number of photons, the photon noise in $Q$ and $U$ is the same as that in $I_Q$ and $I_U$.
We therefore construct the PDFs of $Q$ and $U$ from a Gaussian distribution with the mean equal to the companion signals in $Q$ and $U$, but the variance equal to that of the PDFs of $I_Q$ and $I_U$.
Figure~\ref{fig:dh_tau_photon_noise} (right) shows the resulting PDF in $Q$.
%

\begin{figure}[!htbp]
\centering
\includegraphics[width=\hsize, trim={5 5 25 0}, clip]{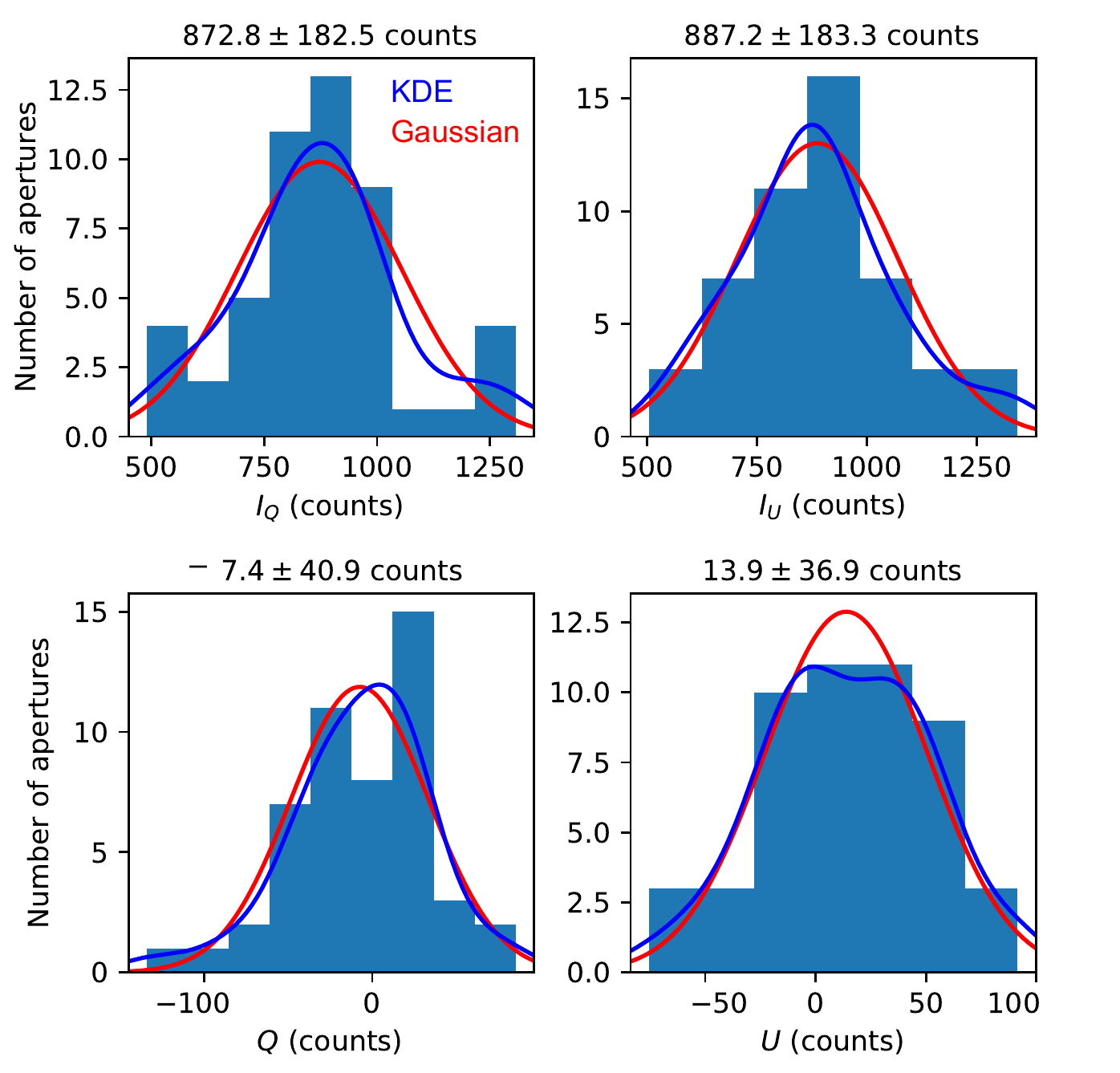}
\caption{Histograms of the background in $I_Q$, $I_U$, $Q$, and $U$ of the \mbox{2019-10-24} data set of DH~Tau, as obtained through summing the flux in the 8-pixel-radius comparison apertures of Fig.~\ref{fig:dh_tau_apertures}. The mean and standard deviation of the samples are shown above the histograms. The blue curves show the PDFs as estimated through KDE and the red curves show the best-fit Gaussian distributions for comparison.}
\label{fig:dh_tau_background_histograms} 
\end{figure}

For the third step, we estimate the PDF of the background in $I_Q$, $I_U$, $Q$, and $U$ using the comparison aperture sums obtained in the first step. 
To not a priori assume a specific functional form of the PDF, we use kernel density estimation (KDE).
In this method, the PDF is obtained by placing a Gaussian kernel of a given bandwidth (i.e.,~a Gaussian distribution with a given standard deviation) at each data point of the sample and summing the resulting kernels. 
We compute the bandwidth of the Gaussian kernel using Scott's rule~\citep{scott_mde}, in this case yielding a bandwidth of ${\sim}84$~counts for $I_Q$ and $I_U$, and ${\sim}18$~counts for $Q$ and $U$.
Histograms of the background samples and the PDFs as estimated via KDE for an aperture radius of 8~pixels are shown in Fig.~\ref{fig:dh_tau_background_histograms}.
We note that for very close-in companions such as PDS~70~b, the number of comparison apertures is low enough that KDE does not produce accurate results. 
When there are fewer than 21 comparison apertures, we therefore account for the small-sample statistics by fitting the background samples with a Student's $t$-distribution with the empirical standard deviation equal to $s = s_\mathrm{bg}\surd(1 + 1 / n)$, with $s_\mathrm{bg}$ the standard deviation of the comparison aperture sums and $n$ the number of comparison apertures~\citep[see][]{mawet_contrast}.
%
%
\begin{figure*}[!htbp]
\centering
\includegraphics[width=\hsize, trim={5 5 5 5}, clip]{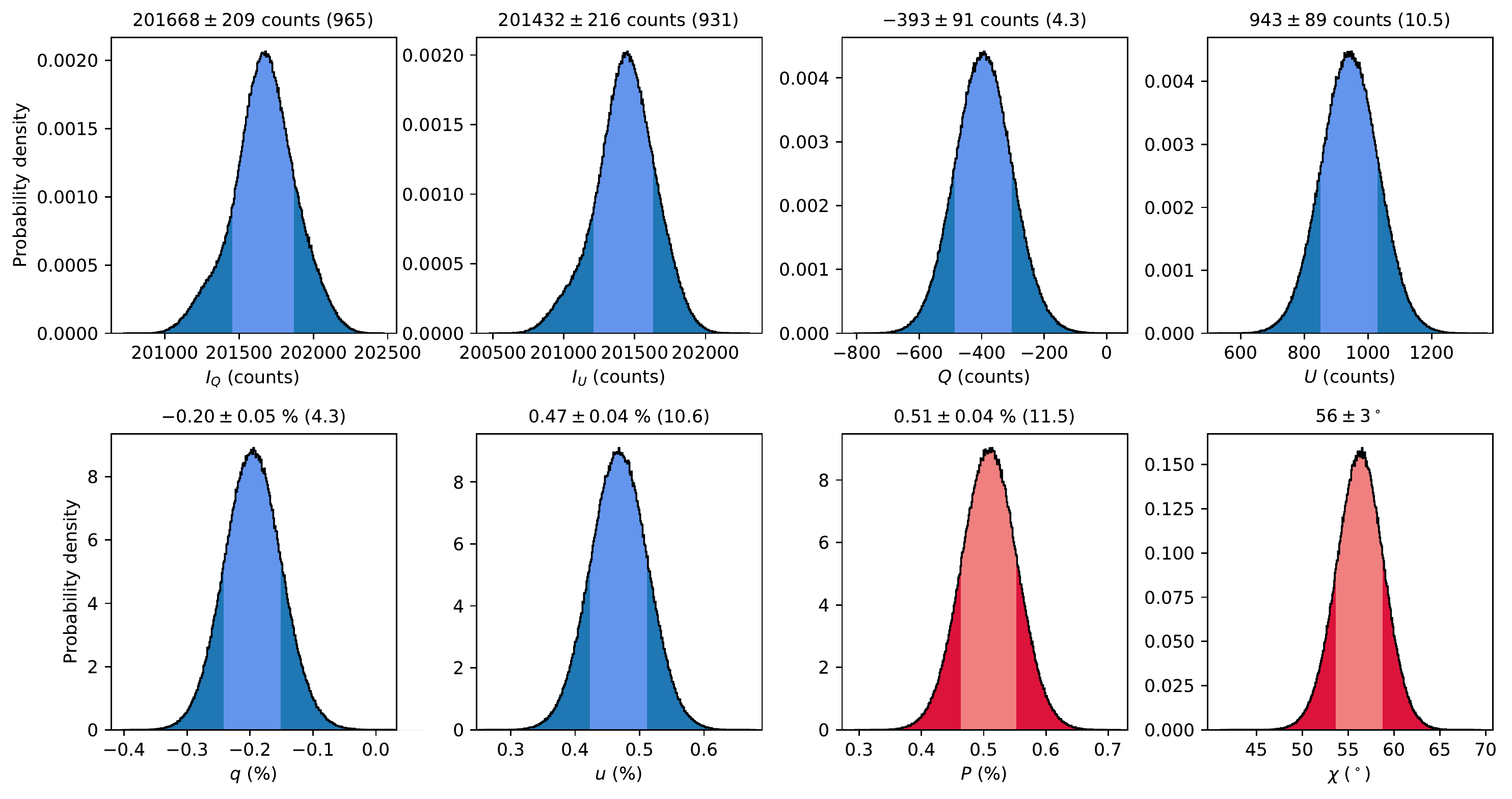}
\caption{Final probability distributions of the signals of DH~Tau~B in $I_Q$, $I_U$, $Q$, and $U$ (top row), and in normalized Stokes $q$ and $u$, degree of linear polarization, and angle of linear polarization (bottom row) from the \mbox{2019-10-24} data set of DH~Tau, using an aperture radius of 8~pixels. The median values of the distributions, as well as the uncertainties computed from the two-sided 68.27\% equal-tailed interval around the median, are shown above the graphs. The S/N, i.e.,~the median value divided by the largest uncertainty, is shown within parentheses. The 68.27\% intervals are also indicated by the light-blue and light-red shaded areas.}
\label{fig:dh_tau_polarization_distributions}
\end{figure*} 

In step four, we compute the final probability distributions in $I_Q$, $I_U$, $Q$, and $U$ that include both the photon noise of the companion and the uncertainty of the background.
For this, we draw $10^6$ random samples from the previously constructed PDFs of the companion signal (step two) and the background (step three).
Because we already subtracted the background when computing the PDF of the companion signal, we first subtract the mean background from the drawn background samples.
We then compute the final distribution by subtracting the resulting background samples from the samples of the companion signal.
Next, we compute the median values of the final distributions and determine the uncertainties from the two-sided 68.27\% equal-tailed interval around the median, corresponding to the $1\sigma$ (one standard deviation) confidence interval of the Gaussian distribution.
The resulting probability distributions for an aperture radius of 8~pixels, including the median values, uncertainties, and S/Ns (i.e.,~the median value divided by the largest uncertainty), are shown in Fig.~\ref{fig:dh_tau_polarization_distributions} (top row).
The data are clearly photon-noise limited in $Q$ and $U$ because the distributions are nearly Gaussian and the uncertainties are close to the standard deviation shown in Fig.~\ref{fig:dh_tau_photon_noise} (right).
It follows that we detect DH~Tau~B with a very high S/N in total intensity and also have significant detections of polarization, especially in Stokes $U$.

As the fifth and final step, we use the $I_Q$-, $I_U$-, $Q$-, and $U$-samples to compute the distributions of normalized Stokes $q = Q / I_Q$, normalized Stokes $u = U / I_U$, the degree of linear polarization $P = \surd(q^2 + u^2)$, and the angle of linear polarization $\chi = \nicefrac{1}{2}\arctan(u / q)$. We compute the median values and uncertainties in the same way as we did for $I_Q$, $I_U$, $Q$, and $U$.
The results of these computations for an aperture radius of 8~pixels are shown in Fig.~\ref{fig:dh_tau_polarization_distributions} (bottom row).

\begin{figure}[!htbp]
\centering
\includegraphics[width=\hsize, trim={5 5 5 5}, clip]{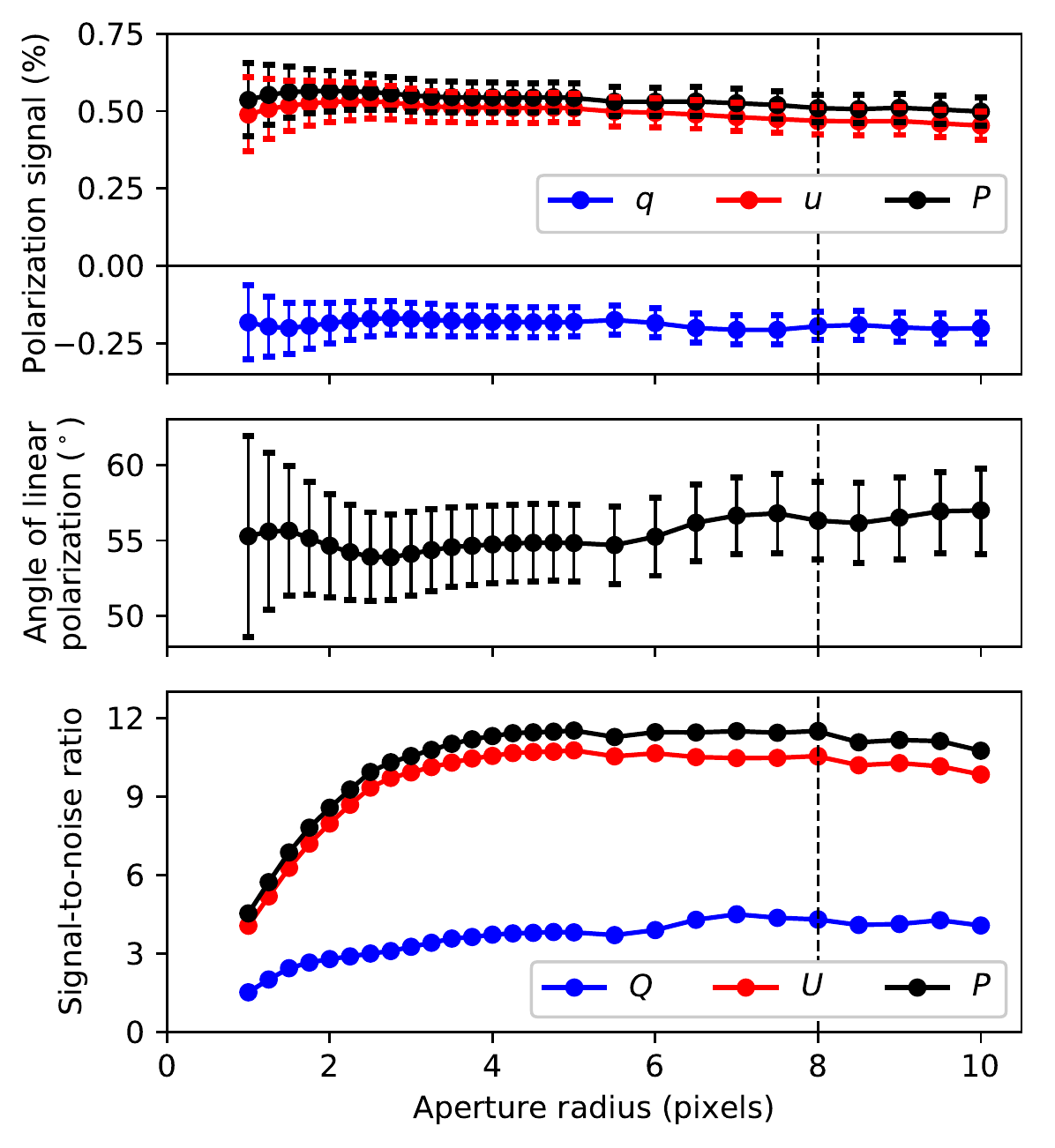}
\caption{Normalized Stokes parameters $q$ and $u$ and degree of linear polarization (top), angle of linear polarization (center), and S/N in $q$, $u$, and the degree of linear polarization (bottom) of DH~Tau~B as a function of aperture radius for the \mbox{2019-10-24} data set of DH~Tau. The uncertainties of the measured values are shown with error bars. The final selected aperture radius of 8~pixels is indicated with the dashed vertical lines.}
\label{fig:dh_tau_polarization_snr_radius}
\end{figure}

After performing the five steps above for each defined aperture radius, we plot the median values and uncertainties of $q$, $u$, the degree and angle of polarization, and the S/N in $q$, $u$, and the degree of polarization as a function of aperture radius in Fig.~\ref{fig:dh_tau_polarization_snr_radius}.
From this figure we see that, within the uncertainties, the polarization of the companion is constant with changing aperture radius.
We select a final aperture radius of 8~pixels, as indicated by the vertical dashed lines in Fig.~\ref{fig:dh_tau_polarization_snr_radius}, because at this radius the S/N in $q$ and $u$ is maximized and the aperture is sufficiently large to suppress (average out) the spurious signals resulting from incompletely removed bad pixels (see Appendix~\ref{app:bad_pixels}).
We conclude that for this \mbox{2019-10-24} data set, 
we measure DH~Tau~B to have a degree of polarization of $0.51 \pm 0.04$\% and an angle of polarization of $56 \pm 3^\circ$ (east of north) in $H$-band.
%
%


%
%

\section{Results}
\label{sec:results}

After careful analysis of our data with the methods as described in Sect.~\ref{sec:method_dh_tau} and Appendices~\ref{sec:method_beta_pic} and \ref{sec:method_hd_19467}, we detected unresolved polarization originating from DH~Tau~B and GSC~6214~B. 
We consider these measurements detections because the measured polarization signals are significant (i.e.,~have an S/N of at least 5 in $q$ or $u$) and are very likely intrinsic to the companions (i.e.,~are not due to interstellar dust).
We present these results in Sects.~\ref{sec:detection_dh_tau} and \ref{sec:detection_gsc_6214}.
We also marginally detected polarization from 1RXS~J1609~B, but we show in Sect.~\ref{sec:detection_1rxs_j1609} that this polarization is best explained by interstellar dust.
For the other 17 companions we do not detect significant polarization.
In Sect.~\ref{sec:upper_limits}, we place upper limits on the degree of polarization of 1RXS~J1609~B and these other companions. 
Finally, in Sect.~\ref{sec:circumstellar_disks}, we briefly describe five circumstellar disks that we detected in our survey and of which two had not been imaged in polarized scattered light before.

\subsection{Detection of intrinsic polarization of DH~Tau~B}
\label{sec:detection_dh_tau}

In this section we present the detection of polarization originating from DH~Tau~B.
Table~\ref{tab:detection_dh_tau} shows the measured $H$-band degree and angle of polarization of DH~Tau~B, including the uncertainties and the attained S/Ns, for each of the various data sets and the data set created by mean-combining the final images of the three data sets taken at favorable atmospheric conditions (i.e.,~the \mbox{2019-10-24}, \mbox{2019-10-25A}, and \mbox{2019-10-25B} data sets; see Table~\ref{tab:observations}).
For each data set the measured $q$- and $u$-signals are within the uncertainties constant with aperture radius.
We determined the final values of the polarization signals using apertures of radius 8~pixels, which is at, or close to, the radius where the S/N in $q$ and $u$ is maximized for the various data sets (see Sect.~\ref{sec:method_dh_tau}). 
As shown in Table~\ref{tab:detection_dh_tau}, we detect significant polarization from DH~Tau~B, reaching S/Ns of around 10 for the three data sets taken at favorable atmospheric conditions. 
The measured degree and angle of polarization for the different data sets are overall consistent.
From visual inspection of the images, we find that the small differences among the data sets are primarily due to small biases caused by incompletely removed bad pixels (see Appendix~\ref{app:bad_pixels}). 
These differences can additionally be caused by time-varying atmospheric conditions and AO performance, the limited accuracy of the Mueller matrix model with which the data have been corrected~\citep[see][]{vanholstein_irdis2}, and other unknown systematic effects. 
From the mean-combined images, we measure DH~Tau~B to have a degree and angle of polarization of $0.48 \pm 0.03\%$ and $58 \pm 2^\circ$ (east of north), respectively, with an S/N of 7.7 in $q$ and 16.1 in $u$.
%
\begin{table*}
\caption{Degree and angle of linear polarization, including the uncertainties, of the parent star DH~Tau~A and the companion DH~Tau~B as measured in $H$-band for each of the five data sets and the data set created by mean-combining the final images of the \mbox{2019-10-24}, \mbox{2019-10-25A}, and \mbox{2019-10-25B} data sets.}
\centering
\begin{threeparttable}
\begin{tabular}{l c c c c c c}
\hline\hline
\T\B
Data set & $P_\mathrm{star}$ (\%) & $\chi_\mathrm{star}$ ($^\circ$) & $P_\mathrm{com}$ (\%) & $\chi_\mathrm{com}$ ($^\circ$) & S/N $q_\mathrm{com}$ & S/N $u_\mathrm{com}$ \\
\hline
2019-08-17 & $0.08 \pm 0.01$ & $83 \pm 10$ & $0.4 \pm 0.1$ & $46 \pm 9\phantom{0}$ & 0.1 & \phantom{1}3.1 \T \\
2019-09-16 & $0.23 \pm 0.01$ & $114 \pm 2\phantom{0}$ & $0.6 \pm 0.2$ & $51 \pm 9\phantom{0}$ & 0.6 & \phantom{1}3.3 \T \\
2019-10-24 & $0.11 \pm 0.01$ & $119 \pm 3\phantom{0}$ & $0.51 \pm 0.04$ & $56 \pm 3\phantom{0}$ & 4.3 & 10.6 \T \\
2019-10-25A & $0.16 \pm 0.01$ & $145 \pm 2\phantom{0}$ & $0.49 \pm 0.05$ & $51 \pm 3\phantom{0}$ & 2.1 & \phantom{1}9.9 \T \\
2019-10-25B & $0.27 \pm 0.02$ & $123 \pm 2\phantom{0}$ & $0.48 \pm 0.05$ & $66 \pm 3\phantom{0}$ & 6.4 & \phantom{1}7.4 \T \\
Mean combined & $0.172 \pm 0.009$ & $128 \pm 1\phantom{0}$ & $0.48 \pm 0.03$ & $58 \pm 2\phantom{0}$ & 7.7 & 16.1 \B \\
\hline
\end{tabular}
\begin{tablenotes}
\vspace{5pt}
\item \hspace{-3pt}\textbf{Notes.} $P_\mathrm{star}$ and $\chi_\mathrm{star}$ are the degree and angle of linear polarization of the parent star DH~Tau~A, respectively, and $P_\mathrm{com}$ and $\chi_\mathrm{com}$ are the degree and angle of polarization of the companion DH~Tau~B. S/N $q_\mathrm{com}$ and S/N $u_\mathrm{com}$ are the S/Ns with which the $q$- and $u$-signals of DH~Tau~B are detected.
\end{tablenotes}
\end{threeparttable}
\label{tab:detection_dh_tau}
\end{table*}

Table~\ref{tab:detection_dh_tau} also lists the stellar degrees and angles of polarization as measured with an annulus at the location of the AO residuals (see Sect.~\ref{sec:data_reduction}).
For the mean-combined data set we determined the uncertainty on the stellar polarization by propagating the uncertainties from the individual data sets using a Monte Carlo calculation and assuming Gaussian statistics.
The measurements of the stellar polarization are very likely affected by some systematic effects because the signals are less consistent than those of the companion and show differences among the data sets that are much larger than the calculated (statistical) uncertainties.
The most likely explanation for these differences is that time-varying atmospheric conditions and AO performance cause the effective coronagraphic extinction to vary from frame to frame.
Because the companion is not affected by the coronagraph, this can also explain why the polarization measured for the companion is more consistent among the data sets.
The stellar polarization measurements show that the star could be truly polarized because the angles of polarization for the three data sets taken at favorable conditions (\mbox{2019-10-24}, \mbox{2019-10-25A}, and \mbox{2019-10-25B}) are quite similar.
Importantly, the measured polarization of the companion differs significantly from that of the star in all data sets, with the companion having a significantly larger degree of polarization and a very different angle of polarization.


DH~Tau, at a distance of 135~pc\footnote{All distances in this paper are retrieved from \citet{bailerjones_gaia}.}, is located at the front side of the Taurus molecular cloud complex that extends from at least 126~pc to 163~pc~\citep{galli_tauruscloud}.
To determine whether DH~Tau~B is intrinsically polarized, we therefore need to determine the contribution of interstellar dust to the measured polarization.
The interstellar polarization is a result of dichroism by elongated dust grains that are aligned with the local (galactic) magnetic field.
Because interstellar dust creates the same polarization for the companion and the star, this contribution can often be determined from the measured stellar polarization (e.g.,~for 1RXS~J1609, see Sect.~\ref{sec:detection_1rxs_j1609}, and ROXs~42B, see~\citealp{jensenclem_exopol2}). 
However, we cannot do that in this case because the star hosts a disk that we spatially resolve in our images (see Sect.~\ref{sec:circumstellar_disks} and Fig.~\ref{fig:circumstellar_disks}, top left) and therefore the stellar polarization is likely a combination of intrinsic and interstellar polarization.

\begin{figure*}[!htbp]
\centering
\includegraphics[width=\hsize, trim={5 7 5 5}, clip]{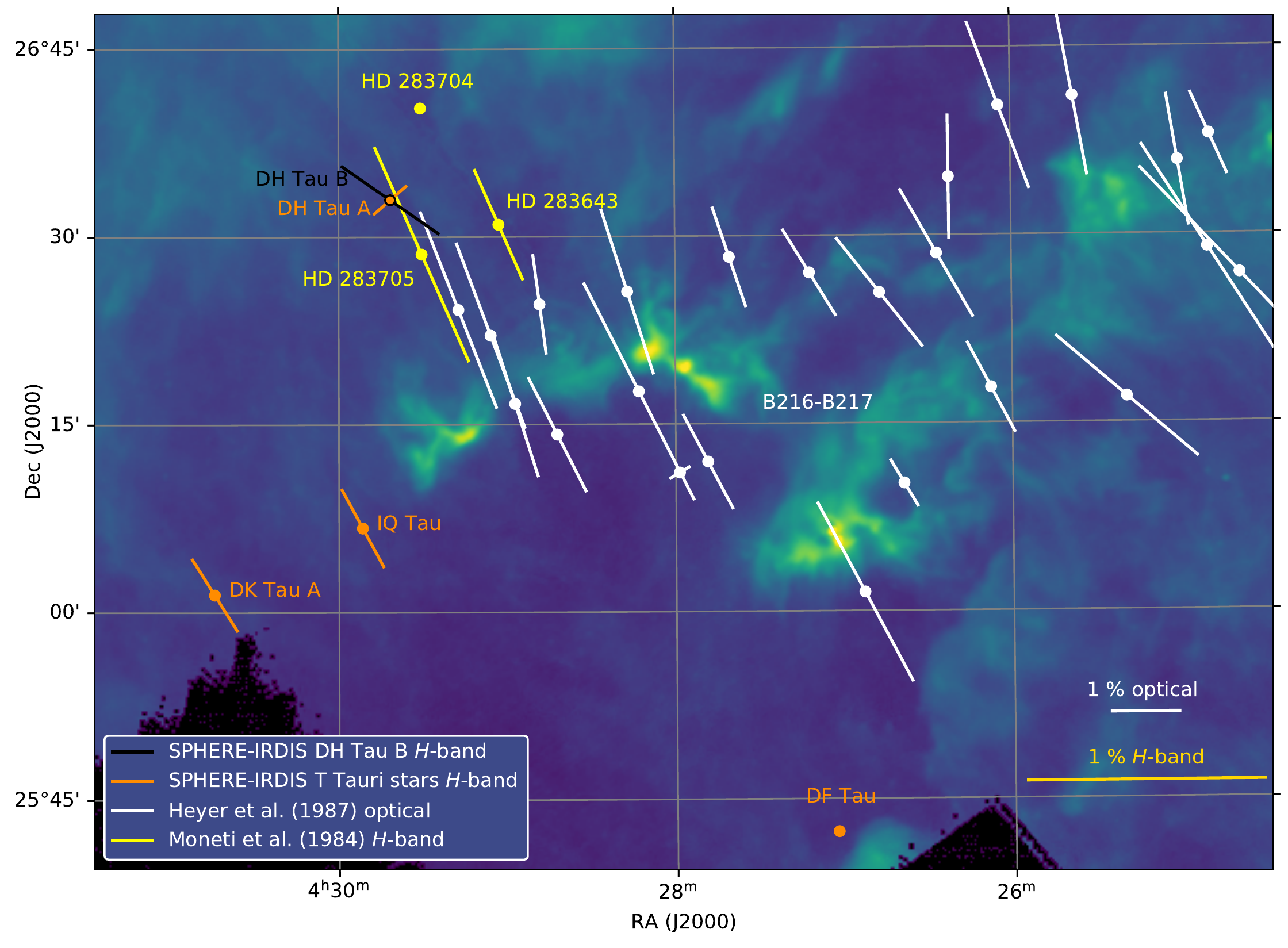}
\caption{Map of the linear polarization of DH~Tau~B and nearby stars superimposed on a Herschel-SPIRE map at \SI{350}{\micro\meter}. 
The length and orientation of the lines indicate the degree and angle of linear polarization, respectively. 
The black line shows the $H$-band polarization we measure for DH~Tau~B, and the orange lines display the SPHERE-IRDIS $H$-band measurements of DH~Tau~A and three other nearby T~Tauri stars whose archival data we analyzed.
White lines show optical measurements by~\protect\citet{heyer_tauruspol}.
Yellow lines indicate the $H$-band polarization of three bright stars closest to DH~Tau as derived from optical measurements by~\protect\citet{moneti_tauruspol}.
The length of the $H$-band vectors are scaled by a factor of four with respect to the optical vectors.}
\label{fig:dh_tau_spire}
\end{figure*}

To investigate the contribution of interstellar dust to the polarization of DH~Tau~B, we show in Fig.~\ref{fig:dh_tau_spire} a map of the polarization of DH~Tau~A and B and a few dozen nearby stars. 
The map is superimposed on a Herschel-SPIRE~\citep{pilbratt_herschel} image at \SI{350}{\micro\meter} that shows the concentrations of interstellar dust in the region.
White lines show optical measurements of stars at the periphery of the B216-B217 dark cloud from~\citet{heyer_tauruspol}. 
Yellow lines display measurements from~\citet{moneti_tauruspol} of the three nearest bright stars to DH~Tau.
Of these stars, HD~283704 (58~pc) is unpolarized as it is located in front of the clouds, whereas HD~283705 (170~pc) and HD~283643 (396~pc) are located behind the clouds and are both polarized with an angle of polarization of $26 \pm 1^\circ$.
Because the stars from~\citet{heyer_tauruspol} and \citet{moneti_tauruspol} are generally much older than DH~Tau and are therefore not expected to have a circumstellar disk that significantly polarizes their light, their polarization must primarily originate from interstellar dust.
Comparing the angles of polarization of DH~Tau~A ($128 \pm 1\degr$) and B ($58 \pm 2\degr$) with those of the reference stars in Fig.~\ref{fig:dh_tau_spire}, we conclude that the polarization of both DH~Tau A and B must include an intrinsic component.
%
%

We now set limits on the interstellar degree of polarization of the DH~Tau system.
To this end, we convert the optical measurements of the degree of polarization of the nearby stars HD~283705 and HD~283643 (2.48\% and 1.27\%) from~\citet{moneti_tauruspol} to $H$-band.
For this conversion we use Serkowski's law of interstellar polarization~\citep{serkowski_interstellarpol}:
\begin{equation}
    P = P_\mathrm{max}\exp\left[-K\ln^2\left(\lambda_\mathrm{max} / \lambda\right)\right],
\label{eq:serkowski}
\end{equation}
where $\lambda$ is the wavelength of the light, $P_\mathrm{max}$ is the maximum degree of polarization, and $\lambda_\mathrm{max}$ is the wavelength at which this maximum occurs. 
The parameter $K$ is computed following~\citet{whittet_interstellarpol}:
\begin{equation}
    K = 0.01 + 1.66 \lambda_\mathrm{max},
\end{equation}
with $\lambda_\mathrm{max}$ in micrometers. 
Because the observations were taken without color filter, we retrieve the spectral response of a Ga-As photomultiplier tube similar to that used for the measurements\footnote{RCA Photomultiplier Manual, \url{http://www.decadecounter.com/vta/pdf/RCAPMT.pdf}, consulted on June 2, 2020.} and multiply it with the transmission of the Earth's atmosphere.
With the resulting spectral transmission, we can compute the degree of polarization that the instrument measures from the transmission-weighted average over the curve from Serkowski's law.
Assuming $\lambda_\mathrm{max} = \SI{0.55}{\micro\meter}$, which is the average value for the 16 bright stars in Taurus observed by~\citet{whittet_interstellarpol}, we fit $P_\mathrm{max}$ for both stars.
From the fitted curves we then compute the degree of polarization at $H$-band, yielding 0.9\% for HD~283705 and 0.5\% for HD~283643.
Because DH~Tau is located at the front side of the clouds (rather than behind the clouds as are the comparison stars), the interstellar polarization of DH~Tau is most likely below 0.9\%, probably below 0.5\%.
This is in agreement with the $H$-band degrees of polarization of three nearby T~Tauri stars whose archival SPHERE-IRDIS polarimetric data we analyzed (see Fig.~\ref{fig:dh_tau_spire}).
Of these stars, DF~Tau (125~pc) is unpolarized, and DK~Tau~A (128~pc), which does not have a disk, and IQ~Tau (131~pc), which has a very faint disk, are 0.33\% and 0.34\% polarized, respectively, both with an angle of polarization of ${\sim}30^\circ$.

\begin{figure}[!htbp]
\centering
\includegraphics[width=\hsize]{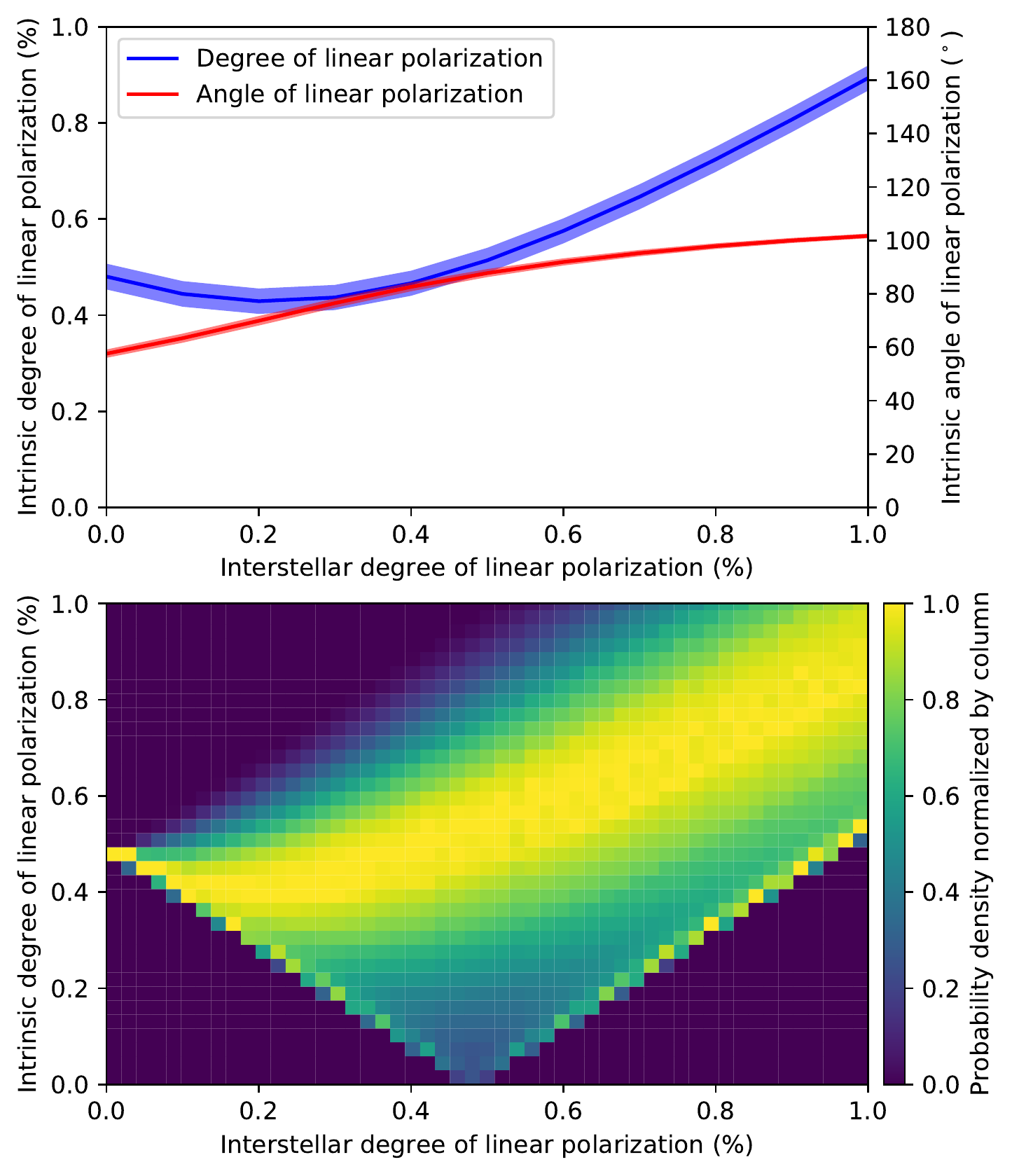}
\caption{Intrinsic polarization of DH~Tau~B after subtracting interstellar polarization signals from the measured polarization of the companion.
Top: Intrinsic degree and angle of linear polarization of DH~Tau~B as a function of the degree of polarization due to interstellar dust, assuming an angle of $26^\circ$ for the interstellar polarization. The bands around the curves show the uncertainties of our measurements. 
Bottom: Probability distributions of the intrinsic degree of polarization of DH~Tau~B for a range of degrees of polarization due to interstellar dust, assuming the angle of the interstellar polarization to have the same distribution as that determined by~\citet{goodman_tauruspol} for the B216-B217 dark cloud adjacent to DH~Tau. The probability distribution of each column is normalized to one.} 
\label{fig:dh_tau_companion_intrinsic}
\end{figure}

Although we do not know the exact interstellar degree of polarization for DH~Tau, the angle of polarization is likely close to $26^\circ$, which is the angle of both HD~283705 and HD~283643.
To see whether DH~Tau~B is intrinsically polarized, we take the polarization signal that we measured in the mean-combined images ($0.48 \pm 0.03\%$ at $58 \pm 2\degr$; see Table~\ref{tab:detection_dh_tau}) and subtract interstellar polarization signals with an angle of polarization of $26^\circ$ and a range of degrees of polarization.
The resulting intrinsic degree and angle of polarization of DH~Tau~B versus the interstellar degree of polarization is shown in Fig.~\ref{fig:dh_tau_companion_intrinsic} (top).
We see that the intrinsic polarization decreases for interstellar degrees of polarization between 0\% and 0.2\% and increases for larger interstellar polarizations. 
The intrinsic polarization increases because an ever larger interstellar polarization needs to be canceled to produce the measured polarization.
For the range plotted, the intrinsic angle of polarization increases from $60^\circ$ to $100^\circ$.
Most importantly, the intrinsic degree of polarization is always higher than 0.4\%, showing that DH~Tau~B should be intrinsically polarized if the interstellar polarization indeed has an angle of polarization of $26^\circ$.
%

From the measurements by \citet{heyer_tauruspol} (white lines in Fig.~\ref{fig:dh_tau_spire}), we see that there are slight variations in the angle of polarization of the stars in the region. 
\citet{goodman_tauruspol} determined that the angles of polarization of these stars are Gaussian distributed with a mean of $27^\circ$ and a standard deviation of $15^\circ$.
Using this distribution of angles, we take a more probabilistic approach and perform a Monte Carlo simulation in which we compute for a range of interstellar degrees of polarization the probability distribution of the intrinsic polarization.
The histograms of the resulting distributions for each value of the interstellar degree of polarization are displayed in Fig.~\ref{fig:dh_tau_companion_intrinsic} (bottom).
In this figure we have normalized the distribution of each column to one.
It follows that the curves of Fig.~\ref{fig:dh_tau_companion_intrinsic} (top) are in fact among the most probable scenarios.
We also see that DH~Tau~B must be at least 0.2\% intrinsically polarized 
for interstellar degrees of polarization between 0\% and 0.3\% or higher than 0.7\%, regardless of the interstellar angle of polarization.
Only for interstellar degrees of polarization between 0.3\% and 0.7\% there is a small possibility (${\sim}8\%$) that DH~Tau~B is not intrinsically polarized.
Based on these findings, we conclude that DH~Tau~B is very likely intrinsically polarized. 


\subsection{Likely detection of intrinsic polarization of GSC~6214~B}
\label{sec:detection_gsc_6214}

\begin{figure}[!htbp]
\centering
\includegraphics[width=\hsize, trim={15 10 10 0}, clip]{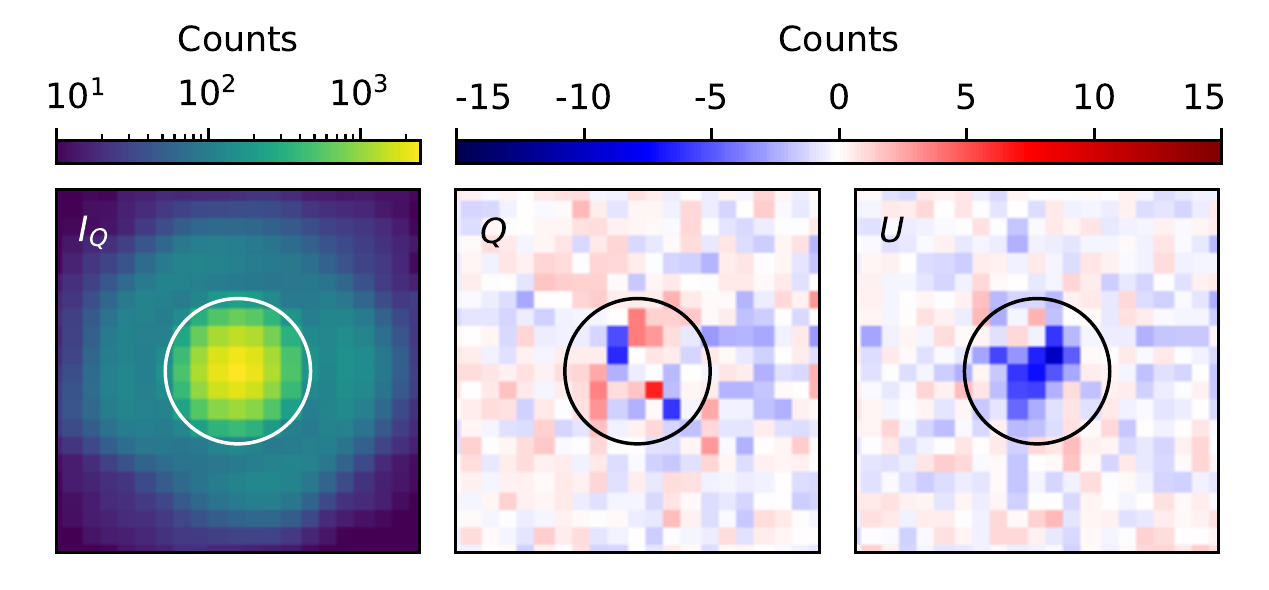}
\caption{Reduced mean-combined $I_Q$-, $Q$-, and $U$-images (after applying the cosmetic correction described in Appendix~\ref{app:yinyang}) at the position of the companion GSC~6214~B, showing an aperture of radius 4~pixels centered on the companion. The $I_U$-image, which is not shown, is very similar to the $I_Q$-image.}
\label{fig:gsc_6214_signals} 
\end{figure}

In this section we present the likely detection of intrinsic polarization originating from GSC~6214~B.
Figure~\ref{fig:gsc_6214_signals} shows the reduced $I_Q$-, $Q$-, and $U$-images in $H$-band at the position of the companion of the data set created by mean-combining the final images of the three data sets.
Table~\ref{tab:detection_gsc_6214} shows the measured polarization of GSC~6214~B for the three individual data sets and the mean-combined one. 
Similar to the DH~Tau data, the measured polarization signals of each data set are within the uncertainties constant with aperture radius.
We select a final aperture radius of 4~pixels, corresponding to the (approximate) radius where the S/N in $q$ and $u$ is maximized in each of the data sets.
Overall the measured degree and angle of polarization of the data sets are consistent within the uncertainties.
The slightly different results of the 2019-02-22 data set compared to the other two data sets could be caused by the relatively strong time-varying atmospheric conditions that the observations were taken under (see Table~\ref{tab:observations}).
Whereas the $q$- and $u$-measurements of the three data sets individually do not reach the required $5\sigma$-limit for a detection, the mean-combined measurement does, reaching an S/N of 5.2 in $u$.
From the mean-combined data we therefore conclude that we detect significant polarization from GSC~6214~B, with a degree and angle of polarization of $0.23 \pm 0.04\%$ and $138 \pm 5^\circ$, respectively.
%
%
%
\begin{table*}
\caption{Degree and angle of linear polarization, including the uncertainties, of the parent star GSC~6214~A and the companion GSC~6214~B as measured in $H$-band for each of the three data sets and the data set created by mean-combining the final images of the three data sets.}
\centering
\begin{threeparttable}
\begin{tabular}{l c c c c c c}
\hline\hline
\T\B
Data set & $P_\mathrm{star}$ (\%) & $\chi_\mathrm{star}$ ($^\circ$) & $P_\mathrm{com}$ (\%) & $\chi_\mathrm{com}$ ($^\circ$) & S/N $q_\mathrm{com}$ & S/N $u_\mathrm{com}$ \\
\hline
2019-02-22 & $0.17 \pm 0.05$ & $27 \pm 9\phantom{0}$ & $0.18 \pm 0.07$ & $143 \pm 16$ & 0.5 & 1.9 \T \\
2019-08-06 & $0.18 \pm 0.06$ & $72 \pm 13$ & $0.26 \pm 0.07$ & $137 \pm 8\phantom{0}$ & 0.2 & 3.6 \T \\
2019-08-07 & $0.08 \pm 0.04$ & $70 \pm 17$ & $0.24 \pm 0.07$ & $139 \pm 9\phantom{0}$ & 0.4 & 2.9 \T \\
Mean combined & $0.10 \pm 0.03$ & $54 \pm 9\phantom{0}$ & $0.23 \pm 0.04$ & $138 \pm 5\phantom{0}$ & 0.5 & \hspace{-1pt}5.2 \B \\
\hline
\end{tabular}
\begin{tablenotes}
\vspace{5pt}
\item \hspace{-3pt}\textbf{Notes.} The meaning of the column headers is described in the notes of Table~\protect\ref{tab:detection_dh_tau}.
\end{tablenotes}
\end{threeparttable}
\label{tab:detection_gsc_6214}
\end{table*}

Table~\ref{tab:detection_gsc_6214} also shows the stellar degrees and angles of polarization.
Because we do not spatially resolve a disk around GSC~6214~A, we used a star-centered aperture extending up to and including the AO residuals to maximize the S/N.
The measured signals show significant differences and are overall inconsistent among the data sets.
The signals average to a degree of polarization of only 0.10\%.
The measurements of the stellar polarization are therefore most likely dominated by spurious signals.
To determine whether the companion is truly polarized, we need to investigate the potential origins of these spurious signals and the effect they have on the measurement of the companion polarization.

If the stellar polarization primarily results from uncorrected instrumental polarization, which to first order equally affects the star and the companion, we would need to subtract these signals from the images.
Using the mean-combined images with the stellar polarization subtracted, we measure for the companion a degree and angle of polarization of $0.32 \pm 0.04\%$ and $141 \pm 4^\circ$, respectively, with an S/N of 1.4 in $q$ and 7.2 in $u$.
This polarization signal is larger and more significant than that measured from the images without the stellar polarization subtracted (see Table~\ref{tab:detection_gsc_6214}).
However, the measured signals are less consistent among the data sets, suggesting that uncorrected instrumental polarization may not be the principal cause of the stellar polarization.

A more likely scenario seems that the stellar polarization signals are dominated by systematic effects due to time-varying atmospheric conditions and AO performance in combination with the coronagraph, similar to the case of DH~Tau (see Sect.~\ref{sec:detection_dh_tau}).
Also in the case of GSC~6214, the systematic effects do not affect (as much) the companion measurements because those measurements are overall consistent among the data sets.
This suggests that the measurements of the companion are more reliable than those of the star.
Because the companion polarization is significantly different from the stellar polarization in all data sets, particularly in the angle of polarization (see Table~\ref{tab:detection_gsc_6214}), and we measure significant polarization from the companion for both the reduction with and without the stellar polarization subtracted (reaching S/Ns of 7.2 and 5.2 in $u$, respectively), we conclude that the companion is most likely truly polarized. 

\begin{figure}[!htbp]
\centering
\includegraphics[width=\hsize, trim={5 5 5 5}, clip]{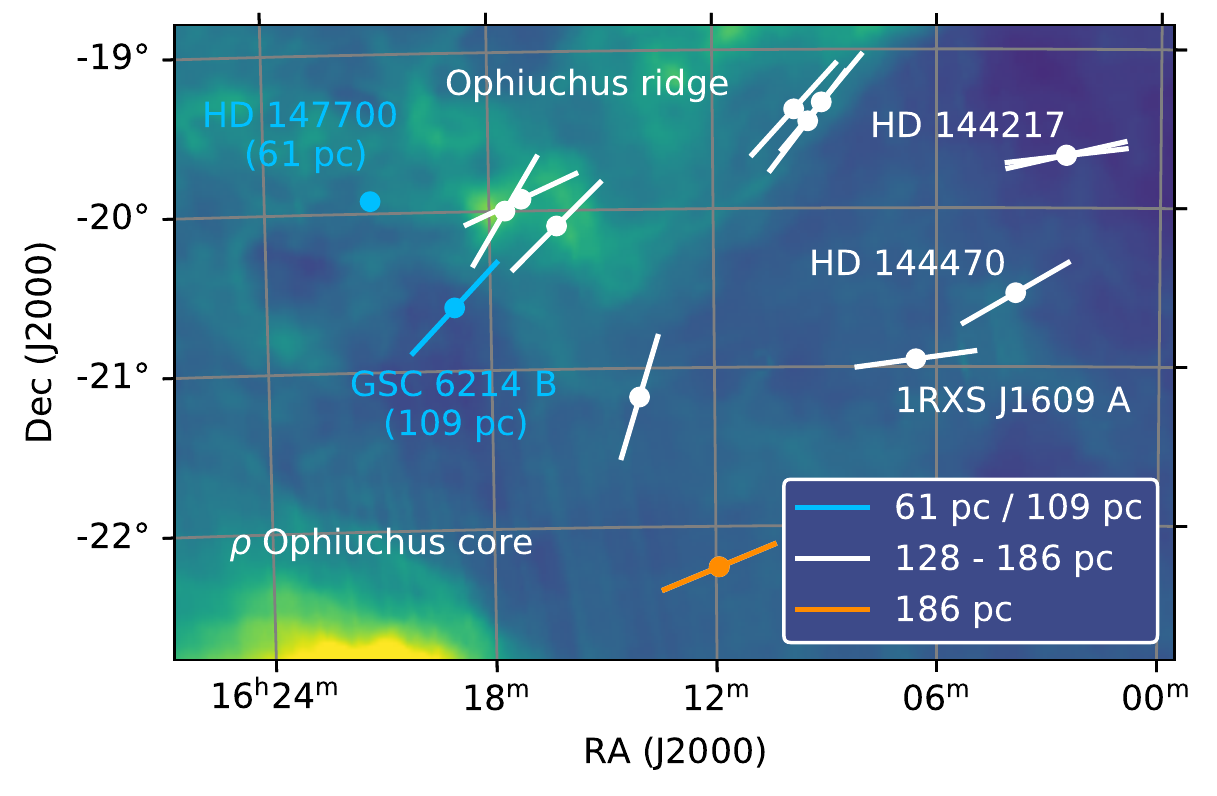}
\caption{Map of the angle of linear polarization of the companion GSC~6214~B, the star 1RXS~J1609~A, and other nearby bright stars superimposed on an IRAS map at \SI{100}{\micro\meter}. 
The angles of GSC~6214~B and 1RXS~J1609~A are from the SPHERE-IRDIS $H$-band measurements from this work, whereas for the other stars the angles are taken from the catalog of optical measurements by~\protect\citet{heiles_standardstars}.
The length of the lines is arbitrary and contrary to Fig.~\protect\ref{fig:dh_tau_spire} does not indicate the degree of polarization. 
HD~ 147700 in unpolarized.
White lines indicate stars at a distance between 128~pc and 142~pc, and blue and orange lines show objects closer or farther away, respectively.
We note that the region shown is much larger than that of Fig.~\protect\ref{fig:dh_tau_spire}, and therefore the angular separation among the stars is much larger as well.}
\label{fig:gsc_6214_iras}
\end{figure}

To determine whether the polarization of GSC~6214~B is intrinsic to the companion or caused by interstellar dust, we show in Figure~\ref{fig:gsc_6214_iras} a map of the angles of polarization of nearby bright stars from the catalog by~\citet{heiles_standardstars}.
The map is displayed over an IRAS~\citep{neugebauer_iras} \SI{100}{\micro\meter} map that shows the dust concentrations in the region of the Ophiuchus molecular cloud complex where GSC~6214 is located. 
Comparing the angle of polarization of GSC~6214~B and the nearby stars, it may seem that the companion is polarized by interstellar dust.
However, GSC~6214 is located at 109~pc, whereas estimates for the distance of the Ophiuchus molecular cloud complex range from approximately 120 to 150~pc~\citep[e.g.,][]{mamajek_ophdistance, lombardi_ophlupusdistance, ortizleon_distances1, yan_molecularclouddistances}. 
Indeed, the three stars closest to GSC~6214 in Fig.~\ref{fig:gsc_6214_iras} are located at 128 to 131~pc.
We therefore consider it more likely that GSC~6214 is located in front of the main concentrations of dust. 
In addition, if the companion were polarized by interstellar dust, we would expect to measure in all data sets a stellar polarization with the same angle of polarization as the companion (which is the case for 1RXS~J1609; see Sect.~\ref{sec:detection_1rxs_j1609}).
In principle it is possible that GSC~6214~A is not significantly polarized because the interstellar polarization is canceled by intrinsic polarization due to an unresolved circumstellar disk.
However, this scenario seems very unlikely because \citet{bowler_gsc6214} do not detect a disk with ALMA and put an upper limit on the disk's mass as low as 0.0015\% of the mass of the star.
Taking into account all considerations, we conclude that it is likely that the polarization we measure for GSC~6214~B is intrinsic to the companion, but we stress that we are less confident than for DH~Tau~B.
%
%


\subsection{Detection of interstellar polarization from 1RXS~J1609~B}
\label{sec:detection_1rxs_j1609}

In this section we present the detection of polarization in the 1RXS~J1609 system.
In all six data sets of 1RXS~J1609, we consistently measure within the uncertainties the same degree and angle of polarization for the central star 1RXS~J1609~A.
In the mean-combined data set, which uses the four highest-quality data sets (2019-08-06, 2019-08-29, 2019-09-17B, and 2019-09-23), we measure for the star a degree and angle of polarization of $0.21 \pm 0.01 \%$ and $97 \pm 2^\circ$, respectively. 
In the same data set, we measure for the companion $0.2 \pm 0.1 \%$ and $95 \pm 26^\circ$, using an aperture radius of 5~pixels.
Although 
the measurement of the companion polarization is not a significant detection, it is striking that it is within the uncertainties the same as the measured stellar polarization.
In the six individual data sets we also measure the polarization of the companion to be consistent with that of the mean-combined data, although with higher uncertainties. 
Finally, using the mean-combined data, we measure for the relatively bright background object that is also visible in the field of view a degree and angle of polarization of $0.3 \pm 0.1 \%$ and $103 \pm 50^\circ$, respectively.
Because all three objects have within the uncertainties the same degree and angle of polarization, their polarization likely originates from the same source, that is, from interstellar dust. 

To confirm this scenario, we turn to Fig.~\ref{fig:gsc_6214_iras}, which shows that 1RXS~J1609 is located in the Ophiuchus molecular cloud complex a few degrees west from GSC~6214.
Contrary to GSC~6214, 1RXS~J1609, at a distance of 139~pc, is definitely located within the dust clouds that are located at a distance of approximately 120 to 150~pc (see Sect.~\ref{sec:detection_gsc_6214}).
Indeed, the measured angles of polarization of 1RXS~J1609~A, 1RXS~J1609~B and the background object agree well with those of the nearby bright stars located at a similar distance (see Fig.~\ref{fig:gsc_6214_iras}).
\citet{serkowski_interstellarpol} have fitted their multiwavelength optical measurements of the stars HD~144470 and HD~144217 (142 and 129~pc; see Fig.~\ref{fig:gsc_6214_iras}) to Serkowski's law of interstellar polarization (see Eq.~(\ref{eq:serkowski})) and determined the values of $P_\mathrm{max}$ and $\lambda_\mathrm{max}$ for both stars.
Using these values, we find that in $H$-band the degrees of polarization are equal to approximately 0.4\% and 0.3\%, respectively. 
These values are similar to the degree of polarization we measure for the star, the companion, and the background object in our images of 1RXS~J1609, where the slight differences are likely due to the inhomogeneous spatial distribution of the interstellar dust.
We conclude that the polarization we measure for 1RXS~J1609~B originates from interstellar dust and therefore set an upper limit on the degree of polarization in Sect.~\ref{sec:upper_limits}.


\subsection{Upper limits on polarization of other companions}
\label{sec:upper_limits}

In this section we present upper limits on the degree of polarization of the 18 companions for which we do not reach the $5\sigma$-limit in $q$ or $u$ to claim a detection. 
For the majority of the companions, the S/N in $q$ and $u$ is typically ${\lesssim}$2--3 for any aperture radius.
For four companions the maximum S/N in $q$ or $u$ reaches a value of almost 4.
However, in these four cases the signals in the $Q$- and $U$-images (after the cosmetic correction described in Appendix~\ref{app:yinyang}) do not resemble scaled-down positive or negative versions of the total-intensity PSF as one would expect for real signals, but show strong pixel-to-pixel variations caused by incompletely removed bad pixels (see Appendix~\ref{app:bad_pixels}).

Table~\ref{tab:upper_limits} shows for each target the upper limits determined from the 68.27\% and 99.73\% intervals, as described in Appendix~\ref{sec:method_beta_pic}. 
For targets for which we obtained multiple data sets, we computed the upper limits from the mean-combined images. 
For the majority of the companions, which are generally the fainter ones, we determined the upper limits using an aperture radius equal to half times the full width at half maximum (FWHM) of the stellar PSF.
This aperture radius is on average 1.9~pixels in $H$-band and 2.6~pixels in $K_s$-band, and is at, or close to, the radius at which the upper limit is minimized.
For seven, generally brighter companions (CD-35~2722~B, AB~Pic~b, HD~106906~b, GQ~Lup~B, GSC~8047~B, PZ~Tel~B in $J$-band, and 1RXS~J1609~B) we used an aperture radius of 5~pixels to average out and suppress the spurious signals created by incompletely removed bad pixels (see Appendix~\ref{app:bad_pixels}).
However, the bad pixels generally still create a bias in the $q$- and $u$-signals, and so we have to accept that this increases the upper limits.
For the data sets where this bias is really strong (i.e., CD-35~2722, PZ~Tel in $J$-band, and TYC~8998), we excluded from the data reduction those frames that contribute strong bad pixels at the position of the companion in the final images.
Because HD~106906~b is located at an angular separation of $7.1\arcsec$ from the central star, which is larger than the isoplanatic angle during the observations, its PSF is strongly elongated in the radial direction from the star.
To account for this, we used an elliptically shaped aperture.
Finally, for the companions of HR~8799, HD~206893, and $\beta$~Pic, we computed the upper limits using the polarimetric images from the reduction with the added classical ADI step (see Appendix~\ref{sec:method_beta_pic}).
%
\begin{table*}
\caption{68.27\% and 99.73\% upper limits on the degree of linear polarization of the companions ($P_\mathrm{com}$), as well as the measured degree and angle of linear polarization of the central star ($P_\mathrm{star}$ and $\chi_\mathrm{star}$), for the targets for which we do not detect significant polarization.}
\centering
\begin{threeparttable}
\begin{tabular}{l c c c c c}
\hline\hline
\T\B
Target & Filter & $P_\mathrm{star}$ (\%) & $\chi_\mathrm{star} (^\circ)$ & 68.27\% upper                  & 99.73\% upper \\
       &        &                        &                               & limit on $P_\mathrm{com}$ (\%) & limit on $P_\mathrm{com}$ (\%) \\
\hline
HR 8799 b & BB\_H\phantom{$_\mathrm{s}$} & $0.057 \pm 0.006$ & $126 \pm 3\phantom{00}$ & $0.6\phantom{0}$ & $1.2$ \T \\
HR 8799 c & BB\_H\phantom{$_\mathrm{s}$} & $0.057 \pm 0.006$ & $126 \pm 3\phantom{00}$ & $0.5\phantom{0}$ & $1.1$ \T \\
HR 8799 d & BB\_H\phantom{$_\mathrm{s}$} & $0.057 \pm 0.006$ & $126 \pm 3\phantom{00}$ & $0.5\phantom{0}$ & $1.2$ \T \\
HR 8799 e & BB\_H\phantom{$_\mathrm{s}$} & $0.057 \pm 0.006$ & $126 \pm 3\phantom{00}$ & $0.6\phantom{0}$ & $1.3$ \T \\
PZ Tel B & BB\_H\phantom{$_\mathrm{s}$} & $0.05 \pm 0.03$ & $17 \pm 24$ & $0.06$ & $0.1$ \T \\
PZ Tel B & BB\_J\phantom{$_\mathrm{ss}$} & $0.13 \pm 0.01$ & $159 \pm 2\phantom{00}$ & $0.1\phantom{0}$ & $0.2$ \T \\
HR 7672 B & BB\_H\phantom{$_\mathrm{s}$} & $0.104 \pm 0.007$ & $138 \pm 2\phantom{00}$ & $0.2\phantom{0}$ & $0.3$ \\
GSC 8047 B & BB\_H\phantom{$_\mathrm{s}$} & $0.04 \pm 0.02$ & $160 \pm 39\phantom{0}$ & $0.2\phantom{0}$ & $0.3$ \\
HD 19467 B & BB\_H\phantom{$_\mathrm{s}$} & $0.054 \pm 0.005$ & $7 \pm 3$ & $0.4\phantom{0}$ & $1.0$ \\
GQ Lup B & BB\_H\phantom{$_\mathrm{s}$} & $0.94 \pm 0.02$ & $83 \pm 1\phantom{0}$ & $0.2\phantom{0}$ & $0.3$ \T \\
HD 206893 B & BB\_K$_\mathrm{s}$ & $0.15 \pm 0.06$ & $107 \pm 15\phantom{0}$ & $0.8\phantom{0}$ & $1.7$ \\
HD 4747 B & BB\_K$_\mathrm{s}$ & $0.11 \pm 0.02$ & $71 \pm 7\phantom{0}$ & $0.3\phantom{0}$ & $0.6$ \\
CD-35 2722 B & BB\_H\phantom{$_\mathrm{s}$} & $0.15 \pm 0.03$ & $66 \pm 6\phantom{0}$ & $0.1\phantom{0}$ & $0.3$ \T \\
AB Pic b & BB\_H\phantom{$_\mathrm{s}$} & $0.05 \pm 0.01$ & $6 \pm 8$ & $0.07$ & $0.2$ \T \\
HD 106906 b & BB\_H\phantom{$_\mathrm{s}$} & $0.097 \pm 0.008$ & $68 \pm 2\phantom{0}$ & $0.2\phantom{0}$ & $0.3$ \\
PDS 70 b & BB\_K$_\mathrm{s}$ & $1.1 \pm 0.1$ & $62 \pm 3\phantom{0}$ & $5.0\phantom{0}$ & $12$ \T \\
PDS 70 b & BB\_H\phantom{$_\mathrm{s}$} & $0.97 \pm 0.02$ & $65 \pm 1\phantom{0}$ & $9.2\phantom{0}$ & $22$ \T \\
1RXS J1609 B & BB\_H\phantom{$_\mathrm{s}$} & $0.21 \pm 0.01$ & $97 \pm 2\phantom{0}$ & $0.2\phantom{0}$ & $0.5$ \\
$\beta$~Pic b & BB\_H\phantom{$_\mathrm{s}$} & $0.075 \pm 0.008$ & $163 \pm 4\phantom{00}$ & $0.2\phantom{0}$ & $0.4$ \\
TYC 8998 b & BB\_H\phantom{$_\mathrm{s}$} & $0.12 \pm 0.09$ & $0 \pm 4$ & $0.3\phantom{0}$ & $0.6$ \T \\
\hline
\end{tabular}
\end{threeparttable}
\label{tab:upper_limits}
\end{table*}

Table~\ref{tab:upper_limits} also shows for each target the stellar degree and angle of polarization.
For the majority of the stars the degree of polarization is around 0.1\%. 
To be conservative and because we generally do not know the origin of these low polarization signals (intrinsic, interstellar dust or spurious), we interpret the signals as biases.
For these targets we therefore computed the upper limits on the companion polarization from both the reductions with and without the stellar polarization subtracted, and show the highest values in Table~\ref{tab:upper_limits}.
For three targets we measure a stellar polarization higher than approximately 0.1\%.
In the case of GQ~Lup and PDS~70 this stellar polarization is caused by a circumstellar disk (see~\citealt{keppler_pds70} and Sect.~\ref{sec:circumstellar_disks}).
Although GQ~Lup is located in the Lupus~I cloud, the contribution of interstellar dust is likely small because HD~141294, the nearest bright star to GQ~Lup (at $14.3\arcmin$ and a distance of 153~pc compared to 151~pc for GQ~Lup), is unpolarized at optical wavelengths~\citep{rizzo_lupuspol, alves_lupuspol}.
For PDS~70 and GQ~Lup we therefore determined the upper limits using only the images without the stellar polarization subtracted.
For 1RXS~J1609 on the other hand, the stellar polarization is caused by interstellar dust (see Sect.~\ref{sec:detection_1rxs_j1609}), and we therefore used the reduction where the stellar polarization is subtracted.

Examining the upper limits in Table~\ref{tab:upper_limits}, we see that for 11 companions the 68.27\% upper limits are ${\leq}0.3\%$, with the lowest upper limit equal to 0.06\% for PZ~Tel~B in $H$-band.
These low upper limits are in almost all cases dominated by the photon noise from the companion in the $Q$- and $U$-images or the bias due to incompletely removed bad pixels.
The upper limits are still larger than the (minimum) polarimetric accuracy of the Mueller matrix model with which the data have been corrected~\citep[see][]{vanholstein_irdis2}.
For the companions of HR~8799, HD~19467, and HD~206893, which are fainter or located at a much smaller separation than the other companions, the 68.27\% upper limits are dominated by the uncertainty of the background in $Q$ and $U$ and have values between 0.4\% and 0.8\%.
For the very close-in planet PDS~70~b we reach upper limits of 5.0\% in $K_s$-band and 9.2\% in $H$-band. 
These upper limits are so high because the comparison apertures contain signal from the inner circumstellar disk of PDS~70~A (see Fig.~\ref{fig:circumstellar_disks}) and the Student's $t$-distribution imposes a large statistical penalty for the low number of available comparison apertures (see Appendix~\ref{sec:method_beta_pic}).
We note that for PDS~70~c~\citep{haffert_pds70}, the circumstellar disk prevents us from measuring the polarization altogether.
Finally, we reach the highest polarimetric point-source contrast in the mean-combined data set of $\beta$~Pic, with a $1\sigma$-contrast of $3\cdot10^{-8}$ at a separations of $0.5\arcsec$ and a contrast below $10^{-8}$ for separations ${>}2.0\arcsec$ (see Appendix~\ref{app:contrast_curve}).
Overall, it follows that our measurements are sensitive to polarization signals of around a few tenths of a percent.


\subsection{Detection of circumstellar disks of DH~Tau, GQ~Lup, PDS~70, $\beta$~Pic, and HD~106906}
\label{sec:circumstellar_disks}

\begin{figure*}[!htbp]
\centering
\includegraphics[width=\hsize]{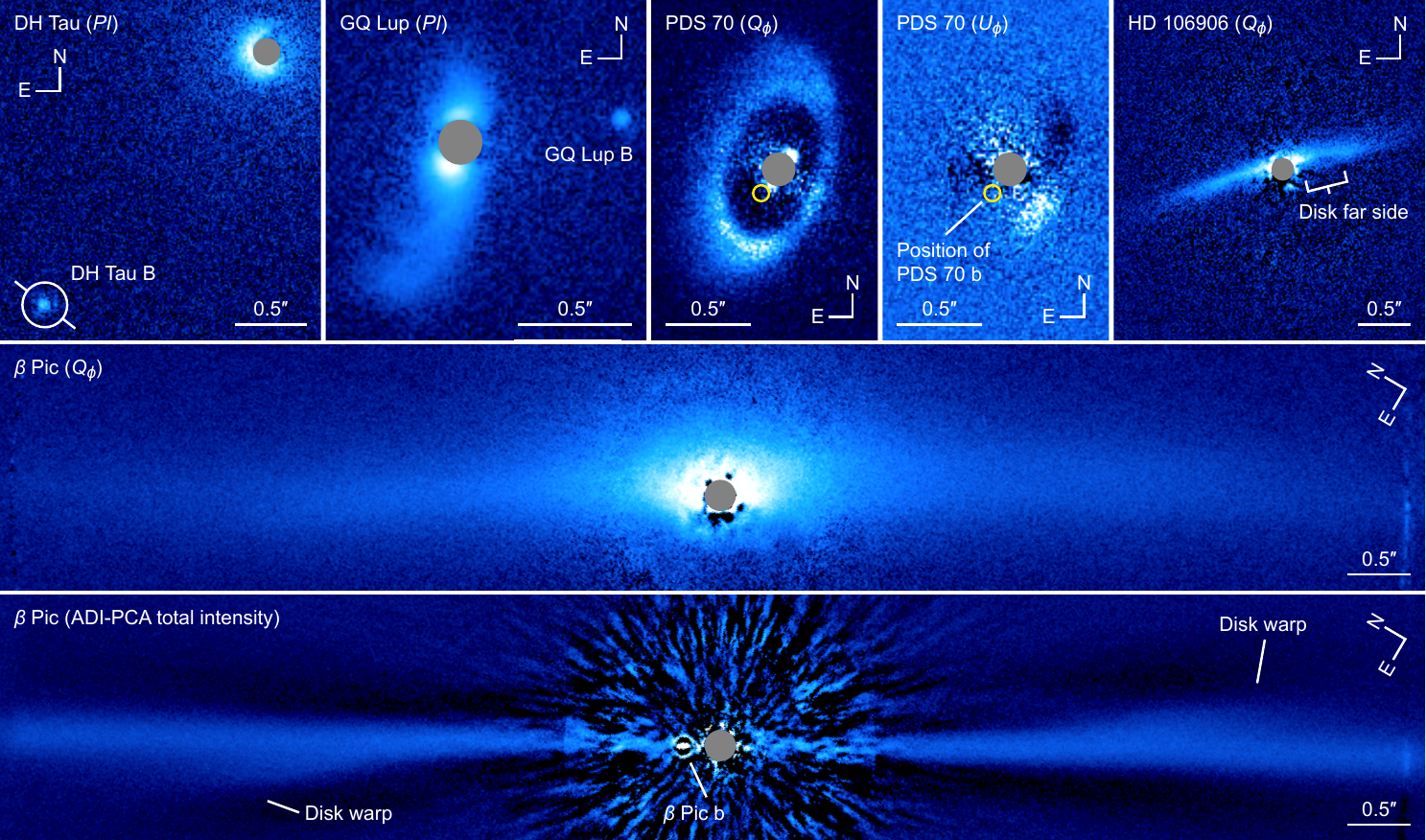}
\caption{$H$-band images of the five circumstellar disks detected in our sample, showing the linearly polarized intensity ($PI$) for DH~Tau and GQ~Lup, $Q_\phi$ and $U_\phi$ for PDS~70, and $Q_\phi$ for HD~106906 and $\beta$~Pic. 
A total-intensity image after applying ADI with PCA (subtracting four principal components) is additionally shown for $\beta$~Pic. 
The $U_\phi$-image of PDS~70 is shown on a linear scale, whereas all other images are shown on a logarithmic scale.
All polarimetric images have the polarized stellar halo subtracted.
The angular scale and sky orientation are indicated in each image. 
Gray circles mask regions obscured by the coronagraph.}
\label{fig:circumstellar_disks}
\end{figure*}

In our survey we also detected the five circumstellar disks displayed in Fig.~\ref{fig:circumstellar_disks}. 
Although the disks of DH~Tau and GQ~Lup have already been detected at mm-wavelengths~\citep{wolff_dhtau, macgregor_gqlup, wu_gqlupdisk}, here we present the first images in polarized scattered light, revealing various interesting features.
For PDS~70, HD~106906, and $\beta$~Pic near-infrared polarimetric images already exist~\citep{keppler_pds70, hashimoto_pds70, kalas_hd106906, millar_betapic}, but our images are generally deeper, reveal new features, or confirm features that were previously observed.
In this section, we therefore briefly discuss these disks, although we consider a detailed analysis beyond the scope of this paper.
%

Figure~\ref{fig:circumstellar_disks} (top left) shows the polarized intensity image of the DH~Tau system, with the circumstellar disk visible in the top right corner of the panel. 
The relatively small disk has a diameter of approximately $0.50\arcsec$ or 67~au at 135~pc.
From ALMA measurements of the Keplerian rotation of the disk, \citet{sheehan_diskrotation} have determined an inclination of $48^\circ$ 
and a position angle of $2.5^\circ$ 
(east of north), with the northern side of the disk rotating toward us (i.e.,~blue shifted).
In our images the disk has a smooth intensity profile with no visible disk gap, rings, or spirals.
A strong brightness asymmetry is visible between the eastern and western sides of the disk, which could be related to the viewing angle of the disk and the dust scattering properties.
This asymmetry is compatible with the position angle derived from ALMA: if the side inclined toward the Earth appears brightest due to enhanced forward scattering, then the eastern side is the forward-scattering near side of the disk.
Alternatively, this brightness asymmetry could result from shadowing by an unresolved inner disk component because the brightness changes quite abruptly with azimuth.
The brightness asymmetry might extend toward the inner (coronagraphically masked) parts of the disk because the angle of polarization that we measure for the average stellar polarization ($128^\circ$, see Table~\ref{tab:detection_dh_tau}) agrees well with the angle of polarization one obtains when integrating over the non-obscured parts of the disk.
In the bottom left corner of the panel the polarization signal of DH~Tau~B is visible, where the angle of polarization is indicated with the two lines protruding from the circle around the companion.

Figure~\ref{fig:circumstellar_disks} (top row, second column) shows the polarized intensity image of the circumstellar disk and companion of GQ~Lup.
From ALMA images~\citep{macgregor_gqlup, wu_gqlupdisk}, which show a rather featureless disk, the disk inclination and position angle are known to be $60^\circ$ and $346^\circ$, respectively.
Our scattered light images show a remarkable north-south asymmetry in the circumstellar disk, with the southern part of the disk extending out to $0.84\arcsec$ (127~au at 151~pc) and the northern part only out to $0.49\arcsec$ (74~au).
Two spiral-like features can be seen protruding eastward from the southern part of the disk.
The disk asymmetry and spiral-like features are reminiscent of those of the disk around RY~Lup~\citep{langlois_rylup} and could be the result of periodic close passes of GQ~Lup~B~\citep[see e.g.,][]{dong_diskspiralsgaps, cuello_diskflybys1, cuello_diskflybys2}.
The orbital analyses presented by~\citet{schwarz_gqlup} and \citet{wu_gqlupdisk} indeed show that the orbit of GQ~Lup~B is almost certainly eccentric and that it is quite likely that the inclinations of the orbit and the disk are similar.
However, \citet{wu_gqlupdisk} argue that although the inclinations may be similar, the disk and companion orbit are likely not in the same plane.
We also find that the starlight of GQ~Lup is polarized due to the unresolved part of the circumstellar disk, with an angle of polarization ($83 \pm 1\degr$; see Table~\ref{tab:upper_limits}) approximately perpendicular to the position angle of the disk.
GQ~Lup~B appears to be polarized in Figure~\ref{fig:circumstellar_disks} (top row, second column), but this polarization is spuriously created by subtracting the stellar polarization from the image.
We will present a dynamical analysis of the complete system and detailed radiative transfer and hydrodynamical modeling of the circumstellar disk in a future paper (van Holstein et al. in prep.). 

Figure~\ref{fig:circumstellar_disks} (top row, third and fourth columns) show the $H$-band $Q_\phi$- and $U_\phi$-images of the circumstellar disk around PDS~70.
The disk is seen at a position angle of $159\degr$ and an inclination of $50^\circ$, with the southwestern side being the near side~\citep{hashimoto_pds70}. 
The $Q_\phi$-image clearly shows the known azimuthal brightness variations of the outer disk ring, as well as bright features close to the coronagraph's inner edge that most likely originate from the inner disk~\citep[see][]{keppler_pds70}.
The $U_\phi$-image contains significant signal, with the maximum value equal to ${\sim}49\%$ of the maximum in the $Q_\phi$-image, revealing the presence of non-azimuthal polarization.
The pattern in $U_\phi$ agrees well with the radiative transfer models by~\citet{canovas_nonazimuthaldisk}, indicating that part of the photons are scattered more than once.
The $Q_\phi$-image also shows a weak spiral-like feature extending toward the east from the northern ansa of the disk and perhaps a similar feature at the southern ansa.
With these features the disk resembles the model images by~\citet{dong_diskspiralsgaps} for the inclination and position angle of the PDS~70 disk.
We may therefore be seeing the effect of two spiral arms in the outer disk ring, potentially induced by PDS~70~b.

Figure~\ref{fig:circumstellar_disks} (top right) shows the $Q_\phi$-image of the debris disk of HD~106906, which is viewed close to edge-on.
The forward-scattering near side of the disk can be seen passing slightly to the north of the star.
The image clearly shows the known east-west brightness asymmetry of the disk, which had until now only been detected in total intensity~\citep[][]{kalas_hd106906, lagrange_hd106906}.
Because our data are particularly deep (i.e., 120~min total on-source exposure time), we detect the backward-scattering far side of the disk to the west of the star, just south of the brighter near side of the disk~\citep[see][]{kalas_hd106906}.

Finally, Fig.~\ref{fig:circumstellar_disks} (center) shows the $Q_\phi$-image of the nearly edge-on-viewed debris disk of $\beta$~Pic.
The disk extends from one side of the $11\arcsec \times 11\arcsec$ IRDIS field of view to the other. 
Earlier near-infrared scattered light images reported by~\citet{millar_betapic} show the disk only to ${\sim}1.7\arcsec$ or 33~au at 20~pc due to the smaller field of view of GPI.
In our images we see the disk extending to at least $5.8\arcsec$ or 115~au on both sides of the star.
The disk midplane is seen slightly offset to the northwest of the star (up in Fig.~\ref{fig:circumstellar_disks} center) due to the disk's small inclination away from edge-on.
Our image also shows the apparent warp in the disk~\citep[see][and references therein]{millar_betapic} that extends eastward (to the bottom left) in the northeastern (left) part of the disk and westward (to the upper right) in the southwestern (right) part of the disk. 
This warp is particularly well visible in Fig.~\ref{fig:circumstellar_disks} (bottom), which shows a total-intensity image after applying ADI with PCA using IRDAP.


%
%

\section{Modeling of polarization from circumsubstellar disks}
\label{sec:modeling_disks}

As discussed in Sects.~\ref{sec:detection_dh_tau} and \ref{sec:detection_gsc_6214}, we (very) likely detected intrinsic polarization from DH~Tau~B and GSC~6214~B, with a degree of polarization of several tenths of a percent in $H$-band.
The host stars of these two companions are among the youngest in our sample ($\lesssim$20~Myr) and the companions have indicators for the presence of circumsubstellar disks through hydrogen emission lines, red near-infrared colors, and excess emission at mid-infrared wavelengths (see Fig.~\ref{fig:companions_properties} and Table~\ref{tab:companions_properties}). 
Therefore, the most plausible explanation for the polarization in these cases is scattering of the companion's thermal emission by dust within a spatially unresolved circumsubstellar disk. 
However, we note that the late M to early L spectral types of these low-mass companions (see Fig.~\ref{fig:companions_properties} and Table~\ref{tab:companions_properties}) suggest their atmospheres could be dusty.
As a result, the polarization could also originate from rotation-induced oblateness, an inhomogeneous cloud distribution, or a combination of these atmospheric asymmetries and a disk~\citep[see][]{stolker_exopol}.
Still, it seems reasonable to assume that the polarization is solely caused by a disk because the companions have low projected rotational velocities~\citep{bryan_pmcspin, xuan_dhtaub}, and out of the 20 companions observed, we only detect intrinsic polarization for the companions that have hydrogen emission lines. 

In this section we perform (spatially resolved) radiative transfer modeling of a representative example of a circumsubstellar disk to investigate whether our detections of polarization of several tenths of a percent can really be explained by such disks.
To this end, we first describe the setup of the radiative transfer model in Sect.~\ref{sec:model_setup}. 
We then examine the generation of an integrated (i.e., spatially unresolved) polarization signal in Sect.~\ref{sec:integrated_signal} and the dependence of the polarization on the properties of the disk in Sect.~\ref{sec:radius_density}.
We stress that we consider an isolated circumsubstellar disk (i.e.,~it is not embedded in a circumstellar disk) and that our models are general and not tailored to either DH~Tau~B or GSC~6214~B.
Because we only study the degree and angle of polarization produced by the disk, the exact spectrum of the companion has little effect on the results.
In Sects.~\ref{sec:discussion_disks} and \ref{sec:discussion_no_disks} we use the results of our modeling to interpret and discuss our measurements.

\subsection{Setup of the radiative transfer model}
\label{sec:model_setup}

To quantify the expected near-infrared polarization from a self-luminous atmosphere with a circumsubstellar disk, we computed a radiative transfer model with \texttt{MCMax} \citep{min2009}, which is a Monte Carlo radiative transfer code for axisymmetric disks that is optimized for the high optical depths in protoplanetary disks. 
The model considers a passive, irradiated disk around a self-luminous substellar atmosphere (the contribution from the light of the central star is negligible). We selected a synthetic spectrum from the \texttt{BT-Settl} atmospheric models \citep{allard2012} at an effective temperature $T_\mathrm{eff} = 2000$~K and surface gravity $\log{g} = 4.0$~dex. We then scaled the spectrum to a luminosity of $10^{-4}$~$L_\odot$ by assuming a radius for the atmosphere of 2~$R_\mathrm{Jup}$ at an age of $\sim$10~Myr \citep[e.g.,][]{baraffe2015}. 
We then modeled the circumsubstellar disk as a scaled down version of a circumstellar disk~\citep[see e.g.,][]{williams_protoplanetarydisks}.
We parametrized the structure of the circumsubstellar disk with a profile for the dust surface density that is inversely proportional to the radius, $\varSigma \propto r^{-1}$. 
Using a surface density at the inner radius of $\varSigma_\mathrm{in} = 0.07~\mathrm{g}~\mathrm{cm}^{-2}$ and an inner and outer disk radius of $R_\mathrm{in} = 0.003$~au and $R_\mathrm{out} = 0.01$~au, we computed the total mass residing in the solids.
For the pressure scale height, we used a linear dependence with the disk radius, $h \propto r$, with a (constant) aspect ratio of $h/r = 0.1$.
The dust opacities contain by volume 60\% silicates, 15\% amorphous carbon, and 25\% porosity \citep{woitke2016}. Furthermore, we used a maximum hollow volume ratio of 0.8 for the distribution of hollow spheres, which approximates the irregularity of the dust grains \citep{min2016}. The size distribution of the grains was chosen in the range of 0.05--3000~\si{\micro\meter} with a power-law exponent of $-3.5$. Dust settling is included with the prescription from \citet{dubrulle1995}, which assumes an equilibrium between turbulent mixing and gravitational settling. In this way, the dust scale height is a function of disk radius and grain size, which is controlled by the viscosity parameter $\alpha = 10^{-4}$.

\subsection{Origin of the spatially integrated polarization}
\label{sec:integrated_signal}

\begin{figure*}[!htbp]
\centering
\includegraphics[width=0.9\hsize, trim={5 8 5 5}, clip]{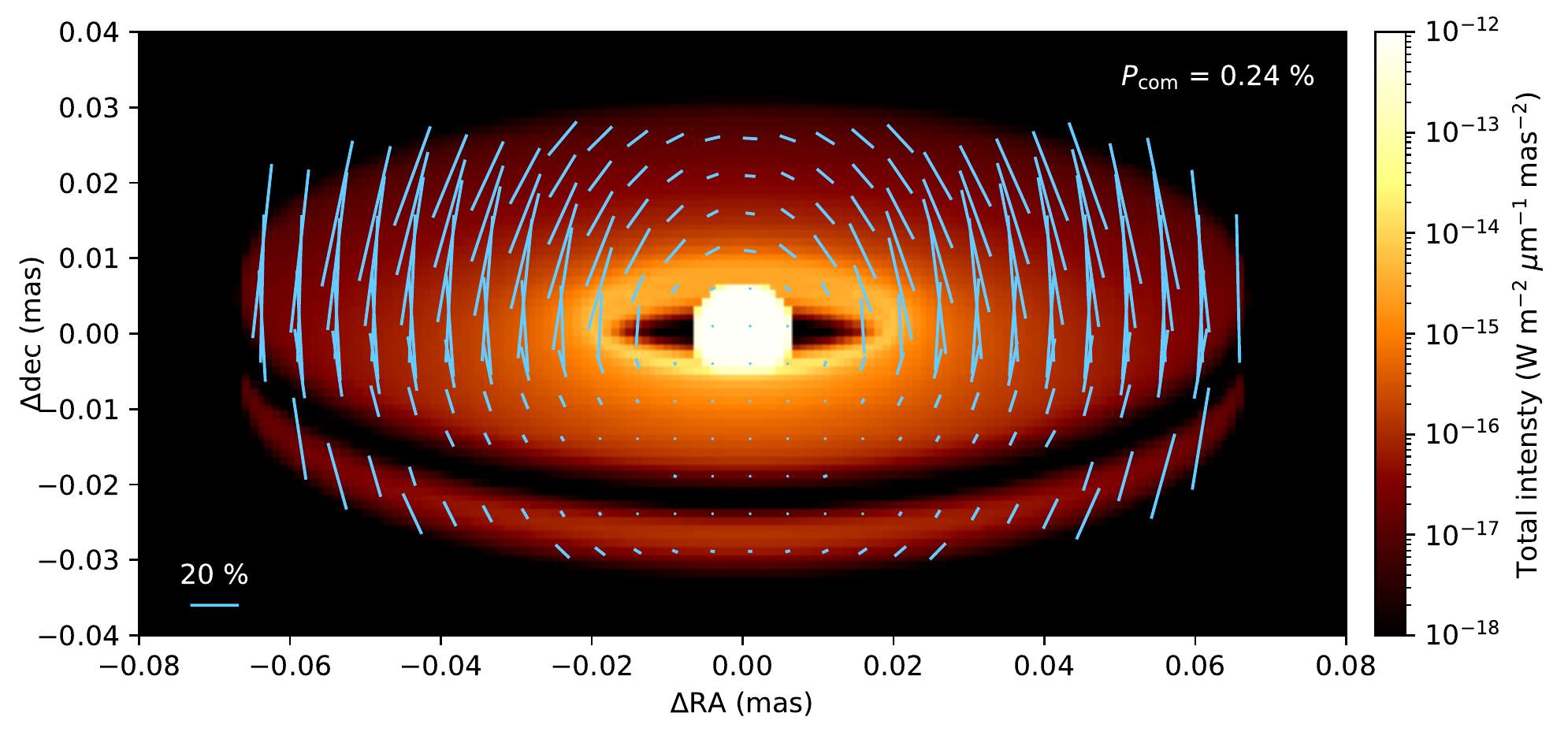} 
\caption{Synthetic image of a self-luminous companion ($T_\mathrm{eff} = 2000$~K) with a compact circumsubstellar disk at a distance of 150~pc. The total intensity surface brightness is shown on a logarithmic scale and the lines indicate the local degree and angle of linear polarization. The spatially integrated degree of polarization is 0.24\% and the angle of polarization is $0\degr$, that is, the spatially unresolved light is linearly polarized along the minor axis of the disk.}
\label{fig:mcmax_image}
\end{figure*}

After setting up the disk and dust properties, we can now perform the radiative transfer computations to study the generation of a spatially integrated polarization signal from the disk. We propagate the Monte Carlo photons through the disk to compute the thermal structure and the local source function. We then run a monochromatic ray tracing at \SI{1.62}{\micro\meter} (the central wavelength of the IRDIS $H$-band filter) to compute the synthetic total-intensity and Stokes $Q$- and $U$-images. Figure~\ref{fig:mcmax_image} displays an example image of the total-intensity surface brightness for a disk inclination of $70\degr$. In this figure, the length and orientation of the lines indicate the local degree and angle of polarization, respectively. Finally, we compute the spatially integrated polarization using the sum of the pixel values in each of the Stokes images. In Fig.~\ref{fig:mcmax_image}, this results in an integrated degree and angle of polarization of 0.24\% and $0\degr$, respectively. Indeed, the polarized flux is largest along the major axis of the disk, where scattering angles are closest to $90\degr$, yielding a net polarization that is oriented perpendicular to the major axis of the disk. In fact, the angle of polarization is always perpendicular to the position angle of the disk, independent of the disk inclination.

For the interpretation of a nonzero integrated polarization, we need to consider various effects that are visible in the spatially resolved image of the disk in Fig.~\ref{fig:mcmax_image}. 
To produce a measurable degree of polarization, the linearly polarized intensity should have a nonzero value while lowering the total intensity will further enhance the degree of polarization. 
In the example of Fig.~\ref{fig:mcmax_image}, most of the polarized flux comes from the inner edge of the disk, where the flux in total intensity is about 10 to 100 times lower than the atmospheric emission. Part of the polarization signal is canceled because there is both horizontally and vertically polarized flux, but a net vertically polarized flux remains. The local degree of polarization increases along the major axis of the disk toward larger separations because of reduced multiple scattering. However, the total intensity is also lower in these regions such that the polarized intensity is also low there. 
This means that the integrated polarization depends primarily on the inner radius and the surface density, whereas the outer radius, and therefore the total disk mass, are much less relevant.
Because the inner radius of the disk is at $\sim$6~$R_\mathrm{Jup}$ and the inclination is $70\degr$, part of the photosphere of the central object is obscured by the near side of the disk. This reduces the total intensity of the system such that the net degree of polarization is enhanced compared to a situation in which the full atmosphere would be visible. 

\subsection{Dependence on the inner radius and surface density}
\label{sec:radius_density}

We now investigate the dependence of the spatially integrated degree of polarization on the inner radius and the surface density at the inner radius. To this end, we run a grid of $10 \times 10$ radiative transfer models with a varying inner radius (5--41~$R_\mathrm{Jup}$) and dust surface density at the inner radius ($10^{-2}$--$10^{2}$~g~cm$^{-2}$). All other parameters are the same as in Sect.~\ref{sec:model_setup}, except for the outer radius which we changed from 0.01~au to 0.4~au. In this way, the disk remains radially sufficiently extended even though the outer radius of the disk has a negligible impact on integrated degree of polarization because most of the polarized flux comes from the inner edge. Because the total disk mass depends on the inner radius and surface density, it is different for each model.
We note that the estimated polarization may rely on additional properties of both the disk structure and the dust grains.

\begin{figure}[!htbp]
\centering
\includegraphics[width=\hsize, trim={5 5 5 5}, clip]{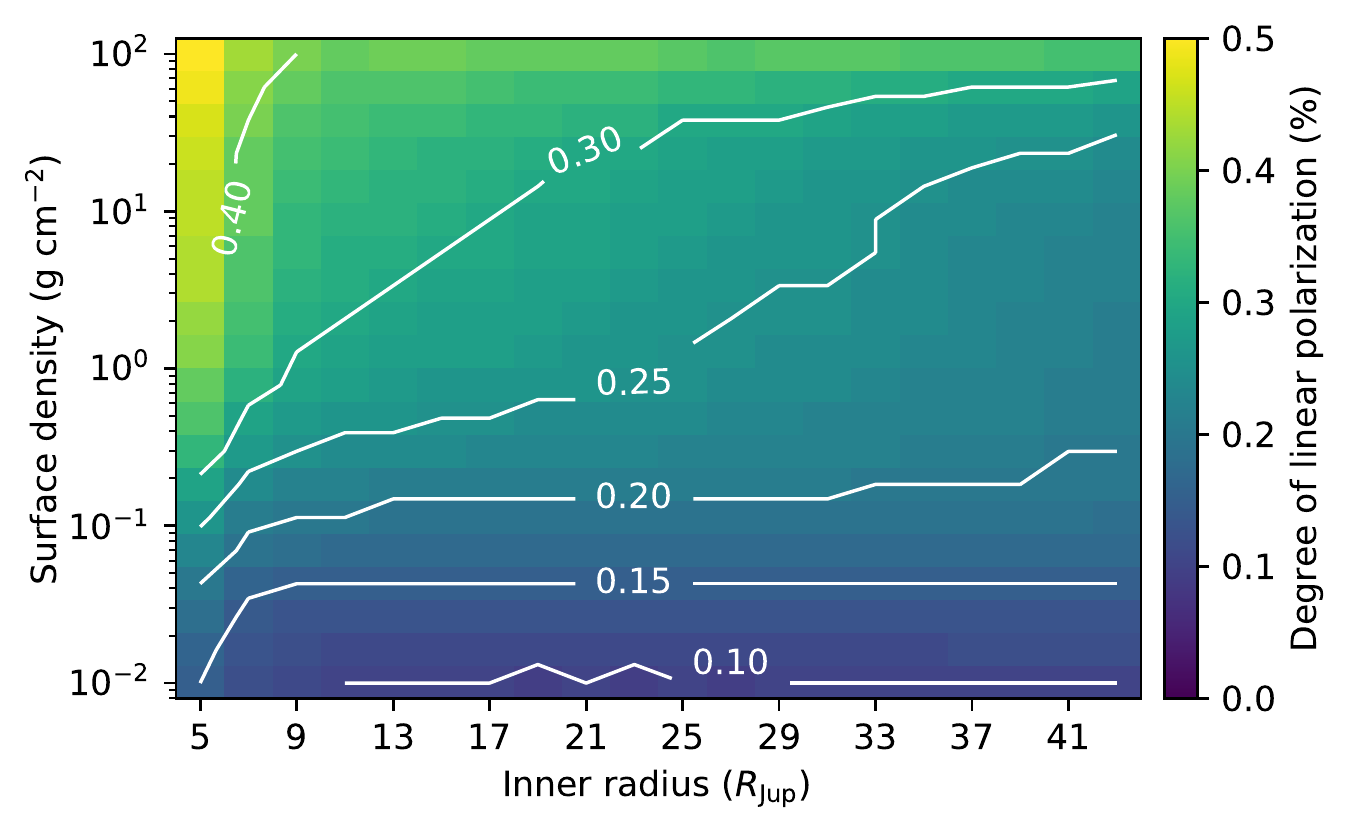}
\includegraphics[width=\hsize, trim={5 5 5 5}, clip]{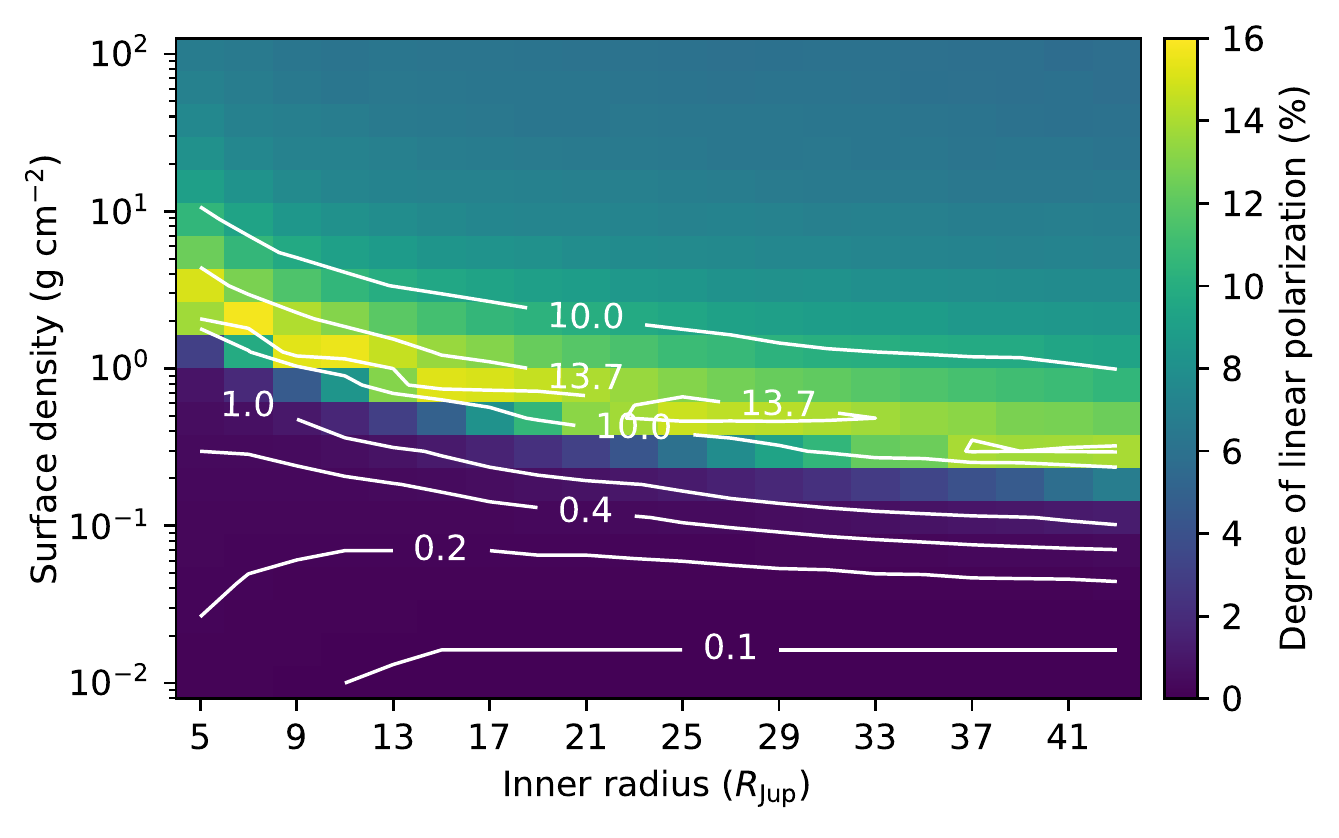}
\caption{Dependence of the integrated degree of linear polarization on the inner radius of the circumsubstellar disk and the surface density of the dust at the inner radius. The grid of radiative transfer models is shown for a disk inclination of $70\degr$ (top) and $80\degr$ (bottom).}
\label{fig:mcmax_grid}
\end{figure}

As discussed in Sect.~\ref{sec:integrated_signal}, the integrated degree of polarization depends strongly on the fractional occultation of the substellar atmosphere by the disk. This effect occurs at a high enough inclination if the projected disk reaches close to the atmosphere and/or the vertical extend of the disk (which scales with the dust surface density) is sufficiently large. 
To resolve with a high precision the obscuration of the atmosphere, we perform the ray tracing at sufficient spatial resolution. We set the disk inclination $i$ to $70\degr$ and $80\degr$ because for geometry reasons detections are biased toward highly inclined disks and, more importantly, a nonzero polarization from a circumsubstellar disk is only to be expected if the disk is sufficiently inclined. For example, we find that for $i < 45\degr$ and $i < 20\degr$, the degree of polarization is ${<}0.15\%$ and ${<}0.03\%$, respectively. We calculate the integrated degree of polarization as before and present the results for each combination of disk inner radius and the surface density at the inner disk radius in Fig.~\ref{fig:mcmax_grid}.

At an inclination of $70\degr$ (see Fig.~\ref{fig:mcmax_grid}, top), the polarization reaches a maximum value of 0.4--0.5\% when the inner radius is 5--10~$R_\mathrm{Jup}$ and the surface density is $\gtrsim$1~g~cm$^{-2}$. At small inner radii, there is a correlation with the surface density because increasing the inner radius can be counteracted by an increase in surface density in order to maintain the same integrated degree of polarization. At a given surface density, the degree of polarization converges to a constant value at larger inner radii because the atmosphere is no longer obscured and most of the polarized flux originates from the cavity edge. For higher surface densities, this turnover point occurs at larger disk radii because the scattering surface is higher.

A more extreme picture emerges when the inclination is increased to $80\degr$ (see Fig.~\ref{fig:mcmax_grid}, bottom). Whereas for surface densities $\lesssim$0.1~g~cm$^{-1}$ the correlation with the inner radius is comparable to the $i = 70\degr$ case, at higher surface densities the substellar atmosphere becomes fully obscured by the disk. There is a peak in the degree of polarization when the vertically extended disk obscures the atmosphere along the minor axis of the disk while there is still some disk surface visible close to the major axis. As a result, the total intensity of the atmosphere is strongly reduced while the polarized flux at scattering angles close to $90\degr$ is less attenuated, leading to a degree of polarization as high as $\sim$15\%. For even higher surface densities, the degree of polarization remains approximately constant at a value of $\sim$8\% because both the substellar atmosphere and the cavity edge are obscured by the vertical extent of the disk. In this case, only light that scatters through the surface layer of the disk will reach the observer, which is therefore no longer dependent on the surface density and the inner radius.


%
%

\section{Discussion}
\label{sec:discussion}

In Sect.~\ref{sec:modeling_disks} we performed radiative transfer modeling of a generic circumsubstellar disk to study the origin of the integrated polarization and the dependence of this polarization on the disk properties.
We use the results of our modeling in Sect.~\ref{sec:discussion_disks} to interpret our likely detections of polarization from DH~Tau~B and GSC~6214~B and our non-detection for GQ~Lup~B.
In Sect.~\ref{sec:discussion_no_disks} we then briefly examine the non-detections of polarization for 1RXS~J1609~B, HD~106906~b, and PDS~70~b, which also have evidence for the existence of a circumsubstellar disk.
Subsequently, we outline the implications of our upper limits on the polarization of the other companions with respect to the presence of atmospheric asymmetries in Sect.~\ref{sec:discussion_atmospheric_asymmetries}.
Finally, in Sect.~\ref{sec:outlook}, we discuss potential measurements with various instruments to confirm and further characterize the circumsubstellar disks of DH~Tau~B and GSC~6214~B.

\subsection{Circumsubstellar disks, rotational periods, and formation of DH~Tau~B, GSC~6214~B, and GQ~Lup~B}
\label{sec:discussion_disks}

As discussed in Sect.~\ref{sec:modeling_disks}, the most plausible explanation for the polarization of DH~Tau~B and GSC~6214~B is the presence of a circumsubstellar disk.
From the radiative transfer modeling in that same section, we see that the integrated degree of polarization of such a disk depends on many parameters and that estimating disk properties is therefore a degenerate problem.
Nevertheless, we can still put constraints on the dust grain sizes and the disk's inclination and position angle, and through that constrain the rotational periods and formation mechanisms of the companions.

Whereas we most likely detect polarization from the disks of DH~Tau~B and GSC~6214~B, no emission has been detected from these companions at mm-wavelengths~\citep{bowler_gsc6214, wu_radiodisk, wolff_dhtau, wu_pmcdisk}.
It is therefore possible that these disks contain mostly micrometer-sized dust grains and only little mm-sized grains, or, as suggested by~\citet{wu_radiodisk}, that the disks are compact and optically thick at mm-wavelengths.
From our polarimetric measurements we cannot determine whether the disks are really compact because most of the polarized flux originates from the inner edge of the disk~(see Sect.~\ref{sec:integrated_signal}).
Because we do not spatially resolve the disks, we can put a limit on the disk size from the measured FWHM of the PSF. The FWHM corresponds to a maximum disk radius of ${\sim}3$~au for both companions.
This radius is much smaller than one-third of the Hill radius (14--20~au, assuming the companions are on a circular orbit), which is the distance at which the disks are suspected to be truncated due to tidal interactions with the primary star~\citep{ayliffe_cpd, martin_cpd, shabram_cpd}.
However, it is possible that the disks extend beyond 3~au, but that we do not reach the sensitivity and contrast to detect the flux at the outer regions.

From the measured degree of polarization we can put constraints on the inclination of the disks.
With degrees of polarization of a few to several tenths of a percent, the disks of DH~Tau~B and GSC~6214~B must have a high inclination because a low-inclination disk will have a very low, nearly zero degree of polarization below the sensitivity that we reach with our measurements (see Sect.~\ref{sec:radius_density}). 
In fact, it could be that GQ~Lup~B hosts such a low-inclination disk because we do not detect significant polarization although the measured hydrogen emission lines are stronger than those of DH~Tau~B and GSC~6214~B~\citep{zhou_pmcaccretion}.
We also see that the inclination of the disks of DH~Tau~B and GSC~6214~B cannot be close to edge-on so that disk completely obscures the companion's atmosphere because in that case we would measure polarization degrees of several to even ten percent.
Such a high degree of polarization of 14\% has been measured for CS~Cha~B in $J$- and $H$-band by~\citet{ginski_cscha}, which the authors indeed interpret as being caused by a highly inclined and vertically extended disk.
This interpretation was recently confirmed by~\citet{haffert_cscha} using medium-resolution optical spectroscopy with MUSE.

The projected rotational velocity, $v \sin{i}$, has been measured for DH~Tau~B, GSC~6214~B and GQ~Lup~B through high-resolution spectroscopic observations~\citep{xuan_dhtaub, bryan_pmcspin, schwarz_gqlup}, finding values of $9.6 \pm 0.7$~km~s$^{-1}$, $6.1^{+4.9}_{-3.8}$~km~s$^{-1}$, and $5.3^{+0.9}_{-1.0}$~km~s$^{-1}$, respectively.
Assuming that the spin axes of the companions are perpendicular to the plane of their disks (the regular moons of our solar system's giant planets, which are believed to have formed in circumsubstellar disks, orbit near the equatorial plane of the planets) and taking the companion radii and uncertainties from~\citet{xuan_dhtaub} and \citet{schwarz_gqlup}, we constrain the rotational period of the companions using a Monte Carlo calculation.
We assume the inclination to be uniformly distributed in $\cos i$, with values between $60\degr$ and $80\degr$ for DH~Tau~B and GSC~6214~B and between $0\degr$ and $45\degr$ for GQ~Lup~B.
We find rotational periods equal to 29--37~h for DH~Tau~B, \mbox{22--77~h} for GSC~6214~B, and 19--64~h for GQ~Lup~B, within the 68\% confidence interval. 
These estimates of the rotational periods are roughly an order of magnitude larger than the average periods expected from the period-mass relation as determined from observations of free-floating low-mass brown dwarfs of similar ages~\citep[e.g.,][]{rodriguezledesma_bdrotation, scholz_bdrotation1, scholz_bdrotation2}.
This discrepancy can be explained by the companions hosting circumsubstellar accretion disks.
The estimated slow rotation of the companions, which is at ${\sim}0.1\%$ of their break-up velocities~\citep[see][]{xuan_dhtaub}, is consistent with a scenario in which the companions lose angular momentum to their disks during accretion and should still spin up as they contract~\citep[see][]{takata_despin, bryan_pmcspin, xuan_dhtaub}.
The long rotational periods we find also show that rotation-induced oblateness does not significantly contribute to the measured polarization because polarization ${>}0.1\%$ is generally expected only for rotational periods of ${\sim}6$~h or less~\citep{sengupta_bdpol2, marley_exopol}.

As we discussed in Sect.~\ref{sec:integrated_signal}, the angle of polarization we measure from an unresolved disk is always oriented perpendicular to the position angle of the disk.
Therefore, the position angle of the disk of DH~Tau~B is likely between $150\degr$ and $190\degr$ (see Fig.~\ref{fig:dh_tau_companion_intrinsic}), whereas that of the disk of GSC~6214~B is around $48\degr$ (see Table~\ref{tab:detection_gsc_6214}).
Because we already found that both disks likely have large inclinations, we have strong constraints on the 3D orientation of the disks.
The disk of DH~Tau~B is most likely misaligned with the circumstellar disk of DH~Tau~A, which has an inclination and position angle of $48\degr$ and $2.5\degr$, respectively (see Sect.~\ref{sec:circumstellar_disks} and Fig.~\ref{fig:circumstellar_disks}), although the position angles could possibly be aligned. 
Such a misalignment of disks is also found for CS~Cha~A and B~\citep{ginski_cscha}.
Although GSC~6214~A is not known to host a circumstellar disk, orbital motion has been detected for GSC~6214~B~\citep{pearce_gsc6214}.
However, the orbital elements are not sufficiently constrained to conclude on possible (mis)alignments of the disk and the orbit.
If a low-inclination disk exists around GQ~Lup~B, it would be misaligned with the circumstellar disk that has an inclination of $60\degr$ (see Sect.~\ref{sec:circumstellar_disks} and Fig.~\ref{fig:circumstellar_disks}) and the orbit of GQ~Lup~B that likely has a similar inclination.
However, the circumsubstellar disk could be aligned with the spin axis of GQ~Lup~A that has an inclination of ${\sim}30\degr$ \citep{donati_gqlup}.

The misalignment of the disks of DH~Tau~A and B, and possibly also of GQ~Lup~A and B, suggests that the companions may have formed in situ through direct collapse in the molecular cloud, akin to the formation mechanism of binary stars.
Indeed, CS~Cha~B, with its misaligned disk, was initially thought to be of substellar nature~\citep{ginski_cscha}, but was recently found to actually be a low-mass star~\citep{haffert_cscha}.
However, formation through gravitational instabilities in the circumstellar disk is also possible because this mechanism can form companions at separations of up to at least 300~au~\citep{tobin_gi}.
Although one might expect the circumstellar and circumsubstellar disks to be coplanar in this scenario, misalignments can result if the companion formed away from the midplane of the original disk, the original disk was asymmetric, or the circumstellar disk or other objects disturbed the circumsubstellar disk~\citep{stamatellos_fragmentation, bryan_obliquity}. 
It seems unlikely, however, that DH~Tau~B, GSC~6214~B, and GQ~Lup~B formed close to their stars and were subsequently scattered to a higher orbit through dynamical encounters with massive inner bodies.
This is because tidal interactions would most likely severely disturb or even destroy the circumsubstellar disks~\citep[see][]{stamatellos_fragmentation, bailey_companions, bowler_gsc6214nir}, and no massive objects at small separations, nor the gaps they would create in the circumstellar disks, have been detected~\citep[see Fig.~\ref{fig:circumstellar_disks};][]{pearce_gsc6214, wu_gqlupdisk}.


\subsection{Circumsubstellar disks of 1RXS~J1609~B, HD~106906~b, and PDS~70~b}
\label{sec:discussion_no_disks}

There is also evidence for disks around 1RXS~J1609~B, HD~106906~b, and PDS~70~b (see Table~\ref{tab:companions_properties}), but we do not detect intrinsic polarization from these companions (see Sect.~\ref{sec:upper_limits}).
It could be that these companions host a disk but that the properties and geometry of these disks is such that they do not produce a measurable degree of polarization.
However, for 1RXS~J1609~B no hydrogen emission lines are detected.
Instead, the evidence for the existence of a disk is based on red near-infrared colors, weak mid-infrared excess that is spatially unresolved between the primary star and the companion, and a moderate extinction~\citep{bailey_companions, wu_1rxsj1609}.
Because we find that the companion is polarized by interstellar dust (see Sect.~\ref{sec:detection_1rxs_j1609}), it seems more likely that these properties are caused by interstellar dust rather than a circumsubstellar disk.

As discussed in Sect.~\ref{sec:upper_limits}, we placed an upper limit of 0.2\% on the degree of polarization of HD~106906~b, with a 68.27\% confidence level.
Because also no hydrogen emission lines are detected for this companion, a possible explanation for the non-detection is that the companion simply does not host a circumsubstellar disk.
In the case of PDS~70~b we do not reach a very high sensitivity and placed a 68.27\% upper limit of 5.0\% on the degree of polarization in $K_s$-band.
Therefore, we can conclude that if PDS~70~b hosts a disk, the inclination is probably not so high that it completely obscures the planet's atmosphere.
Because we only detected polarization for companions with hydrogen emission lines, it seems that these lines are the best non-polarimetric indicators for the existence of a circumsubstellar disk.

\subsection{Atmospheric asymmetries of the companions}
\label{sec:discussion_atmospheric_asymmetries}

Of the 18 companions for which we do not detect significant polarization, 14 show no clear evidence of hosting a circumsubstellar disk (see Fig.~\ref{fig:companions_properties} and Table~\ref{tab:companions_properties}).
Because the majority of the companions have late-M to mid-L spectral types and are therefore expected to have dusty atmospheres, we could expect to detect polarization due to rotation-induced oblateness or an inhomogeneous cloud distribution.
Indeed, polarization between several tenths of a percent to a percent has been detected at near-infrared wavelengths (in particular in $J$-, $Z$-, and $I$-band) for more than a dozen late-M to mid-L field brown dwarfs~\citep{milespaez_bdpol1, milespaez_bdpol2}. 
In our survey, we reached sensitivities (upper limits) ${\leq}0.3\%$ for 11 companions (see Sect.~\ref{sec:upper_limits} and Table~\ref{tab:upper_limits}), and so we might have expected to detect polarization for a few of the companions.
Because we do not detect polarization due to atmospheric asymmetries for any of the companions, these asymmetries either do not exists for the companions observed or they produce polarization below the sensitivity reached.

In the majority of cases, the polarization of field brown dwarfs is interpreted to be caused by rotation-induced oblateness. 
In that sense our non-detections are particularly surprising because the companions observed are generally young and have a low surface gravity (see Table~\ref{tab:companions_properties}), which should result in a more oblate atmosphere for a given rotation rate and therefore more polarization~\citep{sengupta_bdpol2, marley_exopol}.
It is important to note, however, that the field brown dwarfs observed by \citet{milespaez_bdpol1} are old (ages 0.5--5~Gyr) and have measured projected rotational velocities $v \sin{i} > 30$~km~s$^{-1}$.
Indeed, in their sample of several dozen field brown dwarfs, \citet{zapatero_bdrotation} found that about half of the very young field brown dwarfs (1--10~Myr) have $v \sin{i} \leq 10$~km~s$^{-1}$ whereas all old brown dwarfs ($\geq 1$~Gyr) have $v \sin{i} \approx 30$~km~s$^{-1}$.
Very young brown dwarfs rotate slowly because they are still in the process of spinning up as they cool and contract.
Looking at Fig.~\ref{fig:companions_properties}, we can divide our sample roughly into a large group of young, high-temperature companions with late-M to mid-L spectral types, and a smaller group of older, lower-temperature companions of mid-L to T spectral types.
A possible explanation for our non-detections is that while the companions of the first group may have dusty atmospheres, they rotate too slowly to produce a measurable level of polarization. 
And on the other hand the second group may rotate faster, but due to their later spectral types their upper atmospheres may lack the scattering dust to produce polarization~\citep{allard_bddust, sengupta_tdwarfpol}.
A more in-depth analysis of the degrees of polarization produced due to rotation-induced oblateness is presented in~\citet{jensenclem_exopol2}.

There could also be other explanations for our non-detections.
It could be that the dust grains in the upper atmosphere are submicron sized, as also suggested by studies of the emission spectra of (field) brown dwarfs and planets~\citep{hiranaka_bddust, bonnefoy_hr8799}, and that we therefore need to observe at shorter wavelengths than $H$-band (i.e.,~in $Y$- or $J$-band). 
\citet{milespaez_bdpol2} observed one of the field brown dwarfs in $Z$-, $J$-, and $H$-band and found that the degree of polarization decreases strongly with increasing wavelength, with the maximum polarization in $Z$-band and no detection in $H$-band.
Another possibility, as suggested by \citet{milespaez_bdpol2}, is that the low-gravity atmospheres of young objects have thicker dust clouds, resulting in strong multiple scattering and therefore a low integrated degree of polarization.
Finally, our non-detections may also indicate that the atmospheric dust clouds are homogeneously distributed, or that the inhomogeneities do not produce a measurable degree of polarization.
Indeed, \citet{millar_luhman16} recently detected polarization that is likely due to cloud banding on Luhman~16~A, but the measured degree of polarization is only 0.03\% in $H$-band.

\subsection{Confirmation and further characterization of the disks of DH~Tau~B and GSC~6214~B}
\label{sec:outlook}

To confirm that the polarization from DH~Tau~B and GSC~6214~B is truly intrinsic and not caused by interstellar dust, we should perform follow-up observations.
For this we can use the recently implemented star-hopping technique for SPHERE-IRDIS to quasi-simultaneously measure the stellar polarization from nearby diskless reference stars.
As a reference for DH~Tau we can observe DI~Tau, a very close neighbor to DH~Tau located at a separation of only $15.3\arcsec$ and at the same distance from Earth, and whose spectral-energy distribution~\citep[classified as class~III;][]{luhman_taurus} and very low mass-accretion rate~\citep{alonsomartinez_taurus} indicate it very likely does not host a circumstellar disk that creates significant intrinsic stellar polarization.
For GSC~6214 we can observe BD-20~4481, which is of similar spectral type as GSC~6214~A and is located at a separation of $13.3\arcmin$ and at a distance of 113~pc (compared to 109~pc for GSC~6214).
We can use the measurements of the stellar polarization of the reference stars to subtract the interstellar component of the companions' polarization, and with that accurately determine the intrinsic polarization of the companions.
We can also measure the polarization of DH~Tau~A and GSC~6214~A with a different instrument than SPHERE, for example with the WIRC+Pol near-infrared spectropolarimeter~\citep{tinyanont_wircpol} on the Hale Telescope at Palomar Observatory.
Using WIRC+Pol we can determine the stellar polarization as a function of wavelength, enabling us to quantify the interstellar polarization by comparing the measurements with Serkowski's law of interstellar polarization.

We can further characterize the circumsubstellar disks of DH~Tau~B and GSC~6214~B, as well as the companions themselves, by performing measurements with various current and future instruments.
We can perform IRDIS polarimetric measurements in multiple filters to constrain the distribution of particle sizes in the disks.
By combining these measurements with optical and near-infrared medium-resolution spectroscopy using MUSE and ERIS on the VLT, we can constrain the fundamental parameters of the companions.
If we are able to detect the disks with ALMA, we can derive their dust mass from the continuum emission, their gas mass from the CO spectral-line emission, and their effective temperature from the emission in two different wavelength bands.  
Similar to ALMA observations, the dust mass and effective temperature of the disk can be determined from mid-infrared photometric and spectroscopic observations, for example with MIRI on board the James Webb Space Telescope (JWST), METIS on the Extremely Large Telescope (ELT), or even VISIR on the VLT.
Finally, with the sensitivity of MIRI and METIS we could detect silicate emission features at 10 and \SI{18}{\micro\meter}.

%
%

\section{Summary and conclusions}
\label{sec:conclusions}

We measured the near-infrared linear polarization of 20 young planets and brown dwarf companions using SPHERE-IRDIS.
We reduced the data using the IRDAP pipeline to correct for the instrumental polarization and crosstalk of the optical system with an absolute polarimetric accuracy ${<}0.1\%$ in the degree of polarization.
To retrieve the polarization of the companions, we employed a combination of aperture photometry, ADI, and PSF fitting.
We achieved a best $1\sigma$ polarimetric contrast of $3\cdot10^{-8}$ at an angular separation of $0.5\arcsec$ from the star and a contrast ${<}10^{-8}$ for separations ${>}2.0\arcsec$.

We report the first detection of polarization originating from substellar companions, with a measured degree of polarization of several tenths of a percent for DH~Tau~B and GSC~6214~B in $H$-band.
By comparing the measured polarization with that of nearby stars, we find that this polarization is unlikely to be caused by interstellar dust. 
Because the companions have previously measured hydrogen emission lines and red colors, we conclude that the polarization most likely originates from circumsubstellar accretion disks.
Through radiative transfer modeling we constrain the position angles of the disks and find that the disks must have high inclinations to produce these measurable levels of polarization.
For GQ~Lup~B, which has stronger hydrogen emission lines than DH~Tau~B and GSC~6214~B, we do not measure significant polarization.
This implies that if GQ~Lup~B hosts a disk, this disk has a low inclination. 
Assuming that the spin axes of the companions are perpendicular to the plane of their disks, we use previously measured projected rotational velocities to constrain the rotational periods of DH~Tau~B, GSC~6214~B, and GQ~Lup~B to be \mbox{29--37~h}, \mbox{22--77~h}, and \mbox{19--64~h}, respectively, within the 68\% confidence interval.
Finally, we find 1RXS~J1609~B to be marginally polarized by interstellar dust, which suggests that the red colors and extinction that are thought to indicate the presence of a disk are more likely caused by interstellar dust.

The disk of DH~Tau~B, and possibly that of GQ~Lup~B, are misaligned with the disks around the primary stars.
These misalignments suggest that these wide-separation companions have formed in situ through direct collapse in the molecular cloud, although formation through gravitational instabilities in the circumstellar disk cannot be excluded.
Formation at close separations from the star followed by scattering to a higher orbit is unlikely because dynamical encounters with other bodies would most likely severely disturb or even destroy the circumsubstellar disks.

For 18 companions we do not detect significant polarization and place upper limits of typically ${<}0.3\%$ on their degree of polarization.
These non-detections may indicate that young companions rotate too slowly to produce measurable polarization due to rotation-induced oblateness, or that any inhomogeneities in the atmospheric clouds are limited.
Another possibility is that the upper atmospheres of the companions contain primarily submicron-sized dust grains.
This implies that we should perform future measurements in $Y$- or $J$-band, although these bands are more challenging in terms of companion-to-star contrast and contrast performance of the instrument.

In our survey, we also detected the circumstellar disks of DH~Tau, GQ~Lup, PDS~70, $\beta$~Pic, and HD~106906, which for DH~Tau and GQ~Lup are the first disk detections in scattered light.
The disk of DH~Tau is compact and has a strong brightness asymmetry that may reveal the forward- and backward-scattering sides of the disk or may be caused by shadowing by an unresolved inner disk component.
The disk of GQ~Lup shows a pronounced asymmetry and two spiral-like features that could be the result of periodic close passes of GQ~Lup~B.
The PDS~70 disk shows significant non-azimuthal polarization indicating multiple scattering.
We also detect one or two weak spiral-like features protruding from the ansae of the disk that may be the result of two spiral arms in the outer disk ring, potentially induced by PDS~70~b.

Our measurements of the polarization of companions are reaching the limits of the instrument and the data-processing techniques.
We find that incompletely corrected bad pixels can cause systematic errors of several tenths of a percent in the measured polarization. 
To minimize this effect, we recommend to use the field-tracking mode without dithering for future observations that aim to measure the polarization of companions.
However, for companions at close separations or with large star-to-companion contrasts, pupil-tracking observations are still preferred to retrieve the companions' total intensity with ADI.
These close-in companions can alternatively be observed in field-tracking mode when using the recently implemented star-hopping technique to enable reference star differential imaging.
We also find that the measurements of the stellar polarization are affected by systematic errors related to the use of the coronagraph in combination with time-varying atmospheric conditions.
We therefore recommend to take additional stellar polarization measurements without coronagraph.

To further characterize the circumsubstellar disks of DH~Tau~B and GSC~6214~B, as well as the companions themselves, we can perform follow-up observations with SPHERE-IRDIS, ALMA, JWST-MIRI, MUSE and ERIS on the VLT, and METIS on the ELT.
Our polarimetric detections of the disks of DH~Tau~B and GSC~6214~B are a first step in building a complete picture of the companions, their formation, and evolution, and pave the way to detecting polarization of young planets with for example SPHERE+~\citep{boccaletti_sphere+} and the future planet-characterization instrument EPICS (or PCS) on the ELT.

\begin{acknowledgements}
RGvH thanks ESO for the studentship at ESO Santiago during which part of this project was performed.
TS acknowledges the support from the ETH Zurich Postdoctoral Fellowship Program.
The research of FS, JdB, and AJB leading to these results has received funding from the European Research Council under ERC Starting Grant agreement 678194 (FALCONER).
CP acknowledges financial support from Fondecyt (grant 3190691) and from the ICM (Iniciativa Cient\'ifica Milenio) via the N\'ucleo Milenio  de  Formaci\'on Planetaria grant, from the Universidad de Valpara\'iso.
This research has made use of the SIMBAD database, operated at CDS, Strasbourg, France; NASA’s Astrophysics Data System Bibliographic Services; Scipy, a free and open-source Python library used for scientific computing and technical computing \citep{2020SciPy-NMeth}; and Astropy, a community-developed core Python package for Astronomy \citep{astropy:2013, astropy:2018}.

\end{acknowledgements}

%
%

\bibliographystyle{aa}   
\bibliography{biblio}    

%
%

\begin{appendix}
\section{Cosmetic correction of spurious structure in $Q$- and $U$-images}
\label{app:yinyang}

\begin{figure}[!htbp]
\centering
\includegraphics[width=\hsize, trim={20 10 10 5}, clip]{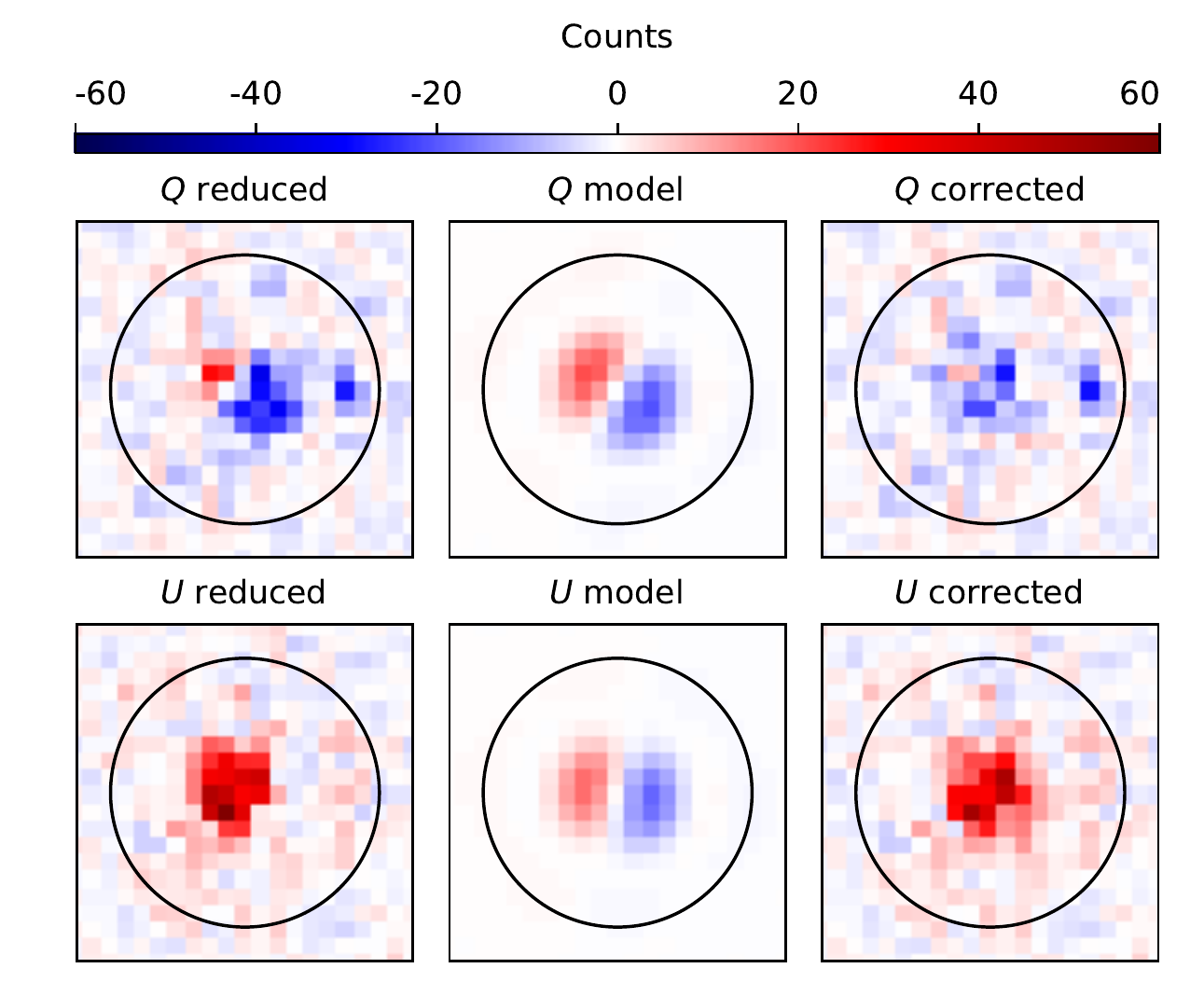}
\caption{Reduced $Q$- and $U$-images (first column), model images of the spurious structures in $Q$ and $U$ (second column), and $Q$- and $U$-images corrected for the spurious structures (third column) at the position of the companion DH~Tau~B of the \mbox{2019-10-24} data set of DH~Tau. An aperture of radius 8~pixels centered on the companion is shown superimposed on the images.}
\label{fig:dh_tau_signals_yinyang} 
\end{figure}

If a companion is polarized, we would expect the polarization signals in the $Q$- and $U$-images to resemble scaled-down positive or negative versions of the corresponding total-intensity images $I_Q$ and $I_U$.
However, for many data sets the $Q$- and $U$-images show spurious structure with adjacent positive and negative signals.
For example, for the \mbox{2019-10-24} data set of DH~Tau, as shown in Fig.~\ref{fig:dh_tau_signals_yinyang} (first column), we see that the $Q$-image contains positive and negative signals and that the signal in $U$ is offset from the center coordinates of the companion. 
These spurious structures result from imperfect relative centering of the images of IRDIS' left and right optical channels and image motion during the observations.
For pupil-tracking observations the spurious structures can additionally originate from image rotation between the two measurements of the double difference.
%

In the case these spurious structures are visible in the $Q$- and $U$-images of a data set, we make to each image individually a (cosmetic) correction similar to that described in~\citet{snik_solarpol}.
For this we retrieve a positive and negative copy of the $I_Q$- or $I_U$-image at the companion position and create a model image in which the two copies are symmetrically shifted in opposite directions from the center coordinates of the companion.
We then subtract this model image from the $Q$- or $U$-image and fit the shifts in the $x$- and $y$-directions by minimizing the sum of squared residuals in an aperture of radius 8~pixels in the resulting image. 
Because the aperture sum in the model images is zero, subtracting these images only suppresses the spurious structures and does not alter the net polarization signals in $Q$ and $U$.

For the data set of DH Tau, we find a total relative shift equal to 0.015~pixels for $Q$ and 0.013~pixels for $U$.
Only small relative shifts are needed because the maximum values of the total-intensity PSFs are more than 100 times larger than the maximum values of the positive and negative signals of the spurious structure.
The model images and the corrected $Q$ and $U$-images are shown in Fig.~\ref{fig:dh_tau_signals_yinyang} (second and third column).
The spurious structure has clearly disappeared in the corrected images, with the $Q$-image only having negative signal and the signal in $U$ being positioned at the companion's center coordinates.

\section{Systematic errors due to bad pixels}
\label{app:bad_pixels}

A few percent of the pixels of the IRDIS Hawaii 2RG detector are bad, that is, they are dead, nonlinearly responding, or hot pixels. When preprocessing the raw frames with IRDAP, bad pixels are identified with a bad pixel map followed by sigma-filtering and then replaced by the median value of the surrounding pixels. These data-reduction steps correct the majority of the bad pixels, but some bad pixels remain uncorrected or are replaced by a value that is not accurate. This results in systematic errors of the pixel values. Whereas these small errors are not a real problem for photometry of point sources in total intensity or imaging of circumstellar disks in polarized light, they become quite problematic when trying to measure the polarization of point sources at a level as low as a few tenths of a percent of the total intensity.

Incomplete correction of bad pixels only marginally affects data taken in field-tracking mode. This is because in field-tracking mode the PSF of the companion is approximately stationary on the detector and only moves very slightly due to variations in AO performance. The bad pixels, which are at a fixed position on the detector, are therefore replaced by approximately the same (median) value in consecutive frames, and so are strongly suppressed when computing the double difference. In addition, any uncorrected or incompletely corrected bad pixels that remain are further averaged out over the various HWP cycles. However, this averaging over HWP cycles only partially applies for our data because we generally observed in field-tracking mode with dithering in which case the detector moves by one to a few pixels each HWP cycle. We note that for total-intensity imaging, for which we compute the median over many exposures rather than differences of exposures, dithering does help suppress bad pixels.

Data taken in pupil-tracking mode are typically more strongly affected by incomplete correction of bad pixels. In pupil-tracking observations the companion moves over the detector, and so in each frame the bad pixels are at a different location with respect to the companion PSF. Therefore, the bad pixels are replaced by very different median values, and relatively large systematic errors remain after the double difference. For data sets with fast parallactic rotation (e.g., the data sets of GSC~8047 and TYC~8998), the bad pixels are more problematic than for data sets with only little rotation (e.g., GSC~6214). For data sets with many HWP cycles the bad pixel effect averages out somewhat, but the systematic errors are still much larger than for field-tracking data.

We attempted to remove the systematic errors due to bad pixels by creating a more aggressive bad pixel map from the dark and flat frames, performing aggressive sigma-filtering, locally replacing the bad-pixel values with cubic spline interpolation rather than with the median filter, and computing the median over the Mueller-matrix-model-corrected $Q$- and $U$-images of each HWP cycle. Unfortunately, we were not able to identify all bad pixels in the data and completely remove the effect. This is primarily because part of the bad pixels cause systematic errors of only several percent or less of the total intensity. Such small deviations from the true value are practically impossible to detect in the images and only become evident when computing differences of images as we do in polarimetry. 

Although we were not able to completely correct for the bad pixels, we can mitigate their effect by excluding those frames that contribute strong bad pixels to the final images or that show bad pixels at the position of the companion in the bad pixel map. In addition we can average out the systematic error due the bad pixels by mean-combining several data sets. We can also use large apertures to perform the photometry with, such that the bad pixels values (which are both positive and negative in polarimetric images) average out and sum to a lower spurious signal. Future observations aimed at measuring the polarization of companions should preferably be performed in field-tracking mode without dithering.

\section{Retrieval of total intensity through ADI: Upper limit on polarization of $\beta$~Pic~b}
\label{sec:method_beta_pic}

\begin{figure}[!htbp]
\centering
\includegraphics[width=\hsize, trim={7 7 7 20}, clip]{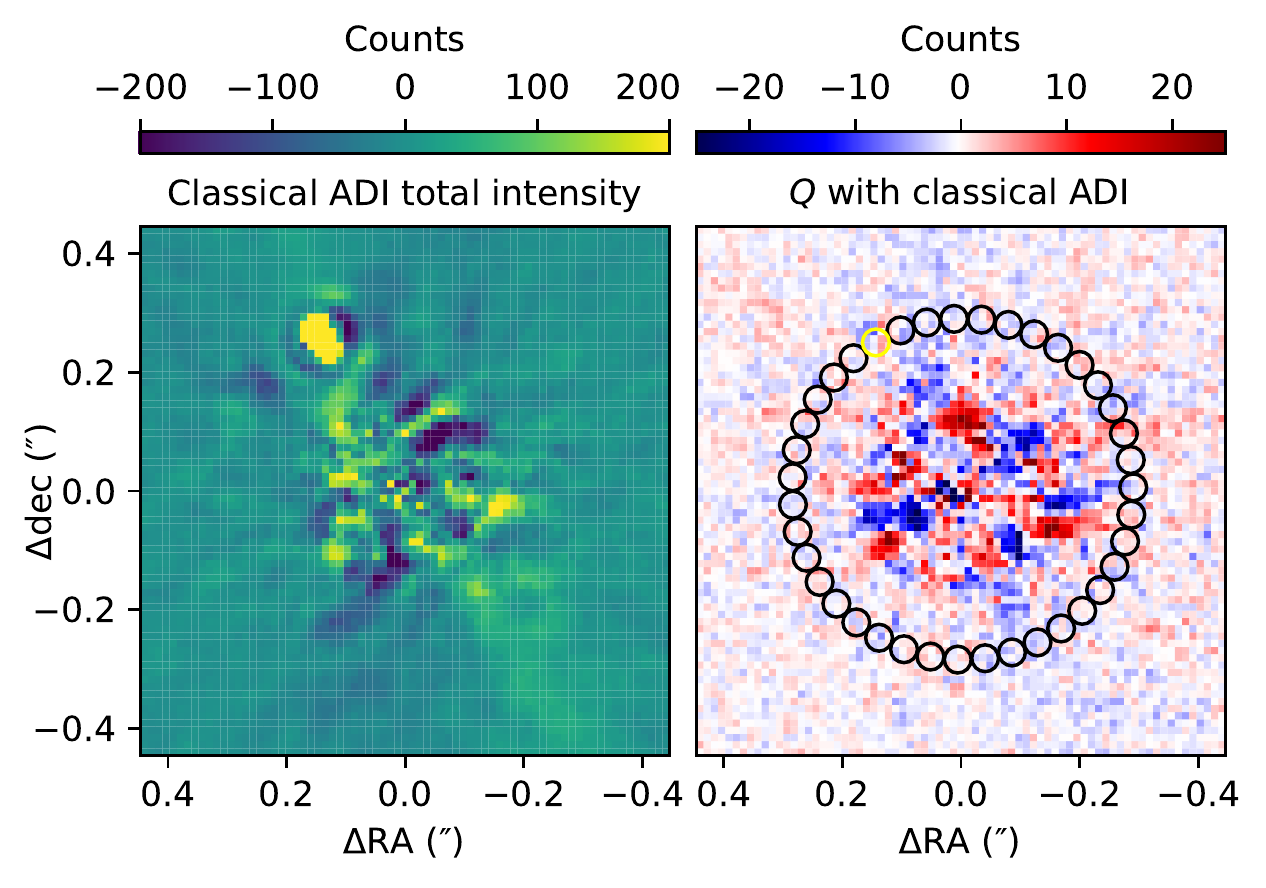}
\caption{Reduced images of the \mbox{2019-11-26} data set of $\beta$~Pic. Left: Total-intensity image after applying classical ADI with IRDAP to reveal the companion $\beta$~Pic~b. Right: $Q$-image after combining polarimetry with classical ADI, showing the aperture of radius 1.85~pixels at the position of the companion (yellow) and the ring of comparison apertures of the same radius around the star (black).}
\label{fig:beta_pic_apertures}
\end{figure}

Companions at small separations or at large star-to-companion contrasts are swamped in the halo of starlight in total intensity.
For data sets that were taken in pupil-tracking mode and have sufficient parallactic rotation, we have therefore slightly adapted the method described in Sect.~\ref{sec:method_dh_tau} and determine the probability distribution of the total intensity of the companion by performing ADI with negative PSF injection.
We still determine the distributions of $Q$ and $U$ using aperture photometry because the stellar speckle halo is almost completely removed in the polarimetric data-reduction steps, in particular for the reductions with the added classical ADI step (see Sect.~\ref{sec:data_reduction}).
We applied this adapted method to the data sets of HR~8799, HD~206893, PDS~70, and $\beta$~Pic.
In this section we demonstrate the method with the \mbox{2019-11-26} $H$-band data set of $\beta$~Pic and show how we calculate upper limits on the degree of polarization of the companion $\beta$~Pic~b.
A total-intensity image of the data after applying classical ADI with IRDAP is shown in Fig.~\ref{fig:beta_pic_apertures} (left).
%

To perform the ADI with negative PSF injection, we use the \texttt{PynPoint}\footnote{\url{https://pynpoint.readthedocs.io}} pipeline, version 0.8.2~\citep{amara_pynpoint, stolker_pynpoint}, and closely follow the steps described in~\citet{stolker_miracles}. 
In short, we fetch the preprocessed science frames 
and the 
stellar PSF image from the reduction with IRDAP in Sect.~\ref{sec:data_reduction}.
Subsequently, we iteratively subtract scaled copies of the stellar PSF from the preprocessed 
science frames at the approximate position of the companion and apply ADI with PCA (in this case subtracting three principal components) to minimize the residuals at that same location.
Using Markov chain Monte Carlo (MCMC) we then sample the posterior distributions of the companion's angular separation, position angle and contrast with respect to the star.
We take the median of the posterior distribution of the contrast as the final contrast value and determine the corresponding statistical uncertainties from the 16th and 84th percentiles.
We also estimate the systematic uncertainty on the contrast by injecting fake companions at various position angles (but the same separation and contrast as the real companion), retrieving them, and computing the distribution of the difference between the retrieved and injected contrasts.
This systematic uncertainty accounts for the azimuthal variations of the noise around the central star and is generally 1 to 5 times larger than the statistical uncertainty~\citep[similar to the results of][]{wertz_hr8799}.
Finally, we compute the overall uncertainty as the quadratic sum of the statistical and systematic uncertainties. 

\begin{figure}[!htbp]
\centering
\subfloat{\includegraphics[width=0.547\hsize, trim={436 5 10 5}, clip]{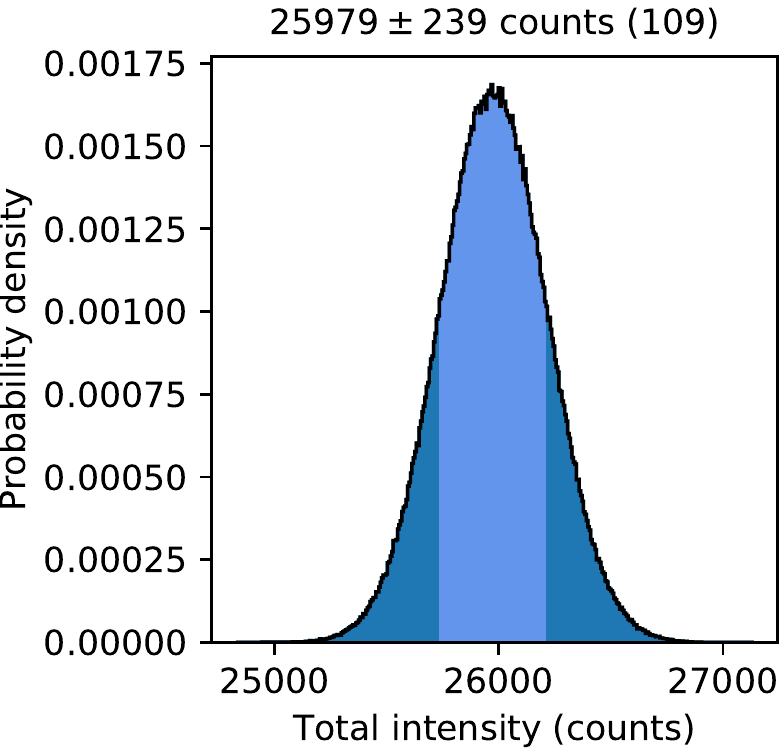}}
\subfloat{\includegraphics[width=0.453\hsize, trim={460 5 215 240}, clip]{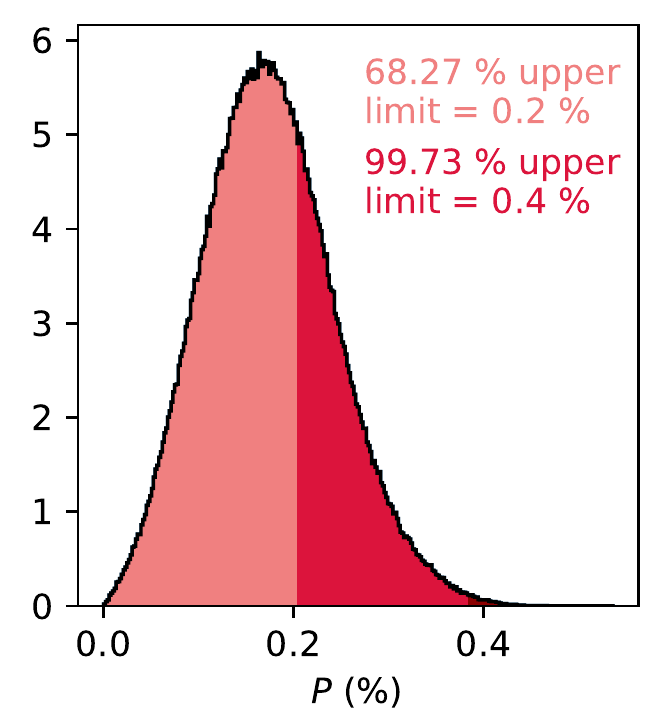}}
\caption{Final probability distributions of the signal of $\beta$~Pic~b from the \mbox{2019-11-26} data set of $\beta$~Pic, using an aperture radius of 1.85~pixels.
Left: Probability distribution of the total intensity. The mean and 68.27\% uncertainties of the distribution are shown above the graph, with the latter also indicated by the light-blue shaded area. The S/N is shown within parentheses. 
Right: Probability distribution of the degree of linear polarization. The upper limits computed from the one-sided 68.27\% and 99.73\% intervals are indicated by the light-red and darker red shaded areas, respectively.}
\label{fig:beta_pic_polarization_distributions}
\end{figure}

After these steps, we determine the probability distribution of the companion's total intensity (expressed in counts) for a range of aperture radii from 1 to 10 pixels.
To this end, we draw $10^6$ samples from a Gaussian distribution with the mean and standard deviation equal to the companion-to-star contrast and uncertainty we retrieved with PynPoint.
We then sum the flux in the stellar PSF image using an aperture of the given radius and multiply the Gaussian samples by this summed flux.
The resulting total-intensity distribution of the companion, which we use for both $I_Q$ and $I_U$, is shown in Fig.~\ref{fig:beta_pic_polarization_distributions} (left) for an aperture radius of 1.85~pixels.
This radius is the final aperture radius we select at the end of this section and corresponds to half times the full width at half maximum (FWHM) we measure on the stellar PSF. 

To determine the probability distributions in $Q$ and $U$, we use the images from the reduction with the added classical ADI step.
In the case of $\beta$~Pic, the classical ADI step does not only further suppress the speckle noise, but also removes most of the signal from the nearly edge-on-viewed circumstellar disk that crosses the position of $\beta$~Pic~b (see Fig.~\ref{fig:circumstellar_disks}, center and bottom).
The classical ADI step suppresses the disk signal because the disk is broad and the parallactic rotation of the observations in only $19.8^\circ$.
Indeed, as can be seen in Fig.~\ref{fig:beta_pic_apertures} (right) for the $Q$-image, the disk is almost completely removed and there are only few speckles left at the separation of the companion.
Any polarization originating from the companion should still be visible in the $Q$- and $U$-images because point sources are much less affected by ADI-induced self-subtraction.

We analyze the $Q$- and $U$-images by following the exact steps as described in Sect.~\ref{sec:method_dh_tau}, but with one exception.
Before performing the aperture photometry, we quantify the throughput of the ADI procedure by performing a simulation in which we inject and retrieve an artificial source at the separation of the actual companion.
We then correct the $Q$- and $U$-images for the self-subtraction by dividing them by the calculated throughput, which for this data set is 49\%.
After performing all the steps, 
we determine the companion’s polarization for each of the defined aperture radii (as in Fig.~\ref{fig:dh_tau_polarization_snr_radius} for DH~Tau~B).
Contrary to the data of DH~Tau, for this data set of $\beta$~Pic we detect no signals with an S/N higher than 0.9 in $Q$ and 1.9 in $U$ for any aperture radius.
Indeed, the reduced $Q$-image (see Fig.~\ref{fig:beta_pic_apertures}, right) and $U$-image contain only noise at the position of the companion.
We therefore conclude that we do not detect significant polarization originating from the companion $\beta$~Pic~b. 

\begin{figure}[!htbp]
\centering
\includegraphics[width=\hsize, trim={5 5 5 5}, clip]{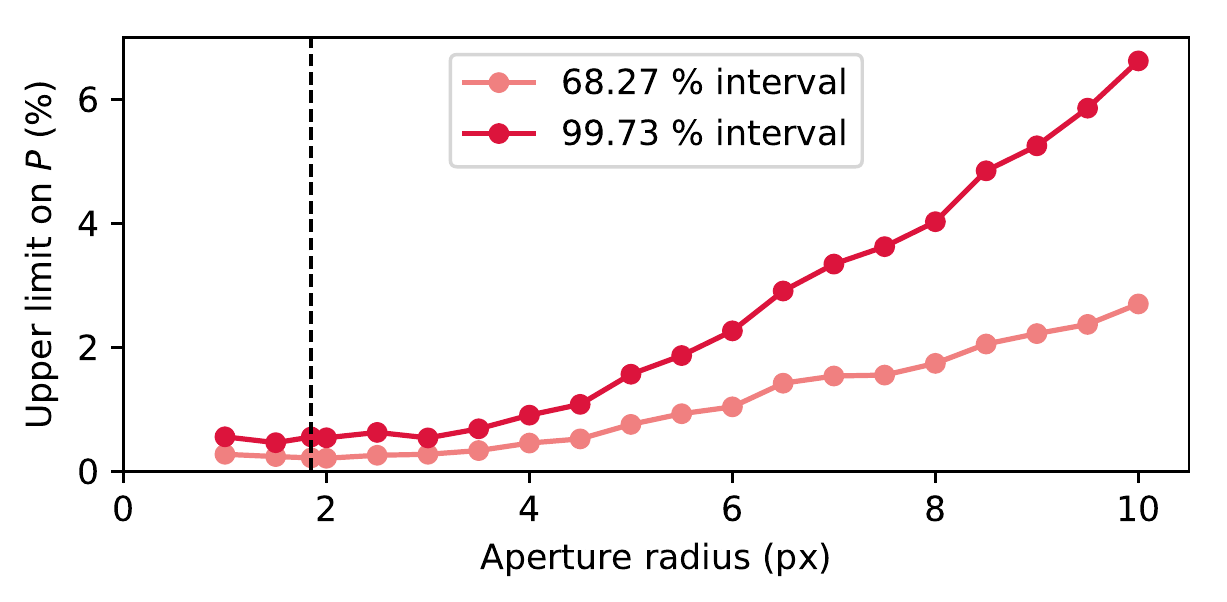}
\caption{Upper limits on the degree of linear polarization of $\beta$~Pic~b computed from the one-sided 68.27\% and 99.73\% intervals as a function of aperture radius from the \mbox{2019-11-26} data set of $\beta$~Pic. The final selected aperture radius of 1.85 pixels, equal to half times the FWHM of the stellar PSF, is indicated with the dashed vertical line.}
\label{fig:beta_pic_upper_limit_radius}
\end{figure}

We now proceed to set limits on the degree of polarization of the companion.
To this end, we determine, for each defined aperture radius, two upper limits from the probability distribution of the degree of polarization. 
We compute these upper limits from the $68.27\%$ and $99.73\%$ intervals, which for a Gaussian distribution would correspond to the $1\sigma$ and $3\sigma$ confidence intervals, respectively.
These intervals are calculated one-sided and starting at zero because the degree of linear polarization is computed as $P = \surd(q^2 + u^2)$ and therefore can only have positive values~\citep[see][]{sparks_polarimetry}.
Figure~\ref{fig:beta_pic_polarization_distributions} (right) shows the distribution of the degree of polarization for an aperture radius of 1.85~pixels with the two intervals indicated.
Figure~\ref{fig:beta_pic_upper_limit_radius} shows the two upper limits as a function of aperture radius.
From this figure we see that the upper limits are relatively constant for an aperture radius smaller than approximately 3.5~pixels.
For larger apertures, the upper limits increase as more noise is included in the companion aperture and the uncertainty of the background due to the low number of background samples increases.
We select our final aperture radius to be 1.85~pixels, equal to half times the FWHM of the stellar PSF, and conclude that the 68.27\% and 99.73\% upper limits on the degree of polarization of $\beta$~Pic~b are equal to 0.2\% and 0.4\%, respectively. 
We note that while for this particular data set the upper limits monotonically increase with aperture radius, for several other data sets this is not the case due to incompletely removed bad pixels (see Sect.~\ref{sec:upper_limits} and Appendix~\ref{app:bad_pixels}).
%
%


\section{Retrieval of total intensity through PSF fitting: Upper limit on polarization of HD~19467~B}
\label{sec:method_hd_19467}

\begin{figure}[!htbp]
\centering
\includegraphics[width=\hsize, trim={5 5 5 5}, clip]{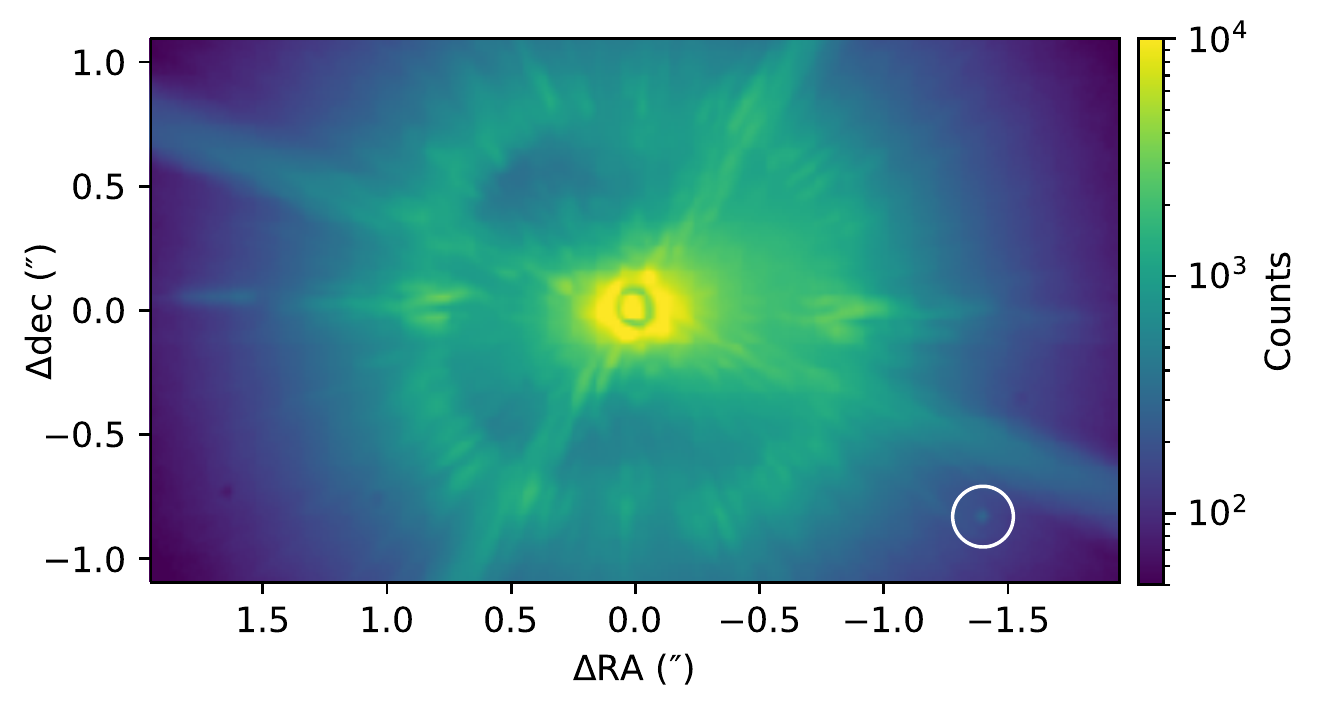}
\caption{Reduced $I_Q$-image of the \mbox{2018-08-07} data set of HD~19467, showing the position of the companion HD~19467~B in the white circle. The asymmetric wind-driven halo and the stellar diffraction spikes are clearly visible.}
\label{fig:hd_19467_I_Q_companion}
\end{figure}

Several observations of faint or close-in companions were not executed in pupil-tracking mode (i.e.,~they were executed in field-tracking mode) or have little parallactic rotation.
For these observations we cannot retrieve the probability distribution of the companion's total intensity through ADI with negative PSF injection as described in Appendix~\ref{sec:method_beta_pic}.
We also cannot use aperture photometry with comparison apertures as outlined in Sect.~\ref{sec:method_dh_tau} because the spatially varying stellar halo at the separations of these companions prevents accurate determination of the background.
We therefore use MCMC to fit the stellar PSF image to the $I_Q$- and $I_U$-images at the companion position and determine the corresponding probability distributions.
We applied this method to the data sets of PZ~Tel, HR~7682, HD~19467, GQ~Lup, and HD~4747.
To confirm that the PSF fitting is accurate, we also used the method to retrieve the total intensities of HR~8799~b, c, and d, and find that the results differ only 0.03 to 0.07~mag with those obtained with PynPoint (see Appendix~\ref{sec:method_beta_pic}).
In this section we demonstrate the PSF fitting method with the \mbox{2018-08-07} $H$-band data set of HD~19467 
and set upper limits on the degree of polarization of the companion HD~19467~B.
Figure~\ref{fig:hd_19467_I_Q_companion} shows the $I_Q$-image of this data set. 
%
%

As the first step in our analysis, we obtain a rough estimate of the companion’s contrast and $x$- and $y$-coordinates.
To this end, we fit a model consisting of a 2D Moffat function and an inclined plane to the reduced $I_Q$-image at the companion position.
The inclined plane accounts for the (approximately) linearly varying local background due to the stellar PSF and the stellar diffraction spikes (see Fig.~\ref{fig:hd_19467_I_Q_companion}) and is described by a constant (the $z$-intercept) and slopes in the $x$- and $y$-direction. 
We then fit a model containing the stellar PSF and an inclined plane to cropped versions of the $I_Q$ and $I_U$-images, using the results from the Moffat fit for the initial estimates of the fit parameters.
We use the Nelder-Mead method as implemented in the Python function \texttt{scipy.optimize.minimize} to minimize the sum of squared residuals ($\mathit{SSR}$):
\begin{equation}
\mathit{SSR} = \sum\limits_{i=1}^N \left[\left(I_{Q,i} - \hat{I}_{Q,i}\right)^2 + \left(I_{U,i} - \hat{I}_{U,i}\right)^2\right],
\label{eq:ssr}
\end{equation}
where $I_{Q,i}$ and $I_{U,i}$ are the flux values in the $i$-th pixel of the cropped $I_Q$- and $I_U$-images, $\hat{I}_{Q,i}$ and $\hat{I}_{U,i}$ are the corresponding modeled flux values, and $N$ is the total number of pixels in each of the cropped images.
We minimize the residuals in the $I_Q$- and $I_U$-images simultaneously to obtain a single set of $x$- and $y$-coordinates for the companion position.
For the other parameters we fit separate values for $I_Q$ and $I_U$.

\begin{figure}[!htbp]
\centering
\includegraphics[width=\hsize, trim={10 10 15 16}, clip]{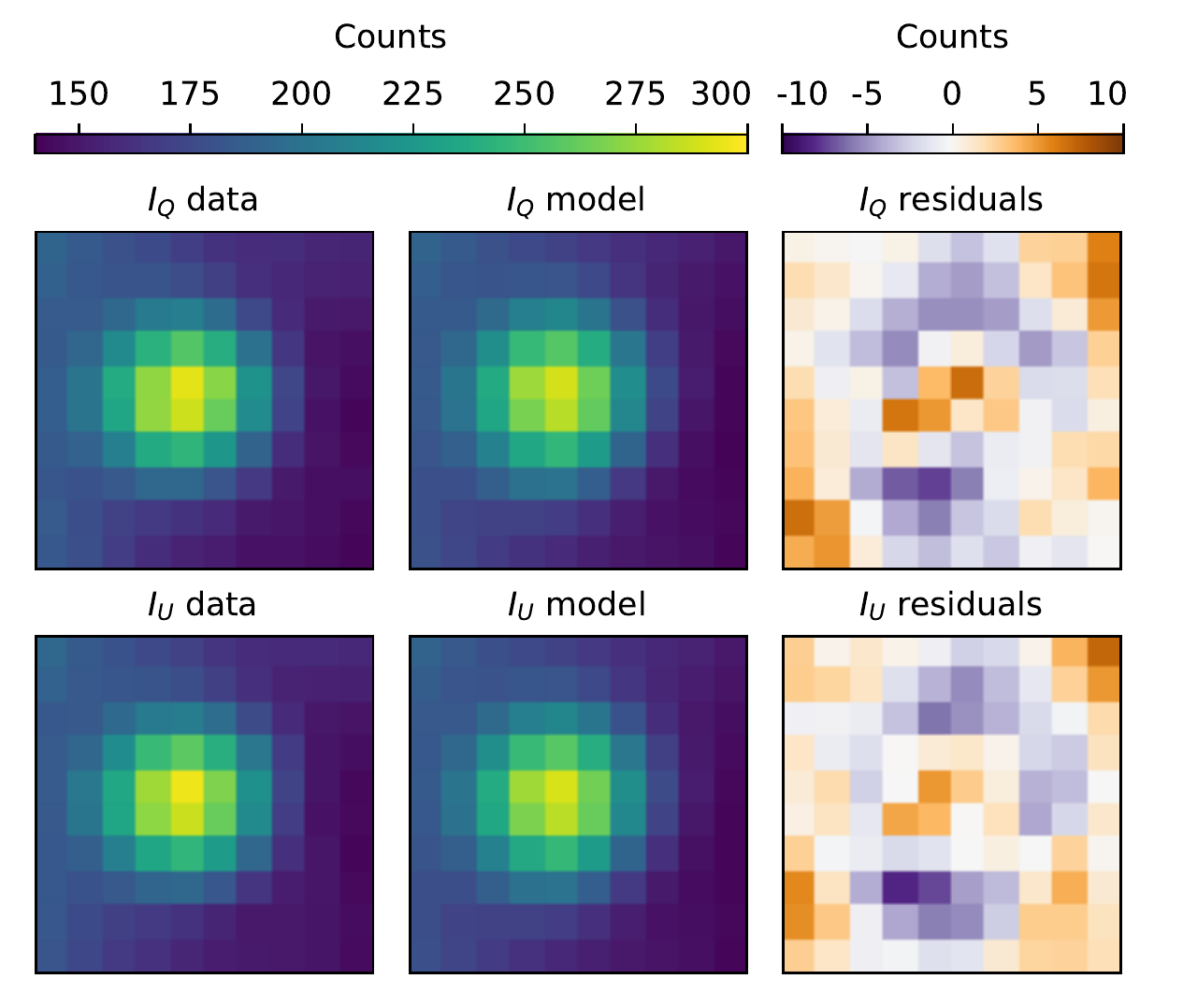}
\caption{Data, best-fit model and residual images of the MCMC fitting of the stellar PSF at position of HD~19467~B to the reduced $I_Q$- and $I_U$-images of the \mbox{2018-08-07} data set of HD~19467.}
\label{fig:hd_19467_mcmc}
\end{figure}

\begin{figure*}[!htbp]
\centering
\includegraphics[width=\hsize]{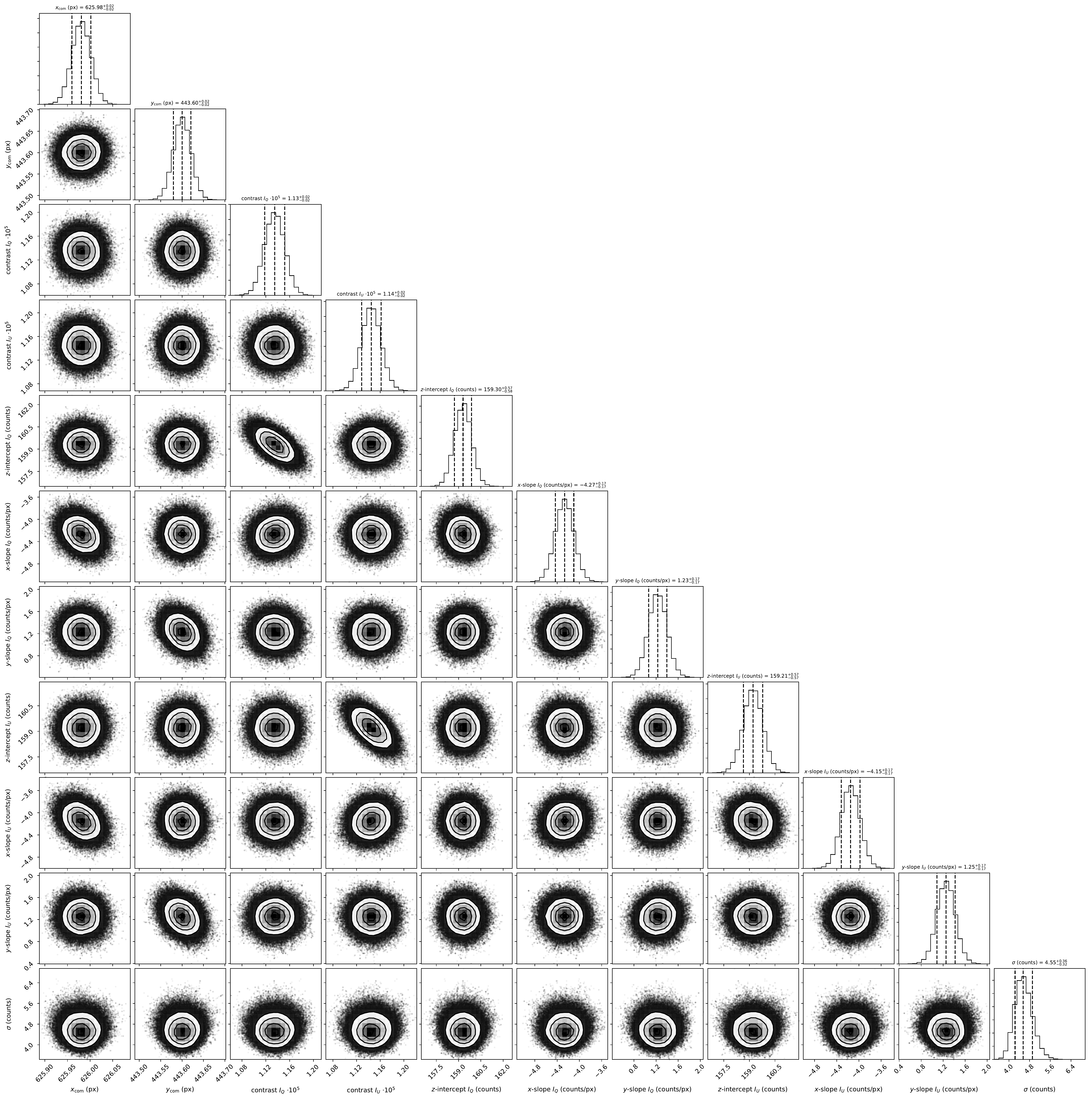}
\caption{Posterior distributions after using MCMC to fit the stellar PSF at the position of the companion HD~19467~B to the reduced $I_Q$- and $I_U$-images of the \mbox{2018-08-07} data set of HD~19467. The fitted parameters are the companion position in $x$ and $y$, the companion-to-star contrast in $I_Q$ and $I_U$, the background's $z$-intercept and slopes in the $x$- and $y$-direction in $I_Q$ and $I_U$, and the standard deviation $\sigma$ that accounts for the noise in the images. The diagonal panels show the marginalized 1D distributions of the fitted parameters and the off-diagonal panels show the 2D projections of the posterior, revealing the covariance of the parameter pairs. The median and uncertainties (computed as the 18th and 84th percentiles) of the distributions are shown above the histograms and are indicated with the dashed vertical lines. The contours superimposed on the off-diagonal panels indicate the $1\sigma$, $2\sigma$ and $3\sigma$ confidence levels assuming Gaussian statistics. The figure is created using the Python package \texttt{corner}~\citep{foremanmackey_corner}.}
\label{fig:hd_19467_corner}
\end{figure*}

We now repeat the PSF fitting using MCMC to obtain the final values of the fit parameters and the corresponding posterior distributions.
We use the MCMC sampler from the Python package \texttt{emcee}~\citep{foremanmacky_emcee} and let 32 walkers explore the probability space with 20,000 steps each (resulting in a total of 640,000~samples).
We randomly generate the starting values of the walkers from Gaussian distributions centered on the best-fit parameter values from our previous fit.
We use a Gaussian distribution for the log-likelihood function:
\begin{equation}
\ln\Lagr \propto -\dfrac{1}{2}\Bigg[N \ln\left(2\pi\sigma^2\right) + \dfrac{\mathit{SSR}}{\sigma^2}\Bigg],
\label{eq:likelihood}
\end{equation}
where $\mathit{SSR}$ is computed from Eq.~(\ref{eq:ssr}) and $\sigma$ is the standard deviation that accounts for the noise in the images.
Because there is no region in the $I_Q$ and $I_U$-images from which we can determine a representative value for $\sigma$, we include it among the parameters to be fitted (i.e.,~we treat $\sigma$ as a nuisance parameter). 
We set the prior for $\sigma$ proportional to $1/\sigma$, that is, Jeffrey's prior, to make sure it is non-informative.
For the other parameters we use uniform priors.
We remove the first 822~steps of each walker, equal to five times the maximum autocorrelation time, and check by visual inspection that the chains of all parameters have converged.
The cropped $I_Q$- and $I_U$-images and the best-fit model and residual images are shown in Fig.~\ref{fig:hd_19467_mcmc}.
Figure~\ref{fig:hd_19467_corner} shows the resulting 1D- and 2D-projections of the posterior distribution of the fitted parameters.
The distributions in Fig.~\ref{fig:hd_19467_corner} are visually very close to being Gaussian and show correlations only between the companion’s contrast in $I_Q$ or $I_U$ and the corresponding $z$-intercept of the background.
%
%
%

We now determine the companion’s probability distributions in $I_Q$ and $I_U$ (expressed in counts) for a range of aperture radii from 1 to 10 pixels. 
Similarly to the method described in Appendix~\ref{sec:method_beta_pic}, we sum the flux in the stellar PSF image using an aperture of the given radius and multiply the MCMC contrast samples in $I_Q$ and $I_U$ by this flux. 
For the remainder of the analysis we follow the steps described in Sects.~\ref{sec:method_dh_tau} and Appendix~\ref{sec:method_beta_pic}, with the only exception that we sample the PDFs in $Q$ and $U$ with the same number of samples as used for the MCMC analysis.
After performing the complete analysis, we detect no signals with an S/N higher than 1.4 in $Q$ and 2 in $U$ for any aperture radius.
Finally, using an aperture radius of 1.86~pixels, equal to half times the FWHM of the stellar PSF, we determine the 68.27\% and 99.73\% upper limits on the degree of polarization of HD~19467~B to be equal to 1.0\% and 2.0\%, respectively.

\section{Contrast curve of $\beta$ Pic data}
\label{app:contrast_curve}

\begin{figure}
\centering
\includegraphics[width=\hsize, trim={5 5 5 5}, clip]{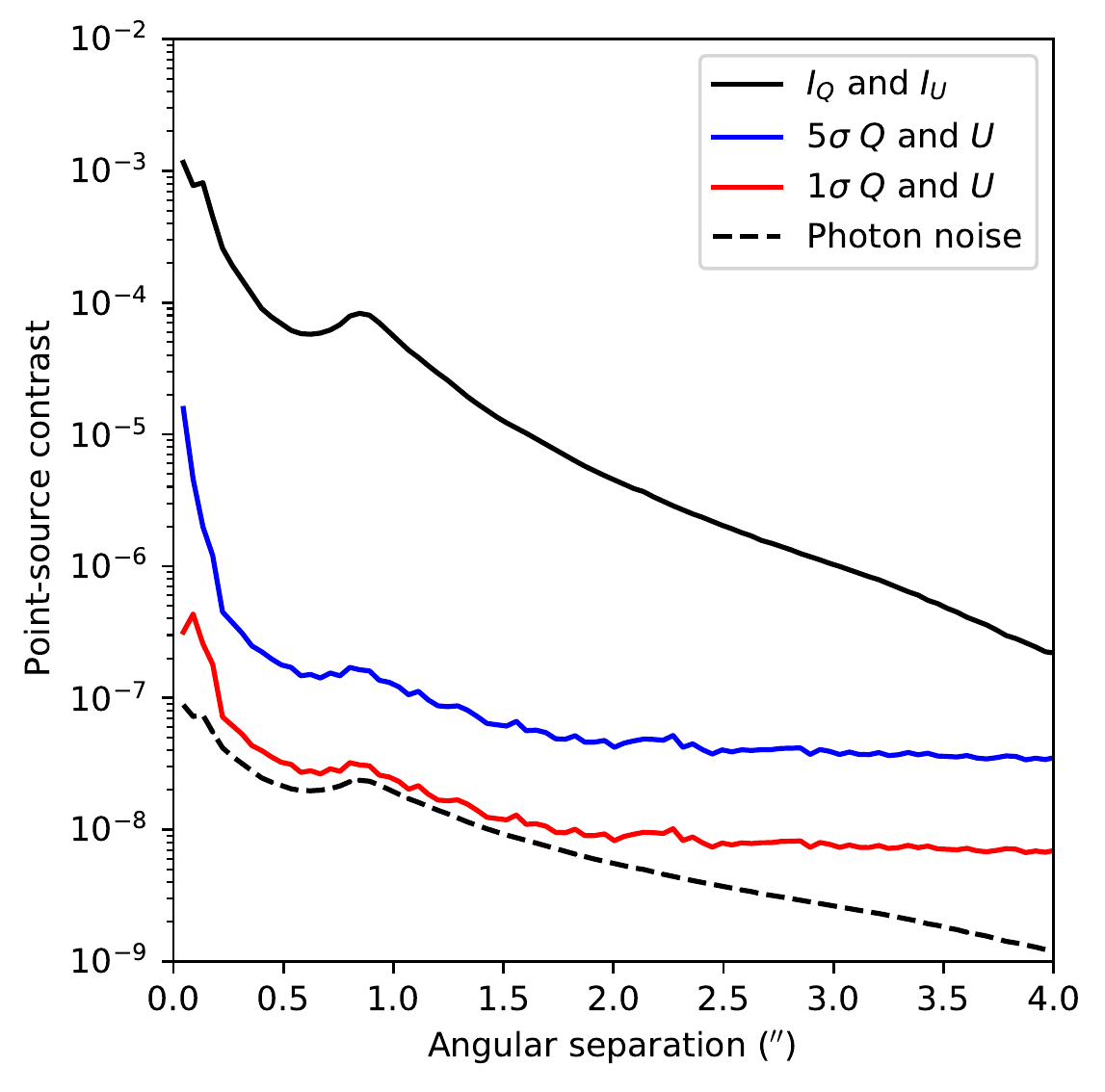}
\caption{$1\sigma$ and $5\sigma$ point-source contrast in $Q$ and $U$ as a function of angular separation from the star for the mean-combined data set of $\beta$~Pic. The azimuthally averaged flux in the total-intensity $I_Q$- and $I_U$-images and the corresponding photon noise are shown for comparison.}
\label{fig:beta_pic_contrast}
\end{figure}

Figure~\ref{fig:beta_pic_contrast} shows the $1\sigma$ and $5\sigma$ point-source contrast in $Q$ and $U$ as a function of angular separation from the star for the mean-combined data set of $\beta$~Pic as constructed with \mbox{IRDAP}.
The curves are computed by summing the flux in rings of apertures around the star, computing the standard deviation over the aperture sums, and normalizing the result with the total stellar flux retrieved from the star flux frames.
At small separations the correction for small-sample statistics is applied~\citep[see][]{mawet_contrast}.
For comparison the figure also shows the azimuthally averaged flux in the total-intensity $I_Q$- and $I_U$-images and the corresponding photon noise. 
At angular separations between ${\sim}0.2\arcsec$ and $2.0\arcsec$ the polarimetric sensitivity is close to the photon-noise limit, with a $1\sigma$-contrast of $7\cdot10^{-8}$ to $1\cdot10^{-8}$ and a $5\sigma$-contrast of $5\cdot10^{-7}$ to $5\cdot10^{-8}$.
At separations larger than $2.0\arcsec$ the sensitivity is limited by read noise or background noise and the $1\sigma$- and $5\sigma$-contrast are ${<}1\cdot10^{-8}$ and ${<}5\cdot10^{-8}$, respectively. 
%

\end{appendix}

\end{document}